\documentclass[11pt]{article}
\usepackage[margin=1in]{geometry}
\usepackage{amsmath,amssymb,amsfonts,amsthm,bbm}
\usepackage{epic,eepic,epsfig,longtable}
\usepackage{multirow,verbatim}
\usepackage{array}
\usepackage{graphicx}
\usepackage{floatrow}
\usepackage{subcaption}
\usepackage{paralist}
\usepackage{latexsym}
\usepackage{comment}
\usepackage{epsfig}
\usepackage{setspace}
\usepackage{CJK}
\usepackage{color}

\usepackage{geometry}
\usepackage{algorithm}
\usepackage{algpseudocode}

\usepackage{pdflscape}

\usepackage{booktabs}
\usepackage{rotating}

\usepackage{tikz}
\usetikzlibrary{arrows.meta, positioning, shapes.geometric}

\usepackage[authoryear, round]{natbib}
\usepackage{lineno,hyperref}

\def\spacingset#1{\renewcommand{\baselinestretch}%
{#1}\small\normalsize} \spacingset{1}

\makeatletter
\def\singlespace{\def\baselinestretch{1}\@normalsize}

\numberwithin{equation}{section}
\renewcommand{\hat}{\widehat}

\renewcommand{\hat}{\widehat}

\newcommand{\bfm}[1]{\ensuremath{\mathbf{#1}}}

   \def\bB{\bfm B}

    \def\FF{\mathbb{F}}

   \def\bI{\bfm I}  
   \def\bJ{\bfm J}

\def\bu{\bfm u}   \def\bU{\bfm U}  
     
   \def\bW{\bfm W}  
   \def\bX{\bfm X}

\def\calX{{\cal  X}} 
\def\calY{{\cal  Y}}

\newcommand{\bfsym}[1]{\ensuremath{\boldsymbol{#1}}}

 \def\bbeta{\bfsym \beta}

 \def\bmu{\bfsym {\mu}}                 
 \def\bnu{\bfsym {\nu}}

 \def\bsigma{\bfsym \sigma}             \def\bSigma{\bfsym \Sigma}
         \def\bLambda {\bfsym {\Lambda}}
           \def\bOmega {\bfsym {\Omega}}

\def\1{\bfsym{1}}

\DeclareMathOperator{\E}{E}

\DeclareMathOperator{\tr}{tr}

\def\newpage{\vfill\eject}

\def\today{\ifcase\month\or
  January\or February\or March\or April\or May\or June\or
  July\or August\or September\or October\or November\or December\fi
  \space\number\day, \number\year}

\newdimen\biblioindent    \biblioindent=30pt

 at 8truept

\newcommand{\beq}{\begin{equation}}
  \newcommand{\eeq}{\end{equation}}
\newcommand{\beqn}{\begin{eqnarray}}
  \newcommand{\eeqn}{\end{eqnarray}}
\newcommand{\beqnn}{\begin{eqnarray*}}
  \newcommand{\eeqnn}{\end{eqnarray*}}

\def\akx{{\rm AKX}}

\allowdisplaybreaks
\usepackage{verbatim}
\def\tilde{\widetilde}

\def\FF{\mathcal{F}}
\def\[{\left [}  \def\]{\right ]} \def\({\left (}  \def\){\right )}
 \def\endpf{$\blacksquare$}
\def\hat{\widehat}

\newtheorem{assumption}{Assumption}
\newtheorem{theorem}{Theorem}

\theoremstyle{definition}

\newtheorem{remark}{Remark}

\title{High-Dimensional Time-Varying Coefficient Estimation in Diffusion Models}
\author{Donggyu Kim$^a$, Minseog Oh$^b$,  and Minseok Shin$^c$\footnote{Corresponding author. E-mail address: minseokshin@postech.ac.kr.} \\
$^a$University of California, Riverside\\
$^b$Sogang University\\
$^c$Pohang University of Science and Technology (POSTECH) \\
}

\begin{document}
\maketitle
\begin{spacing}{1.68}

\begin{abstract}
In this paper, we develop a novel high-dimensional time-varying coefficient estimation method, based on high-dimensional It\^o diffusion processes. 
To account for high-dimensional time-varying coefficients, we first estimate local (or instantaneous) coefficients using a time-localized Dantzig selection scheme under a sparsity condition, which results in biased local coefficient estimators due to the regularization. 
To handle the bias, we propose a debiasing scheme, which provides well-performing unbiased local coefficient estimators.
With the unbiased local coefficient estimators, we estimate the integrated coefficient, and to further account for the sparsity of the coefficient process, we apply thresholding schemes. 
We call this Thresholding dEbiased Dantzig (TED). 
We establish asymptotic properties of the proposed TED estimator.
In the empirical analysis, TED achieves a higher average out-of-sample $R^2$ across assets than benchmark estimators in most periods.
Industry-related factors play a central role in explaining asset returns.
The estimated integrated coefficients show pronounced time variation associated with firm-specific events and seasonal patterns.
\end{abstract}

\section{Introduction} \label{SEC-1}

To explain various data types, numerous regression-based models have been developed. 
Especially,  advances in technology provide us big data, which causes the curse of dimensionality problem.
To tackle this problem in the high-dimensional regression, we usually assume the sparsity of variables,  that is, the number of significant coefficients is small. 
To accommodate the sparsity condition, we often employ the LASSO procedure \citep{tibshirani1996regression}, SCAD \citep{fan2001variable}, and Dantzig selector \citep{candes2007dantzig}.
The works of \citet{belloni2014inference, feng2020taming, yuan2006model, zou2006adaptive} are useful for further reading.
There are numerous related papers that can be found in the above literature.
 These estimation methods result in sparse coefficients, and under the sparsity condition, they are consistent estimators \citep{negahban2012unified}.
Under the diffusion process, \citet{ciolek2022lasso} studied the properties of the LASSO estimator of the drift component  and \citet{gaiffas2019sparse} proposed the estimation procedure for the drift parameter in the high-dimensional Ornstein--Uhlenbeck (OU) process.

On the other hand, in high-frequency finance,  we often observe that coefficients in the regression model are time-varying.
For example, \citet{andersen2021recalcitrant} investigated the intraday variation of the local coefficients, which are called the market betas, between the individual assets and market index.
To account for the time-varying feature, \citet{mykland2009inference} computed the market beta as the aggregation of the market betas estimated over local blocks.
To evaluate the coefficients of multi factor models, \citet{ait2020high} proposed an integrated coefficient approach using the local coefficient.
See also \citet{chen2018inference, oh2024robust}.
We call this high-frequency regression. 
Recently, \citet{chen2024realized} proposed the high-dimensional market beta estimation procedure with large dependent variables and almost finite common factors. 
However, in the field of finance, there are hundreds of potential factor candidates  that  explain the cross section of expected stock returns \citep{cochrane2011presidential, harvey2016and, hou2020replicating, mclean2016does}. 
Thus, we also encounter the curse of dimensionality in high-frequency regressions, so the estimation methods developed for the finite dimension fail to estimate the coefficients consistently.
To overcome this issue, we can consider the high-dimensional regression methods such as the LASSO \citep{tibshirani1996regression}, Dantzig selector \citep{candes2007dantzig}, and SCAD \citep{fan2001variable}.
However, the direct application of these methods cannot explain the time-varying feature of the coefficient process and may suffer from the model errors.
Thus, to fully benefit from the utilization of high-frequency financial data in the high-dimensional regression, we need to develop methodologies that can handle both the curse of dimensionality as well as the time-varying coefficients.

In this paper, we introduce a novel high-dimensional high-frequency regression estimation procedure which can accommodate the sparse and time-varying coefficient processes. 
To model the high-frequency data, we employ diffusion processes whose stochastic difference equations  have a time series regression structure.
We also assume that the coefficient process $\bbeta_t$ follows a diffusion process.  
In this paper, the parameter of interest is the integrated coefficient, $\int_{0}^1 \bbeta_t dt$, that represents the average relationship between variables. 
To handle the curse of dimensionality, we assume that the coefficient processes are sparse, and to account for the sparsity of the time-varying coefficient process,  we employ the Dantzig selector procedure  \citep{candes2007dantzig}.
Specifically, due to the time-varying phenomena, we cannot estimate the integrated coefficient directly, and so we first estimate the instantaneous (or local) coefficients using the  time-localized  Dantzig selector procedure, based on the definition of $\bbeta_t$. 
Alternatively, a LASSO-type estimator can be used instead of the Dantzig selector, as demonstrated in \citet{shin2023robust}, who builds on this study by incorporating robustness against heavy-tailed data distributions.
It is worth mentioning that  \citet{shin2023robust} wrote the paper based on the estimation structure of our paper, and their main contribution is to handle heavy-tailed distribution and to extend the estimation structure to the LASSO-type.
Then, to mitigate the bias coming from the regularization of the Dantzig selector, we propose a debiasing scheme and estimate the integrated coefficient with the debiased Dantzig instantaneous coefficients.
With the debiasing scheme, we can obtain more accurate estimators in terms of the elementwise convergence rate; however, the estimated integrated coefficient is not sparse.
Thus, to accommodate the sparsity, we further regularize the estimated integrated coefficient. 
We call this Thresholding dEbiased Dantzig (TED). 
To our best knowledge, this is the first paper that handles a high-dimensional integrated time-varying coefficient based on high-frequency data.
We also establish its asymptotic properties.

The rest of paper is organized as follows.
Section \ref{SEC-2}  introduces the model set-up.
Section \ref{SEC-3} proposes the TED estimation procedure and establishes its asymptotic properties.
In Section \ref{SEC-4}, we conduct a simulation study to check the finite sample performance of the TED estimation procedure, and in Section \ref{SEC-5}, we apply the TED to the high-frequency financial data.
The conclusion is presented in Section \ref{SEC-6}, and we collect all of the proofs and supplementary materials in Appendix.

\setcounter{equation}{0}
\section{The model set-up} \label{SEC-2}

 We consider the following non-parametric time series regression diffusion model:
\begin{equation} \label{model-1}
dY_t=   \bbeta_{t} ^{\top}  d\bX_{t} +d Z_t,  
\end{equation}
where $Y_t$ is a dependent process,  $\bX_t$ is a $p$-dimensional multivariate covariate process, $\bbeta_t$ is a coefficient process, and $Z_t$ is a residual process.
The $p$-dimensional covariate process $\bX_t$ and residual process $Z_t$ satisfy
\begin{equation}\label{eq:ito-semi}
	d\bX_t= \bmu_t dt+ \bsigma_t   d\bB_t   \quad \text{and} \quad dZ_t= \nu_t d W_t, 
\end{equation}
where $\bmu_t$ is a drift process, $\bsigma_t$ and $\nu_t$ are instantaneous volatility processes,  $\bB_t$ and $W_t$ are $p$-dimensional and one-dimensional standard Brownian motions, respectively, and $\bB_t$ and $W_t$ are independent.
The stochastic processes $\bmu_t$, $\bbeta_t$, $\bsigma_t$, and $\nu_t$ are predictable.
To account for the time-varying coefficient, we further assume that the coefficient $\bbeta_t$ satisfies the following diffusion process:
\begin{equation}\label{diffu_beta}
	d \bbeta_t= \bmu_{\beta, t} dt + \bnu_{\beta,t} d \bW_t^{\beta},
\end{equation}
where $\bmu_{\beta, t}$ and $\bnu_{\beta,t}$ are predictable, and $\bW_t^{\beta}$ is $p$-dimensional standard Brownian motion.
We do not impose any restriction on the dependence structure between $\bW_t^{\beta}$ and $(\bB_t, W_t)$, while we assume independence between $\bB_t$ and $W_t$ for identification.
We note that \eqref{diffu_beta} includes many continuous-time models commonly used in economics and finance, including mean-reverting processes such as the OU process. %
Since the drift and volatility terms are modeled as general predictable stochastic processes, they can incorporate various forms of endogeneity.
To figure out the average relationship between the covariate and dependent processes, we consider the integrated coefficient:
\begin{equation*}	
	I \beta = (I \beta_i)_{i=1,\ldots, p}= \int_{0}^1 \bbeta_t dt.
\end{equation*}
Each $I \beta_i$ represents the average long-run effect of the $i$th covariate process on the dependent process over the observation interval.

In theoretical asset pricing models, such as the CAPM and intertemporal models \citep{breeden1979intertemporal,merton1973intertemporal,sharpe1964capital}, a small number of factors derived from underlying state variables explain asset returns.
In contrast, empirical studies have proposed a large number of return predictors, known as the factor zoo \citep{cochrane2011presidential, mclean2016does}.
As a result, regressions of asset returns on the factor zoo become high-dimensional and suffer from the curse of dimensionality.
On the other hand, empirical evidence suggests that most documented factors fail to remain statistically significant once replication and multiple-testing adjustments are taken into account \citep{harvey2016and,hou2020replicating}.
Therefore, despite the large universe of proposed factors, the effective number of economically meaningful factors is substantially smaller.
Motivated by this, several recent studies impose exact sparsity on factor loadings; that is, they restrict nonzero coefficients to only a small subset of factors \citep{chinco2019sparse,freyberger2020dissecting}.
This approach not only addresses the curse of dimensionality but also enhances economic interpretability.
On the other hand, \citet{he2024no} provide evidence that the exact sparsity assumption is too restrictive for explaining the cross section of returns.
To reflect this, we follow \citet{fan2017estimation} and assume that the coefficient process $\bbeta_t = (\beta_{1t},\ldots,\beta_{pt})^{\top}$ satisfies the following weak sparsity condition:
\begin{equation}\label{sparsity_beta}
	\sup_{0 \leq t \leq 1}\sum_{i=1}^p    |\beta_{it} | ^{\delta}   \leq s_p  \quad \text{and} \quad 	\sum_{i=1}^p|I  \beta_{i} | ^{\delta}  \leq s_p \, \text{ a.s.},
\end{equation}
where $0^0$ is defined as 0, $\delta \in [0,1)$, and $s_p$ is diverging slowly with respect to $p$, for example, $\log p$. 
This allows a few sizable coefficients and many small ones.
See also \citet{bickel2008covariance,cai2011adaptive,cai2011constrained,cai2012optimal}.
In practice, it is often convenient to assume $\delta=0$, which corresponds to the exact sparsity condition with only a few nonzero coefficients.
In this paper, we investigate asymptotic properties under the more general sparsity condition, which includes the above exact sparsity condition as a special case.
This sparsity assumption reflects that not all covariates have strong effects on the dependent process; otherwise, the volatility of the dependent process would diverge.
We note that with the randomness of the coefficient process, in general, the sparsity condition \eqref{sparsity_beta} is satisfied with high probability.
However, for simplicity, we assume that the sparsity condition is satisfied almost surely.
The sparsity condition is also widely employed in the high-frequency finance literature \citep{ciolek2022lasso, gaiffas2019sparse,  kim2016asymptotic,  kim2018adaptive, tao2013optimal, wang2010vast}.
To ensure that the diffusion process satisfies the sparsity condition, additional conditions on $\bmu_{\beta,t}$ and $\bnu_{\beta,t}$ should be imposed.
One sufficient condition is that there exists a set of indices $S$ such that $\left|S\right|\leq s_p$ and $\bmu_{\beta,t,i} = \bnu_{\beta,t,j,k} = 0$ for all $i \notin S$, $(j,k) \notin \left\lbrace (i,i) : i \in S \right\rbrace $, and $t \in [0,1]$.
There might be more general sufficient conditions, such as those allowing for time-varying exact sparsity patterns, but we leave exploring such conditions for a future study.

\section{ High-dimensional high-frequency regression}\label{SEC-3}
\subsection{Estimation procedure} \label{SEC-Estimation}
In this section, we propose an estimation procedure for large integrated coefficients. 
 We first fix some notations.
For any given  $p_1$ by $ p_2$ matrix $\bU = \left(U_{ij}\right)_{i=1,\ldots,p_1, j=1,\ldots,p_2}$, let
 \begin{equation*}
  	 \| \bU \| _{\max} = \max_{i,j} | U_{ij}| , \quad \|\bU\|_1 = \max\limits_{1 \leq j \leq p_2}\sum\limits_{i = 1}^{p_1}|U_{ij}|,\quad \text{ and } \quad   \|\bU\|_\infty = \max\limits_{1 \leq i \leq p_1}\sum\limits_{j = 1}^{p_2}|U_{ij}| .
 \end{equation*}
We denote  the Frobenius norm of $\bU$ by $\|\bU\|_F = \sqrt{ \mathrm{tr}(\bU^{\top} \bU) }$.
The matrix spectral norm $\|\bU\|_2$ is  the square root of the largest eigenvalue of $\bU\bU^\top$.
$C$'s denote  generic constants whose values are free of $n$ and $p$ and may change from appearance to appearance.

From the model \eqref{model-1}, the instantaneous coefficient $\bbeta_t$ satisfies the following equation:
\begin{equation*} 
\frac{d } {dt} [ Y, \bX] _t  = 	\bbeta_t ^{\top}    \frac{d}{dt}   [\bX, \bX] _t \, \text{ a.s.},
\end{equation*}
where $[ \cdot, \cdot]$ denotes the quadratic variation.
The coefficient process $\bbeta_t$ is a function of instantaneous volatilities of $\bX$ and $Y$ as follows: 
\begin{equation}\label{relation}
 \bbeta_t  = \bSigma_t^{-1} \bSigma_{XY, t} \, \text{ a.s.},
\end{equation}
where $\bSigma_t = \bsigma_t \bsigma_t^{\top}$ and $\bSigma_{XY,t} = \frac{d } {dt} [   \bX, Y] _t$. 
Thus, the instantaneous coefficient  can be estimated by the instantaneous volatility estimators.
For the finite dimensional case, the instantaneous volatility-based estimation procedure works well \citep{ait2020high}.
 However, this approach cannot explain the sparse structure \eqref{sparsity_beta}.
 Furthermore, when the dimensionality of the covariate $\bX$ is larger than the sample size, this approach fails to consistently estimate the instantaneous coefficient.  
Therefore, the procedure developed for the finite dimension is neither effective nor efficient. 
In contrast, the direct application of the high-dimensional regression procedure such as the LASSO \citep{tibshirani1996regression} and Dantzig selector  \citep{candes2007dantzig} cannot consistently estimate the integrated coefficients.
Specifically, we can rewrite \eqref{model-1} as follows:
\begin{equation}\label{eq:decomp-high-reg}
dY_t=   {\bbeta}_0 ^{\top}  d\bX_{t} + \(\bbeta_{t} - {\bbeta}_0\)^{\top}  d\bX_{t} + d Z_t,
\end{equation}
where $\bbeta_0$ is a constant vector.
Due to the time-variation of the coefficient process, we have a non-negligible dependent structure between ${\bbeta}_0 ^{\top}  d\bX_{t}$ and $\(\bbeta_{t} -{\bbeta}_0\)^{\top}  d\bX_{t}$.
This produces a bias for the usual high-dimensional regression methods.
To mitigate the dependency and accommodate the sparse structure of the coefficient process in \eqref{sparsity_beta}, we employ the time-localized Dantzig selection method as follows.
Let $\Delta_i ^n A = A_{i \Delta_n} - A_{(i-1) \Delta_{n}}$ for $1 \leq i \leq 1/\Delta_n$, where $\Delta_n=1/n$ is the distance between adjacent observation time points.
Define
\begin{eqnarray*}
 &&\mathcal{Y}_i = 
\left( \Delta_{i+1} ^n Y , \Delta_{i+2} ^n Y , \ldots, \Delta_{i+k_n} ^n Y   \right)^{\top},\quad
 \mathcal{X}_i = 
\left( \Delta_{i+1} ^n \bX  , \Delta_{i+2} ^n \bX  ,  \ldots, \Delta_{i+k_n} ^n \bX   \right)^{\top} , \quad \text{and} \quad \cr
 && \mathcal{Z}_i = \left( \Delta_{i+1} ^n Z, \Delta_{i+2} ^n Z, \ldots, \Delta_{i+k_n} ^n Z \right)^{\top}  ,
\end{eqnarray*}
where $k_n$ is the number of observations in each window to calculate the local regression.
Then, we estimate the sparse instantaneous coefficient as follows:
\begin{equation} \label{Dantzig}
 \hat{\bbeta}_{i \Delta_n} = \arg \min \| \bbeta  \|_1 \quad \text{ s.t. }  \left \|  \frac{1}{k_n \Delta_n }    \mathcal{X}_i  ^{\top} \mathcal{X} _i \bbeta -  \frac{1}{k_n \Delta_n}   \mathcal{X}_i ^{\top} \mathcal{Y}_i  \right \|_{\max} \leq \lambda_n,
\end{equation}
 where $\lambda_n$ is a tuning parameter which converges to zero.
 We specify $\lambda_n$ in Theorem \ref{Thm1}.
With the appropriate $\lambda_n$, we can show that the proposed Dantzig instantaneous coefficient estimator $ \hat{\bbeta}_{i \Delta_n} $ is a consistent estimator (see Theorem \ref{Thm1}).
To estimate the integrated coefficient $I\beta$ with this consistent estimator, we usually consider the sum of the instantaneous coefficient estimators $\hat{\bbeta}_{i \Delta_n}$'s.
However, the Dantzig estimator is biased, so their summation cannot enjoy the law of large numbers properties.
For example, the error of the sum of the Dantzig instantaneous coefficient estimators is dominated by the bias terms, and so it has the same convergence rate as that of $\hat{\bbeta}_{i \Delta_n}$. 
To reduce the effect of the bias, we use a debiasing scheme as follows.
We first estimate the inverse matrix of the instantaneous volatility matrix $\bSigma_{i \Delta_n}$ using the constrained $\ell_1$-minimization for inverse matrix estimation (CLIME) \citep{cai2011constrained}. 
Let $\hat{\bOmega}_{i \Delta_n}$ be the solution of the following optimization problem:
\begin{equation} \label{CLIME}
	\min \| \bOmega\|_1 \quad \text{s.t.} \quad  \|   \frac{1}{k_n \Delta_n }    \mathcal{X}_i  ^{\top}\mathcal{X} _i \bOmega - \bI  \|_{\max} \leq \tau_n, 
\end{equation}
where $\tau_n$ is the tuning parameter  specified in Theorem \ref{Thm2}.
With the CLIME estimator, we adjust the Dantzig instantaneous coefficient estimator as follows:
\begin{equation}\label{debias}
	\tilde{ \bbeta}_{i \Delta_n} = \hat{\bbeta}_{i \Delta_n}  +  \frac{1}{k_n \Delta_n }   \hat{\bOmega}_{i \Delta_n}^{\top} \mathcal{X}_i  ^{\top} ( \mathcal{Y}_i - \mathcal{X}_i  \hat{\bbeta}_{i \Delta_n} ).
\end{equation}
Then, the debiased  Dantzig instantaneous coefficient estimator satisfies
\begin{equation} \label{debias-error}
\tilde{\bbeta}_{i \Delta_n}  - \bbeta_{0,i \Delta_n}=  \frac{1}{k_n \Delta_n} \bOmega_{0,i \Delta_n}  \(   \mathcal{X}_i ^{\top}\mathcal{Z}_i  + \mathcal{A}_i \)+R_i \text{ a.s.},
\end{equation}
where the subscript $0$ represents the true parameter values,  $\mathcal{A}_i$ is a martingale difference defined in \eqref{def-a} in Appendix \ref{SEC-proof}, and  $R_i$ is a negligible  remaining error term (see Theorem \ref{Thm3}).
We note that the debiasing scheme is usually employed to derive asymptotic normality and to conduct the confidence interval construction or hypothesis testing \citep{javanmard2018debiasing, van2014asymptotically, zhang2014confidence}.
However, we adopt the debiasing scheme to improve  the integrated coefficient estimation.
Specifically, the debiasing scheme helps  enjoy the law of large numbers property when averaging the instantaneous coefficient estimators.
The integrated coefficient estimator is defined by 
\begin{equation*}
	\hat{I \beta}  = \sum_{i=0}^{[1/(k_n \Delta_n) ]-1}\tilde{\bbeta}_{i k_n \Delta_n} k_n \Delta_n.
\end{equation*}
As discussed above, the debiasing scheme  helps improve  the elementwise convergence rate of the debiased Dantzig integrated coefficient estimator.
However, the debiased  Dantzig integrated coefficient estimator does not satisfy the sparsity condition \eqref{sparsity_beta} due to the bias adjustment. 
To accommodate the sparsity of the integrated coefficient, we apply the thresholding scheme as follows:
\begin{equation*}
	\tilde{I\beta}_i= s (\hat{I\beta}_i)  \1 \(  |\hat{I\beta}_i  | \geq h_n  \) \quad \text{and} \quad  \tilde{I \beta} = \( \tilde{I\beta}_i \)_{i=1,\ldots,p},
\end{equation*}
where $\1 (\cdot)$ is an indicator function,   the thresholding function $s ( \cdot)$ satisfies that $| s (x)-x| \leq h_n$, and $h_n$ is a thresholding level  specified in Theorem \ref{Thm4}. 
Examples of the thresholding function $s(x)$ include the hard thresholding function $s(x)=x$ and the soft thresholding function $s(x)= x- \mbox{sign}(x) h_n$. 
For the empirical study, we employed the hard thresholding function.
We call this the Thresholded dEbiased Dantzig (TED) estimator.
We summarize the TED estimation procedure in Algorithm \ref{TED}, and additional implementation details including a graphical illustration and computational complexity analyses are provided in Appendix \ref{sec:implementation}.
\begin{remark}\label{remark:why-dantzig}
  The time-localized Dantzig method in \eqref{Dantzig} estimates the instantaneous coefficient process by utilizing the relationship \eqref{relation}.
  This intuitive connection allows the method to be naturally extended to accommodate various stylized features of high-frequency financial data, such as microstructure noise and heavy-tailed distributions.
  The only requirement is that the estimates of instantaneous covariance matrices in the constraint of \eqref{Dantzig} satisfy certain convergence rates, which are discussed in Appendix \ref{sec-extension}.
  Indeed, several existing studies have addressed these features in the context of volatility estimation \citep{ait2017using, barndorff2011multivariate, fan2018robust,kim2018adaptive, shin2023robust, zhang2011estimating}.
  Therefore, our approach can be extended by adopting similar ideas.
  Alternatively, the LASSO method employs a penalized least squares approach to address high-dimensional regression.
  Under locally constant or mildly time-varying coefficients within each time-localized window, the LASSO method can serve as an alternative to the Dantzig selector \citep{bickel2009simultaneous}.
Nevertheless, because of its penalized least squares formulation, the LASSO method does not naturally extend to accommodate these features, in contrast to the Dantzig approach.
In this sense, the proposed method offers greater flexibility.
\end{remark}

\begin{algorithm}
\caption{TED estimation procedure.} \label{TED}
\begin{algorithmic}

\State \textbf{Step 1} Estimate the instantaneous coefficient:
\begin{equation*}
\hat{\bbeta}_{i \Delta_n} = \arg \min \| \bbeta \|_1 \quad \text{ s.t. } \left \| \frac{1}{k_n \Delta_n } \mathcal{X}_i ^{\top} \mathcal{X} _i \bbeta - \frac{1}{k_n \Delta_n} \mathcal{X}_i ^{\top} \mathcal{Y}_i \right \|_{\max} \leq \lambda_n,
\end{equation*}
where $\lambda_n=C_\lambda s_p \sqrt{\log p} \left( \sqrt{ k_n \Delta_n} + k_n ^{-1/2} \right)$ and $k_n = c_k n^{1/2}$ for some large constants $C_\lambda$ and $c_k$.

\State \textbf{Step 2} Debias the Dantzig instantaneous coefficient estimator:
\begin{equation*}
\tilde{ \bbeta}_{i \Delta_n} = \hat{\bbeta}_{i \Delta_n} + \frac{1}{k_n \Delta_n } \hat{\bOmega}_{i \Delta_n}^{\top} \mathcal{X}_i ^{\top} ( \mathcal{Y}_i - \mathcal{X}_i \hat{\bbeta}_{i \Delta_n} ),
\end{equation*}
where
\begin{equation*}
\hat{\bOmega}_{i \Delta_n} = \arg \min \| \bOmega\|_1 \quad \text{s.t.} \quad \| \frac{1}{k_n \Delta_n } \mathcal{X}_i ^{\top}\mathcal{X} _i \bOmega - \bI \|_{\max} \leq \tau_n,
\end{equation*}
with $\tau_n=C_\tau \sqrt{\log p} \left( \sqrt{ k_n \Delta_n} + k_n ^{-1/2} \right)$ for some large constant $C_\tau$.

\State \textbf{Step 3} Estimate the integrated coefficient:
\begin{equation*}
\hat{I \beta} = \sum_{i=0}^{[1/(k_n \Delta_n) ]-1}\tilde{\bbeta}_{i k_n \Delta_n} k_n \Delta_n.
\end{equation*}

\State \textbf{Step 4} Threshold the debiased Dantzig integrated coefficient estimator:
\begin{equation*}
\tilde{I\beta}_i= s (\hat{I\beta}_i) \1 \left( |\hat{I\beta}_i | \geq h_n \right) \quad \text{and} \quad \tilde{I \beta} = \left( \tilde{I\beta}_i \right)_{i=1,\ldots,p},
\end{equation*}
where $\1 (\cdot)$ is an indicator function, the thresholding function $s ( \cdot)$ satisfies that $| s (x)-x| \leq h_n$, $h_n= C_h b_n$ for some constant $C_h$, and $b_n$ is defined in Theorem \ref{Thm3}.

\end{algorithmic}
\end{algorithm}

  \subsection{Asymptotic results}
  
  In this section, we establish the asymptotic properties for the proposed TED estimation procedure.
To investigate the asymptotic properties, we need the following technical conditions.
  \begin{assumption}\label{assumption1}~
  \begin{itemize}
  \item [(a)] The volatility  process  $\bSigma_t = (\Sigma_{ij t} ) _{i,j=1,\ldots, p}$  satisfies the following H\"older condition:
  \begin{equation*}
  	  | \Sigma_{ij t} -\Sigma_{ij s} | \leq  C   |t-s| ^{1/2}     \,  \text{ a.s.}
 \end{equation*}

 \item [(b)]   $\bmu_t$,  $\bmu_{\beta, t}$, $\bbeta_t$,  $\nu_t$, $\bSigma_t$, and $\bSigma_{\beta,t} = \bnu_{\beta,t}\bnu_{\beta,t}^{\top}$ are a.s. bounded, and $ \| \bSigma_t ^{-1}\|_1 \leq C $ a.s.
 
 \item [(c)] The drift process $\bmu_{\beta,t} = \(\mu_{\beta,1t}, \ldots, \mu_{\beta,pt} \)^{\top}$ and the volatility process $\bSigma_{\beta,t}=\(\Sigma_{\beta, ijt}\)_{i,j=1, \ldots, p}$  satisfy the following sparsity condition for $\delta \in [0, 1)$: 
 \begin{equation*}
 \sup_{0 \leq t \leq 1}\sum_{i=1}^p    |\mu_{\beta, it} |^{\delta}   \leq s_p   \quad \text{and} \quad  \sup_{0 \leq t \leq 1}\sum_{i=1}^p |\Sigma_{\beta, iit}|^{\delta/2}    \leq s_p  \text{ a.s.}
 \end{equation*}

  \item [(d)] $ n^{c_1}  \leq p \leq  c_2 \exp ( n^{c_3}) $ for some positive constants  $c_1$, $c_2$, and $c_3<1/8$, and  $\log p (s_p^2 + \log p + s_{\omega,p}^{2/(1-q)})    k_n^{-1}  \rightarrow 0$ as $n,p \rightarrow \infty$.

  \item [(e)] The inverse matrix of the volatility matrix process, $\bSigma_t^{-1} = \bOmega_t = (\omega_{ij t})_{i,j=1,\ldots,p}$, satisfies the following sparsity condition:
  \begin{equation*}
  	 \sup_{0 \leq t \leq 1} \max_{ 1 \leq i \leq p} \sum_{j=1}^p |\omega_{ijt} | ^{q} \leq s_{\omega, p } \,  \text{ a.s.},
 \end{equation*}
 where $q \in [0, 1)$  and  $s_{\omega, p }$ is diverging slowly with respect to $p$, for example, $\log p$.

  \end{itemize}
  \end{assumption}

\begin{remark}
To investigate estimators of time-varying processes, we need continuity conditions such as Assumption \ref{assumption1}(a) and the diffusion process structures for $\bX_t$, $Y_t$, and $\bbeta_t$ in Section \ref{SEC-2}.
Specifically, Assumption \ref{assumption1}(a) controls the discretization error arising from approximating the spot covariance process; see \eqref{Thm1-eq3} in Appendix \ref{SEC-proof}.
Even if Assumption \ref{assumption1}(a) is replaced by the condition that $\bSigma_t$ has a continuous It\^o diffusion process structure with bounded drift and instantaneous volatility processes, we can obtain the same theoretical results with up to $\sqrt{\log p}$, which reflects the high-dimensional setting.
These bounded drift and volatility conditions imply that the maximal change in each element of the volatility matrix process over each local time block has sub-Gaussian tails of order equal to the square root of the block length \citep{dzhaparidze2001bernstein}.
This also yields a discretization error bound of the same order as under Assumption \ref{assumption1}(a), up to $\sqrt{\log p}$.
For simplicity, we impose Assumption \ref{assumption1}(a).
The boundedness condition Assumption \ref{assumption1}(b) provides sub-Gaussian tails  which are often required to investigate  high-dimensional inferences. 
On the other hand, when we investigate the asymptotic behaviors of volatility estimators such as their convergence rate, the boundedness condition can be relaxed to the locally boundedness condition  (see \citet{ait2017using}).
 Specifically,  \citet{jacod2012discretization} showed in Lemma 4.4.9  that if the asymptotic result, such as stable convergence in law  or convergence in probability, is satisfied under the boundedness condition, it is also satisfied under the locally boundedness condition.
Thus, the asymptotic results established in this paper also hold for the locally boundedness condition. 
The sparsity condition for the coefficient process, Assumption \ref{assumption1}(c), is the technical condition for investigating the discretization error of the Dantzig instantaneous coefficient estimator $ \hat{\bbeta}_{i \Delta_n} $. 
Assumption \ref{assumption1}(d) ensures a sufficient number of observations in each time-localized window to estimate and remove the bias at the required convergence rate.
Finally, to investigate asymptotic properties of the CLIME estimator, we need the sparse inverse matrix condition Assumption \ref{assumption1}(e) \citep{cai2011constrained}.
Intuitively, in a Gaussian graphical model, $\bOmega_t$ encodes conditional dependence between variables.
In this sense, it reflects their relationships after controlling for all other variables \citep{fan2016overview}.
When we use factor zoo data as covariates, it is economically plausible that any given factor has direct conditional relationships with only a small number of other factors.
Otherwise, that factor would be approximately spanned by other factors, which undermines its distinct economic interpretation.
For this reason, we expect $\bOmega_t$ to be sparse.
However, financial data usually exhibit pervasive cross-sectional correlations, so the exact sparsity assumption is unlikely to hold, although most off-diagonal elements of the inverse covariance matrix tend to be small in magnitude \citep{lee2024optimal,oh2024property}.
Therefore, as in \eqref{sparsity_beta}, we utilize the $\ell_q$-ball to define a general sparsity class that includes exact sparsity as a special case.
Furthermore, if the smallest eigenvalue of $\bSigma_t$ is strictly bigger than zero, the Frobenius norm of $\bOmega_t$ is bounded by $C\sqrt{p}$. 
This implies that the inverse matrix, $\bOmega_t$, cannot be dense with many non-negligible elements; otherwise, the smallest eigenvalue of $\bSigma_t$ would approach zero, which indicates perfect multicollinearity among the covariates.
Since  the strict positiveness of the smallest eigenvalue is the minimum requirement to investigate the regression-based models, Assumption \ref{assumption1}(e) is reasonable.
\end{remark}

In Theorems \ref{Thm1} and \ref{Thm2} below, we establish the elementwise consistency and $\ell_1$-consistency for the Dantzig instantaneous coefficient and inverse matrix.
Note that we use subscript $0$ for the true parameters. 

\begin{theorem} \label{Thm1}

Under Assumption \ref{assumption1}(a)--(d), let $k_n =    c_k n^{c}$ for some constants $c_k$ and $c \in (1/4,1/2]$.
For any given positive constant $a$, choose $\lambda_n=C_{\lambda,a} s_p \sqrt{\log p} \( \sqrt{ k_n \Delta_n} + k_n ^{-1/2}   \)$ for some large constant  $C_{\lambda,a}$.  
Then, we have, for large $n$,
\begin{equation}\label{Thm1-result1}
	 \max_{i} \| \hat{\bbeta}_{i \Delta_n} - \bbeta_{0,i \Delta_n} \|_{\max} \leq  C \lambda_n \quad \text{and}  \quad \max_{i} \| \hat{\bbeta}_{i \Delta_n} - \bbeta_{0,i \Delta_n} \|_{1} \leq  C s_p \lambda_n  ^{1-\delta},  
\end{equation}
with probability greater than $1-p^{-a}$.
\end{theorem}

\begin{remark}\label{remark:thm1}
In conventional high-dimensional penalized regression, the coefficient vector is typically assumed to be constant over time. 
Under exact sparsity, penalized regression estimators such as the Dantzig selector, LASSO, and SCAD achieve max-norm and $\ell_1$-norm convergence rates of order $\sqrt{\log p / m}$ up to sparsity, where $m$ is the sample size \citep{bickel2009simultaneous,wang2014optimal}. 
Under weak sparsity, the $\ell_1$-norm convergence rate is of order $(\sqrt{\log p / m})^{1-\delta}$ up to sparsity, whereas the max-norm rate remains of order $\sqrt{\log p / m}$ up to sparsity \citep{ye2010rate}.
When we choose $c = 1/2$, the same convergence rate holds for the Dantzig instantaneous coefficient estimators even when the coefficient process is time-varying and satisfies the weak sparsity condition.
However, the leading term of the estimation error of each Dantzig instantaneous coefficient is not a martingale difference; see \eqref{eq:decomp-high-reg}.
As a result, this non-martingale component does not average out when the local estimators are integrated over time, so the aggregated estimator cannot improve on the local convergence rate without an additional debiasing step.
\end{remark}

\begin{theorem} \label{Thm2}

Under Assumption \ref{assumption1}, let $k_n = c_k n^{c}$ for some constants $c_k$ and $c \in (1/4,1/2]$. 
For any given positive constant $a$, choose $\tau_n=C_{\tau,a}   \sqrt{\log p} \( \sqrt{ k_n \Delta_n} + k_n ^{-1/2}   \)$ for some large constant  $C_{\tau,a}$.
Then, we have, for large $n$, 
\begin{equation}\label{Thm2-result1}
	\max_{i} \| \hat{\bOmega}_{i \Delta_n} - \bOmega_{0,i \Delta_n} \|_{\max} \leq  C \tau_n  \quad \text{and} \quad \max_{i} \| \hat{\bOmega}_{i \Delta_n} - \bOmega_{0,i \Delta_n} \|_{1} \leq  C s_{\omega, p}   \tau_n ^{1-q} ,  
\end{equation}
with probability greater than $1-p^{-a}$.  
\end{theorem}
  
\begin{remark}
Theorem \ref{Thm2} shows that by choosing $c=1/2$, the estimators for the instantaneous coefficient and inverse matrix have elementwise convergence rates of $n^{-1/4}$ and  $\ell_1$ convergence rates $n^{-(1-q)/4}$, respectively,  with the $\log$ order term and the sparsity level term. 
We note that when choosing the sub-interval length $k_n = c_k n^{1/2}$ to estimate the instantaneous processes, we have the same order convergence rates of the statistical estimation and time-varying instantaneous process approximation errors.
That is, the order $n^{-1/4}$ is optimal for estimating each element of the instantaneous process; thus, the convergence rates are optimal up to log order.  
\end{remark}

The Dantzig instantaneous coefficient estimator has a near-optimal convergence rate as shown in Theorem \ref{Thm1}. 
However, as discussed in Remark \ref{remark:thm1}, it is a biased estimator, which causes some non-negligible estimation errors when estimating the integrated coefficient. 
To tackle this problem, we employ debiasing schemes with the consistent CLIME estimator as in \eqref{debias}.
The following theorem shows that the debiasing step turns the leading instantaneous coefficient estimation error into a martingale difference and yields an integrated coefficient estimator that attains a near $n^{-1/2}$ elementwise convergence rate up to sparsity and $\log p$ factors.

\begin{theorem} \label{Thm3}
Under the assumptions in Theorems \ref{Thm1}--\ref{Thm2}, we choose $k_n=  c_k n^{1/2}$ for some constant $c_k$.
Then, we have for all $i$,
\begin{equation}\label{Thm3-result1-1}
\tilde{\bbeta}_{i \Delta_n}  - \bbeta_{0,i \Delta_n}=  \frac{1}{k_n \Delta_n} \bOmega_{0,i \Delta_n}  \(   \mathcal{X}_i ^{\top}\mathcal{Z}_i  + \mathcal{A}_i \)+R_i ,
\end{equation}
where $\mathcal{A}_i$ is defined in \eqref{def-a} and 
\begin{equation}\label{Thm3-result1-2}
\max_{i} \| R_i \|_{\max}  \leq  C   \left\{   s_p ^{2-\delta}   ( \log p / n^{1/2}  )^{(2-\delta)/2}+  s_p s_{\omega,p}    ( \log p / n^{1/2}  )^{(2-q)/2}  + s_p  (\log p)^{3/2}  / n^{1/2} \right\},
\end{equation}
 with probability greater than $1-p^{-a}$  for any given positive constant $a$. 
Furthermore, we have, with   probability greater than $1-p^{-a}$ for any given positive constant $a$,
\begin{equation}\label{Thm3-result2}
 \|  \hat{I\beta}  - I\beta_0 \| _{\max}  \leq C  b_n,
\end{equation}
where $b_n=    s_p ^{2-\delta}   ( \log p / n^{1/2}  )^{(2-\delta)/2}+  s_p s_{\omega,p}    ( \log p / n^{1/2}  )^{(2-q)/2}  + s_p  (\log p)^{3/2}  / n^{1/2}  $.
\end{theorem}

\begin{remark}
The debiased Dantzig instantaneous coefficient is decomposed into the martingale difference term $  \mathcal{X}_i ^{\top}\mathcal{Z}_i  + \mathcal{A}_i $ and the non-martingale remaining term $R_i$.
The martingale difference term can enjoy the law of large numbers property, so the integrated coefficient estimator has a faster convergence rate than the Dantzig instantaneous coefficient estimator. 
The remaining non-martingale terms have the same order as those of the martingale terms for the integrated coefficient estimator. 
Unlike the biased Dantzig estimator,  the non-martingale remaining terms do not impact on the integrated coefficient estimator. 
\end{remark}

\begin{remark}
Theorem \ref{Thm3} shows the elementwise convergence rate for the debiased Dantzig integrated coefficient.
When we have the exact sparse coefficient and inverse matrix processes, that is, $\delta= q= 0$, the debiased Dantzig  integrated coefficient estimator has the  convergence rate $ s_p  ( s_p  + s_{\omega,p})  (\log p )^{3/2}/  n^{1/2}$. 
The $n^{1/2}$ term is related with the sample size, which is known as the optimal rate. 
The $ (\log p )^{3/2}$ term comes from handling the high-dimensional error bound. 
Usually, in high-dimensional literature, we have $\sqrt {\log p}$, but the  debiased Dantzig  integrated coefficient estimator has $(\log p)^{3/2}$ due to the handling of the high-dimensional error bounds for estimating two coefficients, such as the instantaneous coefficient and the integrated coefficient, and bounding the random processes.
Finally, the $s_p$ and $s_{\omega,p}$ terms represent the sparsity levels for the coefficient and inverse volatility matrix. 
High-dimensional literature commonly assumes the sparsity level to be negligible; hence, we have the convergence rate $n^{-1/2}$ with up to $\log p$ order.
 \end{remark}

Theorem \ref{Thm3} indicates that, using the debiasing scheme, we obtain well-performing input integrated coefficient estimator $\hat{I\beta}$. 
As described in Section \ref{SEC-Estimation}, we then apply the thresholding scheme to $\hat{I\beta}$ to account for the sparsity and obtain the TED estimator. 
The following theorem shows the benefit of the thresholding scheme for the $\ell_1$-norm convergence rate of the TED estimator.

 \begin{theorem} \label{Thm4}
Under the assumptions in Theorems \ref{Thm1}--\ref{Thm2}, let $k_n=  c_k n^{1/2}$ for some constant $c_k$.
For any given positive constant $a$, choose $h_n= C_{h,a} b_n$ for some constant $C_{h,a}$, where $b_n$ is defined in Theorem \ref{Thm3}.
Then, we have, with   probability greater than $1-p^{-a}$,
\begin{equation}\label{Thm4-result1}
 \|  \tilde{I\beta}  - I\beta_0 \| _{1}  \leq C    s_p  b_n ^{1-\delta}.
\end{equation}
\end{theorem}

Theorem \ref{Thm4} shows that the TED estimator is a consistent estimator in terms of the $\ell_1$ norm under the sparsity condition \eqref{sparsity_beta}.
When estimating the integrated coefficient without the debiasing step, we can obtain the convergence rate 
 $s_p( s_p \sqrt{\log p} n^{-1/4} ) ^{1-\delta} $. 
The benefit of applying the debiasing scheme is the difference between $b_n$ and $s_p \sqrt{\log p} n^{-1/4}$. 
Under the sparsity condition, $b_n$ is $n^{-\{2-(\delta \vee q )\}/4}$ with $\log p$ order for $\delta, q  \in [0,1)$, which is faster than the convergence rate of the Dantzig integrated coefficient estimator.
Therefore, the TED estimator has a faster convergence rate.

 \subsection{Extension to jump diffusion processes} 

In financial practice, we often observe jumps.
To reflect this, we can extend the continuous diffusion process \eqref{model-1} to the jump diffusion process as follows:
\begin{eqnarray}\label{jump_diffusion1}
&& dY_t=   dY^c_{t} + dY^J_{t}, \quad dY^c_{t}=\bbeta_{t}^{\top}  d\bX^c_{t}+d Z_{t}, \quad \text{and} \quad dY^J_{t} = J^y_{t} d \Lambda^y_{t},  
\end{eqnarray}
where $Y_t^{c}$ and $\bX_{t}^c$ are the continuous part of $Y_t$ and $\bX_t$, respectively, $J_t^y$ is the jump size, and $\Lambda_t^y$ is the Poisson process with the bounded intensity. 
The covariate process $\bX_t$ is
\begin{eqnarray}\label{jump_diffusion2}
	&&d\bX_t=  d \bX_t ^c + d \bX_t^J,    \quad  d\bX_t^c= \bmu_t dt+ \bsigma_t d\bB_t , \quad \text{and} \quad   d \bX_t^J =  \bJ_t d \bLambda_t, 
\end{eqnarray}
where $\bJ_t=\(J_{1t}, \ldots, J_{pt}\)^{\top}$  is a jump size process and $\bLambda_t=\(\Lambda_{1t}, \ldots, \Lambda_{pt} \)^{\top}$ is a $p$-dimensional  Poisson process with bounded intensities. 
We assume that the Poisson processes $\Lambda_t^y$ and $\bLambda_t$ are independent of  $\bsigma_t$ and  $\bbeta_t$.
Under this jump diffusion model, we can still use the proposed estimation procedure, but we cannot observe the continuous diffusion process.
To tackle this problem, we first detect the jumps from the observed stock log-return data.
For example, we  use the truncation method as follows.
Define
\begin{equation}\label{jump_adjustment}
 \hat{\mathcal{Y}}_i ^c= 
 \left( \Delta_{i+1}^n \hat{Y}^{c\top}, \Delta_{i+2}^n \hat{Y}^{c\top}, \ldots, \Delta_{i+k_n}^n \hat{Y}^{c\top}\right) ^{\top}
 \text{ and }
 \hat{\mathcal{X}}_i ^c = \left( \Delta_{i+1} ^n \hat{\bX} ^{c}, \Delta_{i+2} ^n \hat{\bX} ^{c}, \ldots, , \Delta_{i+k_n} ^n \hat{\bX} ^{c} \right) ^{\top}
\end{equation}
where $\1_{\{  \cdot \} }$ is an indicator function, $k_n$ is the number of observations in each window used to calculate the local regression,
\begin{equation*}
\Delta_i^n \hat{Y}^c = \Delta_{i} ^n Y \, \1 _{\{|\Delta_{i} ^n Y | \leq u_n \}}, \quad \Delta_{i} ^n \hat{\bX} ^{c} = \( \Delta_{i} ^n X_j\, \1_{\{ | \Delta_{i} ^n X_j | \leq v_{j,n} \}}   \)_{j=1,\ldots, p},   
\end{equation*}
and $u_n$ and $v_{j,n}$, $j=1, \ldots, p$, are the truncation levels.
We employ  $u_n=C_u s_p \sqrt{\log p} n^{-\varrho}$  and  $v_{j,n} = C_{j,v}  \sqrt{\log p} n^{-\varrho}$ for $\varrho \in [15/32 ,1/2)$ and some constants $C_u$ and $C_{j,v}$, $j=1, \ldots, p$.
In the numerical study, we adopt the usual choice in the literature \citep{ait2020high, ait2019principal}.
That is, we use
\begin{equation}\label{jump_adj} 
 u_n = 3 n^{-0.47} \sqrt{BV^Y}  \quad \text{and} \quad  v_{j,n}= 3 n^{-0.47} \sqrt{BV_j},
\end{equation}
where the bipower variations $BV^Y = \dfrac{\pi}{2}\sum_{i=2}^{n} | \Delta_{i-1} ^n Y| \cdot | \Delta_{i} ^n Y|$ and $BV_j = \dfrac{\pi}{2}\sum_{i=2}^{n} | \Delta_{i-1} ^n X_j| \cdot | \Delta_{i} ^n X_j|$.
Then, to estimate the integrated coefficient $I\beta$, we employ the estimation method in Section \ref{SEC-Estimation}  using $\hat{\mathcal{Y}}_i ^c$ and $\hat{\mathcal{X}}_i ^c$ instead of $\mathcal{Y}_i$ and $\mathcal{X}_i$.
We denote the jump-adjusted TED estimator by  $\tilde{I\beta}^c$.
In the following theorem, we investigate the asymptotic property of the jump-adjusted TED estimator.

 \begin{theorem} \label{Thm5}
Under the models \eqref{jump_diffusion1}--\eqref{jump_diffusion2}, let assumptions in Theorem \ref{Thm4} hold.
Then, we have, with probability greater than $1-p^{-a}$  for any given positive constant $a$,
\begin{equation}\label{Thm5-result1}
 \|  \tilde{I\beta} ^c - I\beta_0 \| _{1}  \leq C    s_p  b_n ^{1-\delta}.
\end{equation}
\end{theorem}

Theorem \ref{Thm5} shows that the jump-adjusted TED estimator has the same convergence rate  obtained in Theorem \ref{Thm4}.
Therefore, we conclude that the jumps can be detected well and that their effects can be mitigated.

\subsection{Discussion on the tuning parameter selection}\label{SEC-Tuning}

To implement the TED estimation procedure, we need to choose the tuning parameters.
In this section, we discuss how to select the tuning parameters for the numerical studies. 
We first obtain $\Delta_i^n \hat{Y}^c$ and $\Delta_{i} ^n \hat{\bX} ^{c}$ with the truncation levels defined in \eqref{jump_adj}.
Then, to handle the scale issue, we standardize  each column of  $\hat{\mathcal{Y}}_i ^c$ and $\hat{\mathcal{X}}_i ^c$  to have a mean of 0 and a variance of 1.
The re-scaling is conducted after obtaining the TED estimator.
For the local regression stage \eqref{Dantzig}, we choose $k_{n}=[n^{1/2}]$.
Also, we select
\begin{equation}\label{tuning} 
\lambda_n = c_{\lambda}n^{-1/4}\(\log p\)^{3/2}, \,\,\, \tau_n=c_{\tau}n^{-1/4}\sqrt{\log p}, \,\,\, \text{and} \,\,\, h_n=c_h n^{-1/2} \(\log p\)^{3/2},
\end{equation}
where $c_{\lambda}$, $c_{\tau}$, and $c_h$ are tuning parameters. 
In the simulation and empirical studies, we choose $c_{\lambda} \in [0.1, 10]$ that minimizes the 5-fold cross-validation (CV) error, defined as the average squared prediction error on the validation sets.
Also, we select  $c_{\tau} \in [0.1, 10]$ by minimizing the following loss function:
\begin{equation*}
\tr\[\( \frac{1}{k_n \Delta_n }    \hat{\mathcal{X}}_i ^{c\top} \hat{\mathcal{X}}_i ^c \hat{\bOmega}_{i \Delta_n} - \bI_{p}\)^2\],
\end{equation*}
where $\bI_{p}$ is the $p$-dimensional identity matrix.
Finally, we choose $c_h$ using a leave-one-out cross-validation scheme as follows.
We define the $j$th leave-one-block-out integrated coefficient as follows:
\begin{equation*}
	\hat{I \beta}_{(-j)}  = \sum_{i=0, i \neq j}^{[1/(k_n \Delta_n) ]-1}\tilde{\bbeta}_{i k_n \Delta_n} \frac{k_n \Delta_n}{1 - k_n \Delta_n} .
\end{equation*}
We then define the TED estimator with the tuning parameter $c_h$ and $\hat{I \beta}_{(-j)}$ as follows:
\begin{eqnarray*}
  \tilde{I \beta}_{(-j)} (c_h) = s(\hat{I \beta}_{(-j)}) \1 (| \hat{I \beta}_{(-j)} | \geq h_n)
  ,
\end{eqnarray*}
where $h_n = c_h n^{-1/2} (\log p)^{3/2}$.
We then select $c_h \in [0.01,10]$ that minimizes the following loss function:
\begin{equation*}
  k_n \Delta_n \sum_{j=0}^{[1/(k_n \Delta_n) ]-1} \left( \mathcal{Y}_{j} -  \hat{\mathcal{X}}_{j}^{c} \tilde{I \beta}_{(-j)} (c_h) \right) ^{\top}  \left( \mathcal{Y}_{j} -  \hat{\mathcal{X}}_{j}^{c} \tilde{I \beta}_{(-j)} (c_h) \right) 
  .
\end{equation*}

\section{A simulation study} \label{SEC-4}

In this section, we conducted simulations to check the finite sample performance  of the proposed TED estimator.
We generated data from the time series regression jump diffusion model \eqref{jump_diffusion1}--\eqref{jump_diffusion2}.
We considered two types of coefficient processes, time-varying and constant coefficients.
Detailed descriptions of the simulation setup are provided in Appendix \ref{sec:sim_setup}.
We set $p=100$ with $s_p = [\log p] = 4$ effective covariates, and varied sample sizes $n=1000, 2000, 4000$.
We repeated the simulation $200$ times.

To implement the TED estimation procedure, we used the hard thresholding function $s\(x\)=x$ and employed the tuning parameter selection method discussed in Section \ref{SEC-Tuning}. 
To illustrate how effectively the TED estimator captures the time-varying behaviors, Appendix \ref{appendix:trajectory} provides trajectory plots of the true coefficients along with the debiased instantaneous coefficient estimates.
The results in Appendix \ref{appendix:trajectory} illustrate that the TED estimator effectively captures the temporal dynamics of the true coefficients.

For the purposes of comparison, we considered the integrated coefficient estimator proposed by \citet{ait2020high}. 
We note that, for small $p$, one can account for the time variation of the coefficient process. 
We call this the  AKX  estimator.
Specifically,  the AKX estimator is calculated as follows:
\begin{equation}\label{AKX}
\hat{\bbeta}^{\akx}_{i\Delta_n}=(\hat{\mathcal{X}}_{i}^{c\top} \hat{\mathcal{X}}_{i}^c)^{-1} \hat{\mathcal{X}}_{i}^{c\top} \hat{\calY}_{i}^c \quad \text{and} \quad
	\tilde{I \beta}^{\akx} = \sum^{[1/(K_n \Delta_n)]-1}_{i=0} \hat{\bbeta}^{\akx}_{iK_n\Delta_n} K_n\Delta_n,
\end{equation}
where $\hat{\calX}_{i}^c$ and $\hat{\mathcal{Y}}_i^c$ are defined in \eqref{jump_adjustment} and we used $K_n=[n^{0.47}]$ instead of $k_n=[n^{0.5}]$. For $\hat{\mathcal{X}}_{i}^{c\top} \hat{\mathcal{X}}_{i}^c$ in \eqref{AKX}, we added $10^{-4} \bI_p$ to avoid the singularity coming from the ultra high-dimensionality.
We also employed regression estimators that can handle the sparsity of the high-dimensional coefficient process (without accounting for time-varying coefficients),
such as LASSO (LAS), Ridge (RID), elastic net (ELA), SCAD (SCA), Bayesian LASSO (BAY), Dantzig (DAN),
and their adaptive versions,
such as adaptive LASSO (ALAS), adaptive Ridge (ARID), adaptive elastic net (AELA), adaptive SCAD (ASCA), and adaptive Dantzig (ADAN),
as benchmarks \citep{candes2007dantzig,fan2001variable,park2008bayesian,tibshirani1996regression,zou2006adaptive,zou2005regularization,zou2009adaptive}.
The tuning parameters for these benchmark estimators were selected to minimize the 5-fold CV errors, consistent with the approach used for the TED estimator.
We calculated the average estimation errors under the max norm, $\ell_1$ norm, and $\ell_2$ norm.

\begin{figure}[!ht]
\centering
\includegraphics[width = 0.8\textwidth]{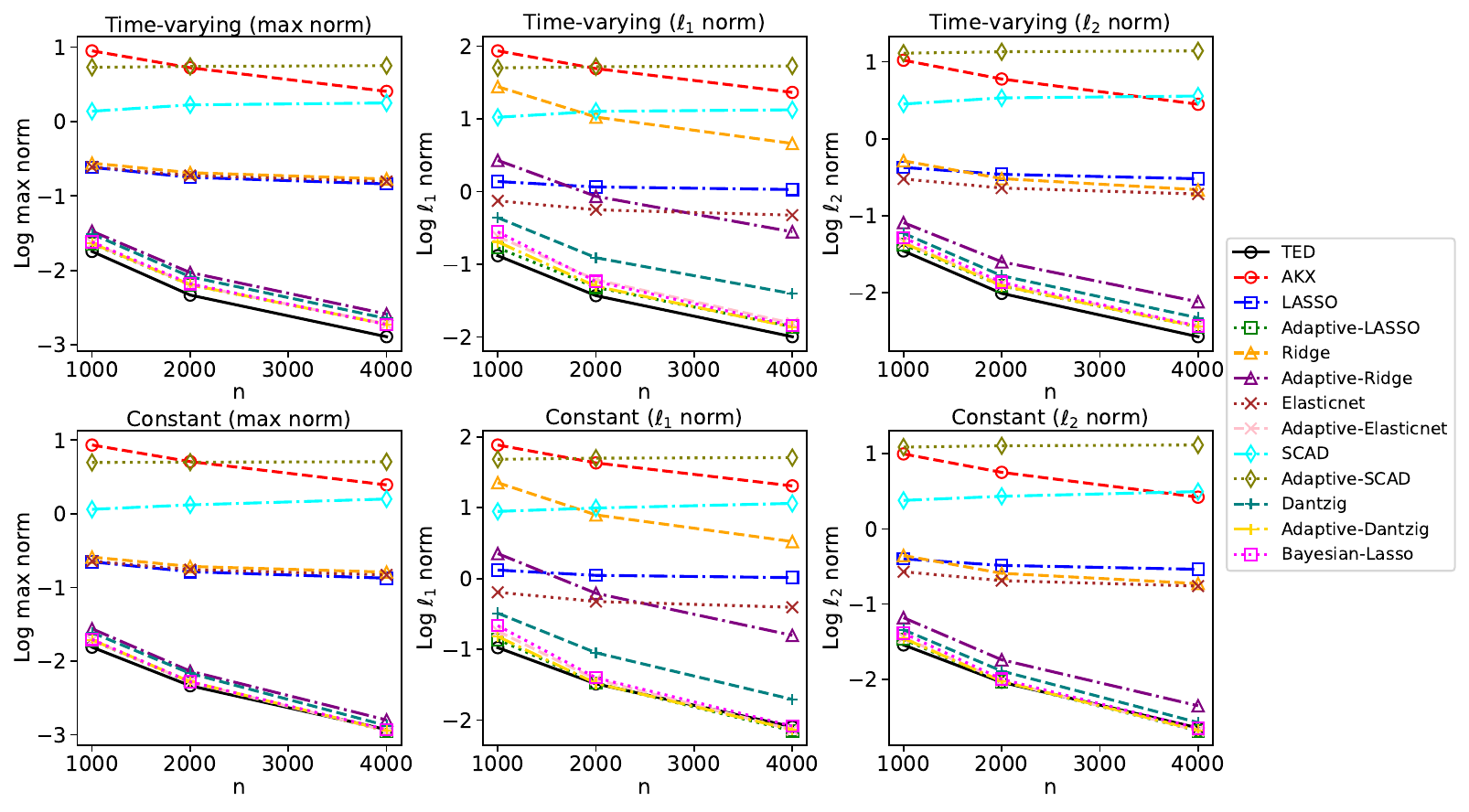}
\caption{The log max, $\ell_1$, and $\ell_2$ norm error plots of TED and benchmarks for $p=100$ and $n=1000, 2000, 4000$.}
\label{Beta_simulation}
\end{figure}

Figure \ref{Beta_simulation} plots the log max, $\ell_1$, and $\ell_2$ norm errors of TED and benchmarks for the time-varying and constant coefficient processes with $p=100$ and $n=1000, 2000, 4000$. 
From Figure \ref{Beta_simulation}, we find that the estimation errors of the TED estimator are decreasing as the number of high-frequency observations increases.
For the time-varying coefficient process, the TED estimator outperforms other estimators.
This may be because the proposed TED estimation method can account for both time variation and the high-dimensionality of the coefficient process, while the other benchmark estimators fail to explain one of them.
For the constant coefficient process, TED and ALAS, ADAN, and AELA show better performance than the other benchmark estimators.
This is probably due to the fact that the other estimators cannot adequately handle the curse of dimensionality or the bias caused by penalties imposed to induce sparsity.
\begin{table}[!ht]
\caption{Average precision and recall of variable selection for TED and benchmarks across different sample sizes $n$.}
\label{tab:precision_recall}
\centering
\resizebox{\textwidth}{!}{%
\begin{tabular}{crrrrrrrrrrrrr}
\toprule
\multicolumn{14}{c}{\textbf{Time-varying}} \\
\midrule
&  \multicolumn{13}{c}{\textit{Precision}} \\
\cmidrule(lr){2-14}
$n$ & TED & AKX & LAS & ALAS & RID & ARID & ELA & AELA & SCA & ASCA & DAN & ADAN & BAY \\
\midrule
1000 & 0.927 & 0.040 & 1.000 & 0.802 & 0.040 & 0.040 & 0.996 & 0.639 & 1.000 & 1.000 & 0.323 & 0.721 & 0.684 \\
2000 & 0.967 & 0.040 & 1.000 & 0.888 & 0.040 & 0.040 & 1.000 & 0.769 & 1.000 & 1.000 & 0.308 & 0.830 & 0.836 \\
4000 & 0.971 & 0.040 & 1.000 & 0.939 & 0.040 & 0.040 & 1.000 & 0.862 & 1.000 & 1.000 & 0.289 & 0.912 & 0.925 \\
\midrule
&  \multicolumn{13}{c}{\textit{Recall}} \\
\cmidrule(lr){2-14}
$n$ & TED & AKX & LAS & ALAS & RID & ARID & ELA & AELA & SCA & ASCA & DAN & ADAN & BAY \\
\midrule
1000 & 0.995 & 1.000 & 0.930 & 0.995 & 1.000 & 1.000 & 0.990 & 0.995 & 0.750 & 0.263 & 0.998 & 0.995 & 0.994 \\
2000 & 0.996 & 1.000 & 0.924 & 0.994 & 1.000 & 1.000 & 0.988 & 0.995 & 0.705 & 0.254 & 0.999 & 0.995 & 0.996 \\
4000 & 0.996 & 1.000 & 0.926 & 0.995 & 1.000 & 1.000 & 0.990 & 0.996 & 0.680 & 0.251 & 0.999 & 0.995 & 0.998 \\
\midrule
\midrule
\multicolumn{14}{c}{\textbf{Constant}} \\
\midrule
&  \multicolumn{13}{c}{\textit{Precision}} \\
\cmidrule(lr){2-14}
$n$ & TED & AKX & LAS & ALAS & RID & ARID & ELA & AELA & SCA & ASCA & DAN & ADAN & BAY \\
\midrule
1000 & 0.926 & 0.040 & 1.000 & 0.805 & 0.040 & 0.040 & 0.999 & 0.706 & 1.000 & 1.000 & 0.337 & 0.772 & 0.714 \\
2000 & 0.962 & 0.040 & 1.000 & 0.911 & 0.040 & 0.040 & 1.000 & 0.813 & 1.000 & 1.000 & 0.324 & 0.901 & 0.870 \\
4000 & 0.982 & 0.040 & 1.000 & 0.978 & 0.040 & 0.040 & 1.000 & 0.937 & 1.000 & 1.000 & 0.320 & 0.960 & 0.939 \\
\midrule
&  \multicolumn{13}{c}{\textit{Recall}} \\
\cmidrule(lr){2-14}
$n$ & TED & AKX & LAS & ALAS & RID & ARID & ELA & AELA & SCA & ASCA & DAN & ADAN & BAY \\
\midrule
1000 & 1.000 & 1.000 & 0.996 & 1.000 & 1.000 & 1.000 & 1.000 & 1.000 & 0.904 & 0.251 & 1.000 & 1.000 & 1.000 \\
2000 & 1.000 & 1.000 & 1.000 & 1.000 & 1.000 & 1.000 & 1.000 & 1.000 & 0.895 & 0.250 & 1.000 & 1.000 & 1.000 \\
4000 & 1.000 & 1.000 & 1.000 & 1.000 & 1.000 & 1.000 & 1.000 & 1.000 & 0.856 & 0.250 & 1.000 & 1.000 & 1.000 \\
\bottomrule
\end{tabular}%
}
\end{table}

Finally, we evaluated the variable selection accuracy of the TED estimator and benchmark methods using precision and recall metrics.
Table \ref{tab:precision_recall} presents the average precision and recall across different sample sizes and benchmarks for the time-varying and constant coefficient processes with $n=1000, 2000, 4000$.
From Table \ref{tab:precision_recall}, for the time-varying coefficient process, we observe that while several methods achieve perfect precision or recall separately, only TED and ELA simultaneously maintain both precision and recall above $0.95$ for $n=4000$.
When comparing TED with ELA, TED shows higher recall.
This indicates that TED is more effective in correctly identifying all significant factors.
In the case of the constant coefficient process, TED achieves reasonably high precision and recall, $0.982$ and $1$, respectively, while LAS and ELA achieve perfect precision and recall.
This may be because, under the constant coefficient process, TED cannot significantly benefit from its complex estimation procedure, which is primarily designed to handle time-varying coefficients.
These results suggest that the TED estimator effectively accounts for the time variation and high-dimensionality of the coefficient process and is robust to the coefficient process structure.

\section{An empirical study} \label{SEC-5}

We applied the proposed TED estimator to real high-frequency trading data from January 2013 to December 2020.
We took stock price data from the End of Day website  (https://eoddata.com) and obtained 5-min log-price data using the previous tick scheme \citep{wang2010vast, zhang2011estimating}, where half trading days were excluded.
We considered the log-prices of the five assets as the dependent processes.
Specifically, we selected Apple Inc. (AAPL), Berkshire Hathaway Inc. (BRK.B), General Motors Company (GM), Alphabet Inc. (GOOG), and Exxon Mobil Corporation (XOM).
These firms are the top market value stocks in five global industrial classification standards (GICS) sectors: information technology, financials, consumer discretionary, communication services, and energy. 
For the covariates, we used the high-frequency factor zoo dataset provided by \citet{aleti2022high}, which is publicly available on the author's website.
This dataset contains 272 portfolios: 218 characteristic-sorted factor portfolios, 48 industry portfolios, and the six Fama-French portfolios.
Here, the six Fama-French portfolios include market (MKT), value (HML), size (SMB), profitability (RMW), investment (CMA), and momentum (MOM) factors.

To examine the practical advantage of allowing time-varying coefficients, we first conducted a test for coefficient constancy based on the kernel estimation approach of \citet{ang2012testing}.
Since the approach of \citet{ang2012testing} cannot handle high-dimensionality, we implemented the test with the six Fama-French factors, which serve as a canonical benchmark in asset pricing \citep{asness2013value, barroso2015momentum, carhart1997persistence,  fama2015five, fama2016dissecting}.
For each asset and month, we computed the test statistic in Theorem 3 of \citet{ang2012testing} to evaluate the joint null hypothesis that the six Fama-French factor coefficients are constant within the month.
Across five assets and 96 months, the total number of tests was 480.
The rejection rates at the 5\%, 1\%, and 0.1\% significance levels are 0.996, 0.994, and 0.979, respectively. 
This result indicates the importance of accommodating time-varying coefficients, which motivates the use of the TED procedure.
The tuning parameters for the TED estimator were selected based on Section \ref{SEC-Tuning}.
We additionally conducted an analysis to examine how the choice of tuning parameters affects the TED estimator in Appendix \ref{additional-empirical-analysis}.
The results indicate that the performance of TED is sensitive to the choice of parameters and also highlight the effectiveness of the proposed tuning parameter selection procedure.
Since the AKX estimator is designed for the finite dimension, we also employed the AKX6 estimator.
The  AKX6 estimator  employs the same estimation method as the AKX  estimator except that it only uses the six Fama-French factors as factor candidates.
We also employed OLS with the six factors (OLS6) along with the benchmark estimators used in the simulation study, namely, LAS, ALAS, RID, ARID, ELA, AELA, SCA, ASCA, BAY, DAN, and ADAN.

\begin{table}[!ht]
\caption{The annual average in-sample and out-of-sample $R^2$ for the TED, AKX, AKX6, and various regression estimators across the five assets.}
\label{Table1}
\centering
\resizebox{\textwidth}{!}{%
\begin{tabular}{l*{16}{r}}
\toprule
 & & \multicolumn{15}{c}{In-sample $R^2$} \\
\cmidrule(lr){3-17}
Period & RV (\%) & TED & AKX & AKX6 & OLS6 & LAS & ALAS & RID & ARID & ELA & AELA & SCA & ASCA & BAY & DAN & ADAN \\
\midrule
whole & 12.0 & 0.539 & 0.309 & 0.066 & 0.208 & 0.465 & 0.591 & 0.602 & 0.613 & 0.512 & 0.603 & 0.475 & 0.257 & 0.571 & 0.606 & 0.610 \\
2013  & 7.9  & 0.667 & 0.322 & 0.038 & 0.191 & 0.554 & 0.728 & 0.749 & 0.753 & 0.633 & 0.742 & 0.562 & 0.301 & 0.732 & 0.745 & 0.751 \\
2014  & 8.8  & 0.525 & 0.272 & 0.045 & 0.184 & 0.445 & 0.585 & 0.597 & 0.607 & 0.495 & 0.599 & 0.457 & 0.223 & 0.556 & 0.601 & 0.605 \\
2015  & 10.2 & 0.573 & 0.348 & 0.072 & 0.241 & 0.510 & 0.624 & 0.636 & 0.648 & 0.547 & 0.637 & 0.521 & 0.331 & 0.616 & 0.640 & 0.645 \\
2016  & 10.0 & 0.570 & 0.324 & 0.068 & 0.218 & 0.494 & 0.616 & 0.627 & 0.635 & 0.540 & 0.625 & 0.506 & 0.333 & 0.598 & 0.628 & 0.632 \\
2017  & 5.5  & 0.453 & 0.205 & 0.018 & 0.143 & 0.384 & 0.508 & 0.520 & 0.530 & 0.428 & 0.522 & 0.391 & 0.209 & 0.494 & 0.524 & 0.528 \\
2018  & 13.0 & 0.586 & 0.391 & 0.112 & 0.269 & 0.518 & 0.625 & 0.642 & 0.647 & 0.556 & 0.636 & 0.525 & 0.292 & 0.606 & 0.639 & 0.644 \\
2019  & 9.1  & 0.515 & 0.288 & 0.050 & 0.199 & 0.441 & 0.559 & 0.570 & 0.581 & 0.486 & 0.569 & 0.452 & 0.243 & 0.541 & 0.574 & 0.576 \\
2020  & 22.8 & 0.425 & 0.325 & 0.124 & 0.220 & 0.374 & 0.483 & 0.479 & 0.505 & 0.408 & 0.492 & 0.383 & 0.122 & 0.428 & 0.500 & 0.502 \\
\midrule
 & & \multicolumn{15}{c}{Out-of-sample $R^2$} \\
\cmidrule(lr){3-17}
Period & RV (\%) & TED & AKX & AKX6 & OLS6 & LAS & ALAS & RID & ARID & ELA & AELA & SCA & ASCA & BAY & DAN & ADAN \\
\midrule
whole & 12.0 & 0.482 & 0.286 & 0.060 & 0.201 & 0.432 & 0.459 & 0.474 & 0.468 & 0.460 & 0.450 & 0.440 & 0.246 & 0.376 & 0.443 & 0.414 \\
2013  & 7.9  & 0.613 & 0.308 & 0.038 & 0.189 & 0.520 & 0.609 & 0.621 & 0.618 & 0.577 & 0.603 & 0.525 & 0.289 & 0.522 & 0.600 & 0.570 \\
2014  & 8.8  & 0.475 & 0.251 & 0.039 & 0.176 & 0.419 & 0.448 & 0.468 & 0.462 & 0.447 & 0.437 & 0.430 & 0.215 & 0.369 & 0.435 & 0.407 \\
2015  & 10.2 & 0.505 & 0.318 & 0.063 & 0.233 & 0.471 & 0.494 & 0.493 & 0.491 & 0.492 & 0.485 & 0.478 & 0.309 & 0.392 & 0.473 & 0.443 \\
2016  & 10.0 & 0.510 & 0.306 & 0.067 & 0.212 & 0.461 & 0.488 & 0.503 & 0.498 & 0.488 & 0.481 & 0.468 & 0.316 & 0.395 & 0.479 & 0.448 \\
2017  & 5.5  & 0.398 & 0.185 & 0.017 & 0.137 & 0.350 & 0.364 & 0.386 & 0.375 & 0.374 & 0.349 & 0.357 & 0.199 & 0.289 & 0.346 & 0.319 \\
2018  & 13.0 & 0.538 & 0.355 & 0.088 & 0.255 & 0.492 & 0.517 & 0.527 & 0.526 & 0.513 & 0.511 & 0.503 & 0.290 & 0.455 & 0.498 & 0.479 \\
2019  & 9.1  & 0.461 & 0.271 & 0.056 & 0.192 & 0.407 & 0.435 & 0.446 & 0.444 & 0.435 & 0.423 & 0.416 & 0.228 & 0.355 & 0.425 & 0.382 \\
2020  & 22.8 & 0.366 & 0.294 & 0.112 & 0.212 & 0.347 & 0.331 & 0.356 & 0.342 & 0.363 & 0.322 & 0.354 & 0.126 & 0.244 & 0.305 & 0.275 \\
\bottomrule
\end{tabular}%
}
\end{table}
To investigate the performance of the TED estimator and benchmarks, we first calculated the monthly in-sample and out-of-sample $R^2$ from the monthly integrated coefficient estimates.
We obtained the out-of-sample $R^2$ using the integrated coefficient estimates from the previous month.
Then, we calculated the annual average $R^2$ over the five assets and available months within each year.
Table \ref{Table1} reports the annual average in-sample and out-of-sample $R^2$ for the TED and benchmarks.
RV is the annualized realized volatility of the MKT factor for each period.
It is computed as the square root of the sum of squared 5-minute returns and reported in percent.
We included RV to examine how the performance of TED varies across different levels of market volatility.
From Table \ref{Table1}, we find that the average out-of-sample $R^2$ of TED exceeds all benchmarks.
Furthermore, its annual average out-of-sample $R^2$ exceeds those of the other benchmarks in every year except 2013, regardless of RV.
This is probably because only the TED estimator can account for both the high-dimensionality and time-varying nature of the coefficient process.
In terms of in-sample $R^2$, TED does not yield the highest value.
This is because conventional penalized regression methods aim to maximize the global penalized in-sample fit, while TED maximizes the penalized in-sample fit within each time-localized window.
Since coefficients vary over time, integrated coefficients obtained by aggregating local estimates do not necessarily guarantee the highest in-sample $R^2$.
Among the regression-based models (without handling time-varying coefficients), RIDGE yields the highest average out-of-sample $R^2$.
One possible reason for this observation is the presence of time-varying sparsity patterns.
If the sparsity patterns vary significantly over time, the support of active covariates aggregated across periods becomes substantially larger, resulting in a denser regime.
In this scenario, RID is expected to outperform sparsity-based penalties such as LAS, because the $\ell_2$ penalty provides more appropriate shrinkage when the aggregated support is enlarged.
We also note that TED effectively captures these time-varying sparsity patterns.

To examine the sensitivity of our results to methodological choices---such as tuning parameter selection, alternative approaches for estimating instantaneous coefficients, and an alternative inverse covariance estimation method---we conducted additional analyses.
Although our baseline setting generally outperformed alternative choices, other methodological choices provided comparable results.
These results confirm the robustness of our main findings.
In addition, we compared the predictive accuracy of TED by evaluating the mean absolute error (MAE) of the one-step-ahead fitted returns.
We confirmed that TED consistently outperformed other approaches in terms of this MAE as well.
We also evaluated the out-of-sample performance at various sampling frequencies (5-, 10-, and 30-min).
Details can be found in Appendix \ref{additional-empirical-analysis}.

\begin{figure}[ht!]
\centering
\includegraphics[width = 1\textwidth]{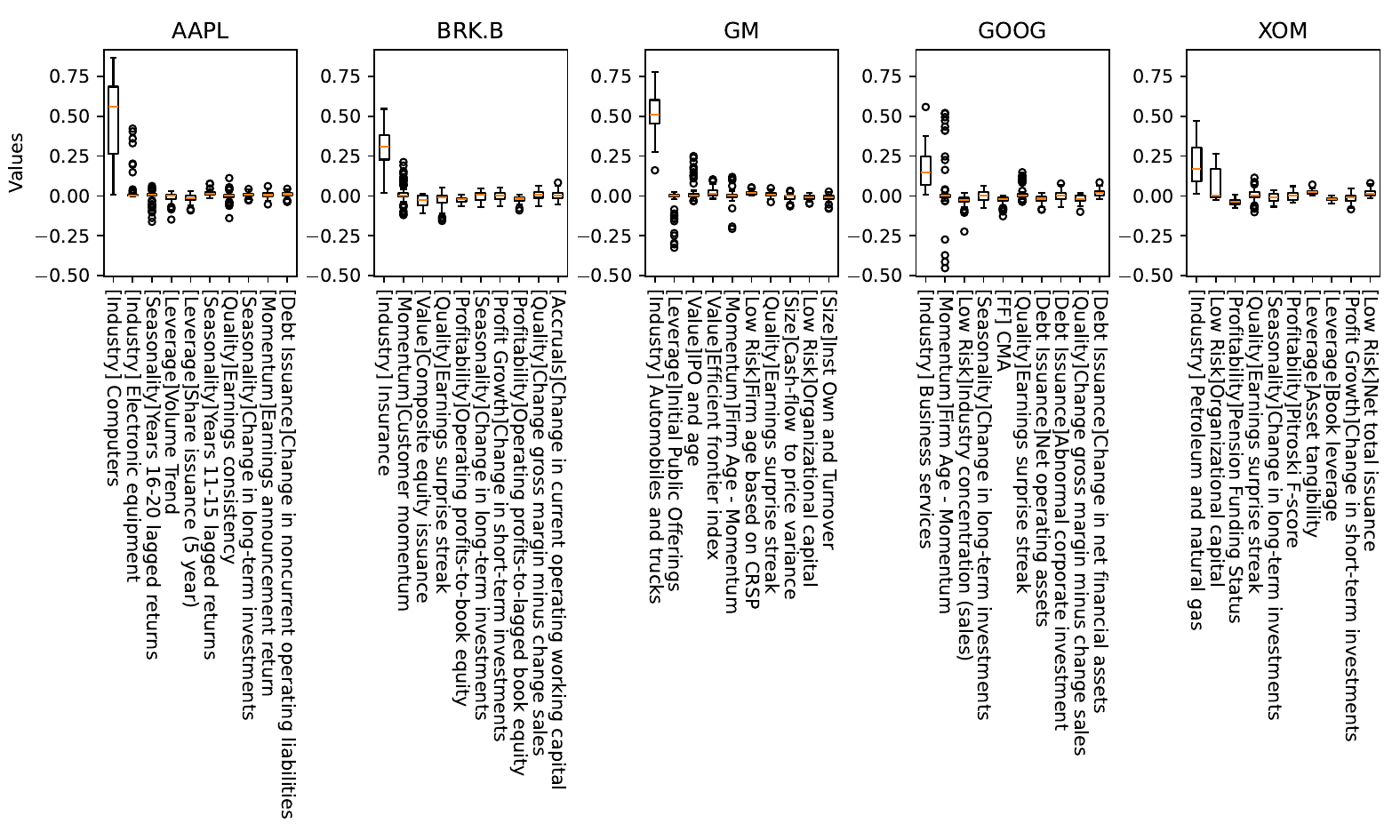}
\caption{Boxplots of integrated coefficients for the top 10 most influential factors in the factor zoo data.} \label{fig:emp_allvalue}
\end{figure}

\begin{figure}[ht!]
\centering
\includegraphics[width = 0.9\textwidth]{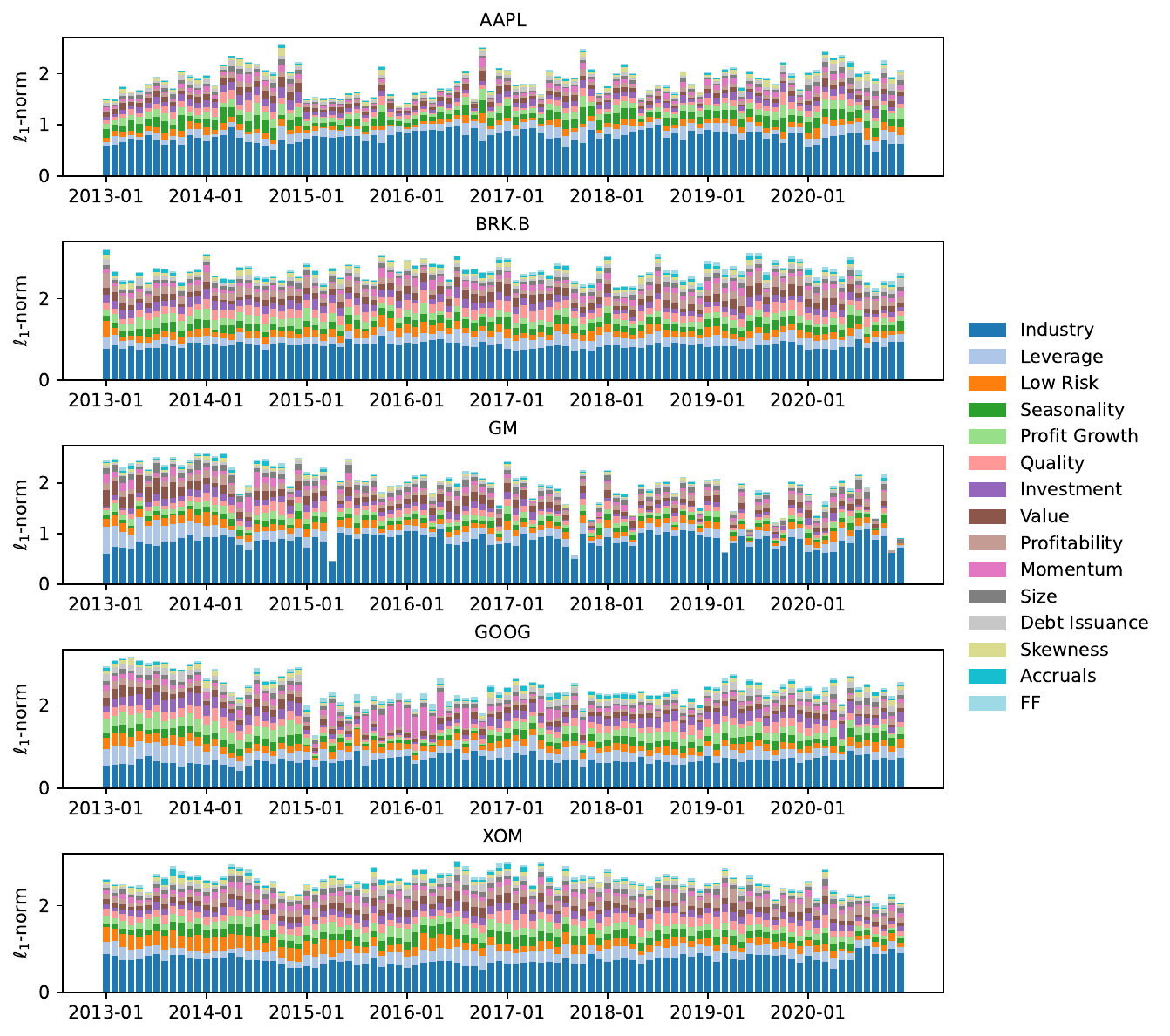}
\caption{Stacked bar charts of the absolute sum of the monthly integrated coefficients from the TED estimation procedure within each factor cluster for the five assets across 15 factor clusters.
    The 15 clusters comprise the 13 factor clusters in \citet{jensen2023there}, the FF cluster including the Fama-French five factors and the momentum factor, and the Industry cluster covering the industry factors of \citet{fama1997industry}.}
\label{fig:emp_nonzero_time}
\end{figure}

\begin{figure}[ht!]
\centering
\includegraphics[width = 1 \textwidth]{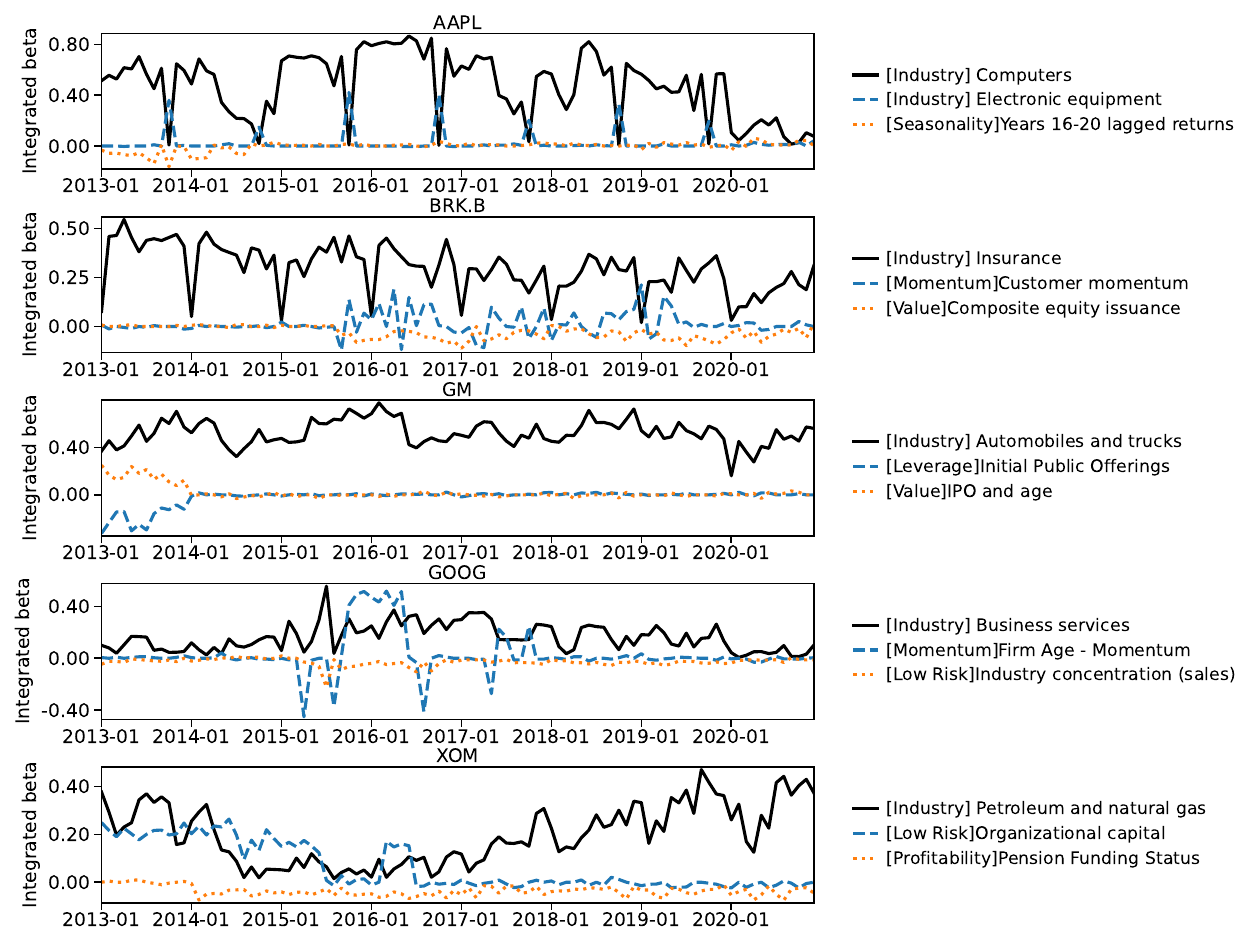}
\caption{Integrated coefficients from the TED estimation procedure for the three factors with the largest absolute sum of integrated coefficients over time from the factor zoo for each of the five assets.}
\label{fig:emp_mainfactors}
\end{figure}

Now, we investigate the TED estimation results.
Figure \ref{fig:emp_allvalue} shows boxplots of the integrated coefficients for the ten most influential factors, specifically those selected based on the largest average absolute values for each asset.
Figure \ref{fig:emp_nonzero_time} plots the absolute sum of the monthly integrated coefficients from the TED estimation procedure within each of the 15 factor clusters for the five assets across time.
The 15 clusters consist of the 13 factor clusters in \citet{jensen2023there}, the FF cluster including the Fama-French five factors and the momentum factor, and the Industry cluster covering the industry factors of \citet{fama1997industry}.
From Figures \ref{fig:emp_allvalue}--\ref{fig:emp_nonzero_time}, we find that the largest coefficient corresponds to the asset's own industry.
This may reflect the strong explanatory power of industry factors compared to other factors.
From Figure \ref{fig:emp_nonzero_time}, we also find that the value of the integrated coefficient varies over time and that most of the $\ell_1$-norms of the integrated coefficients are less than 3.
Since sparsity levels were determined in a data-driven manner via cross-validation, this result may suggest that the sparsity condition \eqref{sparsity_beta} holds empirically for some $\delta \in [0,1)$.
Several factors played a significant role in most periods. 
Among these, we focus on the three factors with the largest absolute sum of integrated coefficients over time for each of the five assets, and Figure \ref{fig:emp_mainfactors} illustrates their integrated coefficients.
From Figure \ref{fig:emp_mainfactors}, we find that industry factors generally exhibit the largest integrated coefficients.
Furthermore, the time-varying dynamics of prominent integrated coefficients among the selected factors clearly align with firm-specific events and thus offer an economic interpretation.
Specifically, for AAPL, the Computers industry factor generally dominates, while the Electronic Equipment industry factor consistently becomes prominent around September each year, except for 2020.
This pattern likely reflects Apple's seasonal new product announcements, suggesting that its performance becomes temporarily aligned with electronic-component supply chains.
For BRK.B, the customer momentum factor loading increases after mid-2015.
The customer momentum factor captures return spillovers from major customers to their suppliers \citep{cohen2008economic}.
This pattern is plausibly linked to Berkshire's acquisition of Precision Castparts Corporation (PCC), a large aerospace-parts supplier \citep{bruner2015warren}.
For GM, the IPO factor (short firms within three years of their IPO and long seasoned firms) and the IPO-and-age factor (long older IPO firms and short younger IPO firms within three years of listing)---both motivated by the long-run IPO performance patterns \citep{aleti2022high,chen2021open,ritter1991long}---have negative and positive loadings until January 2014, respectively.
These loadings suggest that GM initially exhibited return characteristics similar to recent IPO firms while simultaneously reflecting its status as an older firm at its November 2010 re-IPO.
For GOOG, the firm age-momentum factor loading \citep{zhang2006information} becomes prominent over the period from mid-2015 to mid-2016.
This may be because on August 10, 2015, Google announced the creation of Alphabet as a holding company to structurally separate its core internet services from more ambitious and innovation-oriented projects.
Following this reorganization, investors temporarily perceived the company's stock as similar to younger firms, which typically exhibit strong momentum.
Consequently, its returns may have exhibited greater exposure to the firm age-momentum factor.
For XOM, the organizational capital factor loading \citep{eisfeldt2013organization} steadily declines until mid-2015.
This decline likely coincides with Exxon Mobil's sustained reduction in selling, general, and administrative (SG\&A) expenses following the 2014--2016 oil price collapse \citep{prest2018explanations}\footnote{Exxon Mobil Corporation's Form 10-K for the fiscal year ended December 31, 2016 reports consolidated SG\&A expenses of \$12.6 billion in 2014, \$11.5 billion in 2015, and \$10.8 billion in 2016.}.

To compare the time-varying dynamics of the integrated coefficients from TED with those from benchmark estimators, we extend our analysis in Appendix \ref{additional-empirical-analysis}.
Specifically, we provide Figures \ref{fig:top5stack} and \ref{fig:top5ib}, which are analogous to Figures \ref{fig:emp_nonzero_time} and \ref{fig:emp_mainfactors}, respectively.
These figures include the five estimators with the highest average out-of-sample $R^2$ in Table \ref{Table1}, all of which assume constant coefficients except for TED.
The results show that for factors with large average absolute values of TED's integrated coefficients over time, the benchmark estimators reflect TED's substantial time-varying dynamics; however, for factors with smaller integrated coefficients, they may select different factors from those selected by TED.
Details can be found in Appendix \ref{additional-empirical-analysis}.

Finally, to interpret the results from the perspective of canonical asset pricing factors, we examined the integrated coefficients of the Fama-French five factors and the momentum factor.
The results of this analysis are provided in Appendix \ref{additional-empirical-analysis}.

From these results, we can infer that the coefficient processes are sparse and time-varying. 
Hence, incorporating these features is important to account for market dynamics.
The proposed TED procedure can provide an effective tool for dealing with these issues when analyzing market dynamics using high-frequency data.

\section{Conclusion}\label{SEC-6}

In this paper, we proposed a novel  Thresholding dEbiased Dantzig (TED) estimation procedure that can accommodate the sparse and time-varying coefficient process in the high-dimensional set-up. 
Specifically, to account for the sparse and time-varying coefficient process, we applied the Dantzig procedure to the instantaneous coefficient estimator, which results in a biased estimator. 
To reduce the bias, we proposed a debiased estimation procedure.
We estimated the integrated coefficient with this new debiased instantaneous coefficient estimator.
We showed that the Dantzig procedure can handle the sparsity of the instantaneous coefficient and that the debiased scheme mitigates the errors from the bias of the instantaneous coefficient estimator. 
To accommodate the sparsity of the integrated coefficient, we further regularized the coefficient estimator.
Finally, we showed that  the proposed TED estimator can obtain the near-optimal convergence rate.

In the empirical study, the TED estimator outperforms other estimators in terms of predictive accuracy in out-of-sample analysis.
Furthermore, we found that the coefficient process is sparse and time-varying.
These findings revealed that, when analyzing high-dimensional high-frequency regression, the TED estimator is a useful tool that can handle the curse of dimensionality and time-varying coefficients.
That is, in practice, the TED procedure makes it possible to analyze the stock market with a relatively short period using high-frequency data.

Statistical inference---hypothesis tests or confidence intervals---on the integrated coefficient estimator is also important and essential.
However, under the current debiasing scheme, the sum of non-martingale terms, $R_i$, is not negligible compared to the sum of martingale terms, which makes inference infeasible.
To solve this issue, we may need to develop a novel debiasing scheme for the biased integrated coefficient estimator.
This is a demanding task.
On the other hand, high-dimensional covariates often exhibit a factor structure, which can be addressed using a PCA-based approach.
However, applying this more elaborate model within a simple framework introduces substantial estimation error, which outweighs the benefits of alleviating model misspecification.
An interesting avenue for future work is to develop a method that explicitly incorporates the factor structure of covariate processes while providing a clearer explanation of market dynamics.
We leave these issues for future research.

\section*{Disclosure statement}
The authors report there are no competing interests to declare.

\section*{Data availability statement}
The data that support the findings of this study are available from the End of Day website (\url{https://eoddata.com}), firm fundamentals from the Center for Research in Security Prices (CRSP) and Compustat Merged Database, and the high-frequency factor returns from \url{https://www.sakethaleti.com/data}.
\section*{Funding}
The research of Donggyu Kim was supported in part by the National Research Foundation of Korea (NRF) grant funded by the Korea government (MSIT) (RS-2024-00343129).
The research of Minseok Shin was supported in part by the National Research Foundation of Korea (NRF) grant funded by the Korean government (MSIT) (RS-2025-24535699), and in part by the Institute of Information \& Communications Technology Planning \& Evaluation (IITP)-Global Data-X Leader HRD program grant funded by the Korean government (MSIT) (IITP-2025-RS-2024-00441244).

\bibliography{myReferences}

\begin{thebibliography}{}

\bibitem[A{\"\i}t-Sahalia et~al., 2020]{ait2020high}
A{\"\i}t-Sahalia, Y., Kalnina, I., and Xiu, D. (2020).
\newblock High-frequency factor models and regressions.
\newblock {\em Journal of Econometrics}, 216(1):86--105.

\bibitem[A{\"\i}t-Sahalia and Xiu, 2017]{ait2017using}
A{\"\i}t-Sahalia, Y. and Xiu, D. (2017).
\newblock Using principal component analysis to estimate a high dimensional
  factor model with high-frequency data.
\newblock {\em Journal of Econometrics}, 201(2):384--399.

\bibitem[A{\"\i}t-Sahalia and Xiu, 2019a]{ait2019hausman}
A{\"\i}t-Sahalia, Y. and Xiu, D. (2019a).
\newblock A hausman test for the presence of market microstructure noise in
  high frequency data.
\newblock {\em Journal of Econometrics}, 211(1):176--205.

\bibitem[A{\"\i}t-Sahalia and Xiu, 2019b]{ait2019principal}
A{\"\i}t-Sahalia, Y. and Xiu, D. (2019b).
\newblock Principal component analysis of high-frequency data.
\newblock {\em Journal of the American Statistical Association},
  114(525):287--303.

\bibitem[Aleti, 2022]{aleti2022high}
Aleti, S. (2022).
\newblock The high-frequency factor zoo.
\newblock {\em Available at SSRN 4021620}.

\bibitem[Andersen et~al., 2021]{andersen2021recalcitrant}
Andersen, T.~G., Thyrsgaard, M., and Todorov, V. (2021).
\newblock Recalcitrant betas: Intraday variation in the cross-sectional
  dispersion of systematic risk.
\newblock {\em Quantitative Economics}, 12(2):647--682.

\bibitem[Ang and Kristensen, 2012]{ang2012testing}
Ang, A. and Kristensen, D. (2012).
\newblock Testing conditional factor models.
\newblock {\em Journal of Financial Economics}, 106(1):132--156.

\bibitem[Asness et~al., 2013]{asness2013value}
Asness, C.~S., Moskowitz, T.~J., and Pedersen, L.~H. (2013).
\newblock Value and momentum everywhere.
\newblock {\em The Journal of Finance}, 68(3):929--985.

\bibitem[Barndorff-Nielsen et~al., 2011]{barndorff2011multivariate}
Barndorff-Nielsen, O.~E., Hansen, P.~R., Lunde, A., and Shephard, N. (2011).
\newblock Multivariate realised kernels: consistent positive semi-definite
  estimators of the covariation of equity prices with noise and non-synchronous
  trading.
\newblock {\em Journal of Econometrics}, 162(2):149--169.

\bibitem[Barroso and Santa-Clara, 2015]{barroso2015momentum}
Barroso, P. and Santa-Clara, P. (2015).
\newblock Momentum has its moments.
\newblock {\em Journal of Financial Economics}, 116(1):111--120.

\bibitem[Belloni et~al., 2014]{belloni2014inference}
Belloni, A., Chernozhukov, V., and Hansen, C. (2014).
\newblock Inference on treatment effects after selection among high-dimensional
  controls.
\newblock {\em The Review of Economic Studies}, 81(2):608--650.

\bibitem[Bickel and Levina, 2008]{bickel2008covariance}
Bickel, P.~J. and Levina, E. (2008).
\newblock Covariance regularization by thresholding.
\newblock {\em The Annals of Statistics}, pages 2577--2604.

\bibitem[Bickel et~al., 2009]{bickel2009simultaneous}
Bickel, P.~J., Ritov, Y., and Tsybakov, A.~B. (2009).
\newblock {Simultaneous analysis of LASSO and Dantzig selector}.
\newblock {\em The Annals of Statistics}, 37(4):1705--1732.

\bibitem[Boyd and Vandenberghe, 2004]{boyd2004convex}
Boyd, S.~P. and Vandenberghe, L. (2004).
\newblock {\em Convex optimization}.
\newblock Cambridge university press.

\bibitem[Breeden, 1979]{breeden1979intertemporal}
Breeden, D.~T. (1979).
\newblock An intertemporal asset pricing model with stochastic consumption and
  investment opportunities.
\newblock {\em Journal of Financial Economics}, 7(3):265--296.

\bibitem[Cai and Liu, 2011]{cai2011adaptive}
Cai, T. and Liu, W. (2011).
\newblock Adaptive thresholding for sparse covariance matrix estimation.
\newblock {\em Journal of the American Statistical Association},
  106(494):672--684.

\bibitem[Cai et~al., 2011]{cai2011constrained}
Cai, T., Liu, W., and Luo, X. (2011).
\newblock A constrained $\ell_1$ minimization approach to sparse precision
  matrix estimation.
\newblock {\em Journal of the American Statistical Association},
  106(494):594--607.

\bibitem[Cai and Zhou, 2012]{cai2012optimal}
Cai, T.~T. and Zhou, H.~H. (2012).
\newblock {Optimal rates of convergence for sparse covariance matrix
  estimation}.
\newblock {\em The Annals of Statistics}, 40(5):2389 -- 2420.

\bibitem[Candes and Tao, 2007]{candes2007dantzig}
Candes, E. and Tao, T. (2007).
\newblock The dantzig selector: Statistical estimation when p is much larger
  than n.
\newblock {\em The Annals of Statistics}, 35(6):2313--2351.

\bibitem[Carhart, 1997]{carhart1997persistence}
Carhart, M.~M. (1997).
\newblock On persistence in mutual fund performance.
\newblock {\em The Journal of Finance}, 52(1):57--82.

\bibitem[Chen et~al., 2024]{chen2024realized}
Chen, D., Mykland, P.~A., and Zhang, L. (2024).
\newblock Realized regression with asynchronous and noisy high frequency and
  high dimensional data.
\newblock {\em Journal of Econometrics}, 239(2):105446.

\bibitem[Chen, 2018]{chen2018inference}
Chen, R.~Y. (2018).
\newblock Inference for volatility functionals of multivariate {It\^{o}}
  semimartingales observed with jump and noise.
\newblock {\em arXiv preprint arXiv:1810.04725}.

\bibitem[Chinco et~al., 2019]{chinco2019sparse}
Chinco, A., Clark-Joseph, A.~D., and Ye, M. (2019).
\newblock Sparse signals in the cross-section of returns.
\newblock {\em The Journal of Finance}, 74(1):449--492.

\bibitem[Cio{\l}ek et~al., 2025]{ciolek2022lasso}
Cio{\l}ek, G., Marushkevych, D., and Podolskij, M. (2025).
\newblock On lasso estimator for the drift function in diffusion models.
\newblock {\em Bernoulli}, 31(3):1811--1833.

\bibitem[Cochrane, 2011]{cochrane2011presidential}
Cochrane, J.~H. (2011).
\newblock Presidential address: Discount rates.
\newblock {\em The Journal of Finance}, 66(4):1047--1108.

\bibitem[Cohen and Frazzini, 2008]{cohen2008economic}
Cohen, L. and Frazzini, A. (2008).
\newblock Economic links and predictable returns.
\newblock {\em The Journal of Finance}, 63(4):1977--2011.

\bibitem[DuBois and Bruner, 2017]{bruner2015warren}
DuBois, J. and Bruner, R.~F. (2017).
\newblock Warren e. buffett, 2015.
\newblock Darden Case Study UVA-F-1769, University of Virginia Darden School
  Foundation.
\newblock Revised November 8, 2017.

\bibitem[Dzhaparidze and Van~Zanten, 2001]{dzhaparidze2001bernstein}
Dzhaparidze, K. and Van~Zanten, J. (2001).
\newblock {On Bernstein-type inequalities for martingales}.
\newblock {\em Stochastic Processes and their Applications}, 93(1):109--117.

\bibitem[Eisfeldt and Papanikolaou, 2013]{eisfeldt2013organization}
Eisfeldt, A.~L. and Papanikolaou, D. (2013).
\newblock Organization capital and the cross-section of expected returns.
\newblock {\em The Journal of Finance}, 68(4):1365--1406.

\bibitem[Fama and French, 1997]{fama1997industry}
Fama, E.~F. and French, K.~R. (1997).
\newblock Industry costs of equity.
\newblock {\em Journal of Financial Economics}, 43(2):153--193.

\bibitem[Fama and French, 2015]{fama2015five}
Fama, E.~F. and French, K.~R. (2015).
\newblock A five-factor asset pricing model.
\newblock {\em Journal of Financial Economics}, 116(1):1--22.

\bibitem[Fama and French, 2016]{fama2016dissecting}
Fama, E.~F. and French, K.~R. (2016).
\newblock Dissecting anomalies with a five-factor model.
\newblock {\em The Review of Financial Studies}, 29(1):69--103.

\bibitem[Fan and Kim, 2018]{fan2018robust}
Fan, J. and Kim, D. (2018).
\newblock Robust high-dimensional volatility matrix estimation for
  high-frequency factor model.
\newblock {\em Journal of the American Statistical Association},
  113(523):1268--1283.

\bibitem[Fan et~al., 2017]{fan2017estimation}
Fan, J., Li, Q., and Wang, Y. (2017).
\newblock Estimation of high dimensional mean regression in the absence of
  symmetry and light tail assumptions.
\newblock {\em Journal of the Royal Statistical Society Series B: Statistical
  Methodology}, 79(1):247--265.

\bibitem[Fan and Li, 2001]{fan2001variable}
Fan, J. and Li, R. (2001).
\newblock Variable selection via nonconcave penalized likelihood and its oracle
  properties.
\newblock {\em Journal of the American Statistical Association},
  96(456):1348--1360.

\bibitem[Fan et~al., 2016]{fan2016overview}
Fan, J., Liao, Y., and Liu, H. (2016).
\newblock An overview of the estimation of large covariance and precision
  matrices.
\newblock {\em The Econometrics Journal}, 19(1):C1--C32.

\bibitem[Fan et~al., 2013]{fan2013large}
Fan, J., Liao, Y., and Mincheva, M. (2013).
\newblock Large covariance estimation by thresholding principal orthogonal
  complements.
\newblock {\em Journal of the Royal Statistical Society: Series B (Statistical
  Methodology)}, 75(4):603--680.

\bibitem[Feng et~al., 2020]{feng2020taming}
Feng, G., Giglio, S., and Xiu, D. (2020).
\newblock Taming the factor zoo: A test of new factors.
\newblock {\em The Journal of Finance}, 75(3):1327--1370.

\bibitem[Freyberger et~al., 2020]{freyberger2020dissecting}
Freyberger, J., Neuhierl, A., and Weber, M. (2020).
\newblock Dissecting characteristics nonparametrically.
\newblock {\em The Review of Financial Studies}, 33(5):2326--2377.

\bibitem[Ga{\"\i}ffas and Matulewicz, 2019]{gaiffas2019sparse}
Ga{\"\i}ffas, S. and Matulewicz, G. (2019).
\newblock Sparse inference of the drift of a high-dimensional
  ornstein--uhlenbeck process.
\newblock {\em Journal of Multivariate Analysis}, 169:1--20.

\bibitem[Harvey et~al., 2016]{harvey2016and}
Harvey, C.~R., Liu, Y., and Zhu, H. (2016).
\newblock … and the cross-section of expected returns.
\newblock {\em The Review of Financial Studies}, 29(1):5--68.

\bibitem[He et~al., 2024]{he2024no}
He, J., Zhao, L., and Zhou, G. (2024).
\newblock No sparsity in asset pricing: Evidence from a generic statistical
  test.
\newblock {\em Available at SSRN 4730259}.

\bibitem[Hou et~al., 2020]{hou2020replicating}
Hou, K., Xue, C., and Zhang, L. (2020).
\newblock Replicating anomalies.
\newblock {\em The Review of Financial Studies}, 33(5):2019--2133.

\bibitem[Jacod and Protter, 2011]{jacod2012discretization}
Jacod, J. and Protter, P.~E. (2011).
\newblock {\em Discretization of processes}, volume~67.
\newblock Springer Science \& Business Media.

\bibitem[Javanmard and Montanari, 2018]{javanmard2018debiasing}
Javanmard, A. and Montanari, A. (2018).
\newblock Debiasing the lasso: Optimal sample size for gaussian designs.
\newblock {\em The Annals of Statistics}, 46(6A):2593--2622.

\bibitem[Jensen et~al., 2023]{jensen2023there}
Jensen, T.~I., Kelly, B., and Pedersen, L.~H. (2023).
\newblock Is there a replication crisis in finance?
\newblock {\em The Journal of Finance}, 78(5):2465--2518.

\bibitem[Kim et~al., 2018]{kim2018adaptive}
Kim, D., Kong, X.-B., Li, C.-X., and Wang, Y. (2018).
\newblock Adaptive thresholding for large volatility matrix estimation based on
  high-frequency financial data.
\newblock {\em Journal of Econometrics}, 203(1):69--79.

\bibitem[Kim and Wang, 2016]{kim2016sparse}
Kim, D. and Wang, Y. (2016).
\newblock Sparse {PCA}-based on high-dimensional {It\^o} processes with
  measurement errors.
\newblock {\em Journal of Multivariate Analysis}, 152:172--189.

\bibitem[Kim et~al., 2016]{kim2016asymptotic}
Kim, D., Wang, Y., and Zou, J. (2016).
\newblock Asymptotic theory for large volatility matrix estimation based on
  high-frequency financial data.
\newblock {\em Stochastic Processes and their Applications}, 126:3527--3577.

\bibitem[Lee and Seregina, 2024]{lee2024optimal}
Lee, T.-H. and Seregina, E. (2024).
\newblock Optimal portfolio using factor graphical lasso.
\newblock {\em Journal of Financial Econometrics}, 22(3):670--695.

\bibitem[McLean and Pontiff, 2016]{mclean2016does}
McLean, R.~D. and Pontiff, J. (2016).
\newblock Does academic research destroy stock return predictability?
\newblock {\em The Journal of Finance}, 71(1):5--32.

\bibitem[Merton, 1973]{merton1973intertemporal}
Merton, R.~C. (1973).
\newblock An intertemporal capital asset pricing model.
\newblock {\em Econometrica: Journal of the Econometric Society}, pages
  867--887.

\bibitem[Mykland and Zhang, 2009]{mykland2009inference}
Mykland, P.~A. and Zhang, L. (2009).
\newblock Inference for continuous semimartingales observed at high frequency.
\newblock {\em Econometrica}, 77(5):1403--1445.

\bibitem[Negahban et~al., 2012]{negahban2012unified}
Negahban, S.~N., Ravikumar, P., Wainwright, M.~J., and Yu, B. (2012).
\newblock A unified framework for high-dimensional analysis of $ m $-estimators
  with decomposable regularizers.
\newblock {\em Statistical Science}, 27(4):538--557.

\bibitem[Oh and Kim, 2024]{oh2024property}
Oh, M. and Kim, D. (2024).
\newblock Property of inverse covariance matrix-based financial adjacency
  matrix for detecting local groups.
\newblock {\em arXiv preprint arXiv:2412.05664}.

\bibitem[Oh et~al., 2024]{oh2024robust}
Oh, M., Kim, D., and Wang, Y. (2024).
\newblock Robust realized integrated beta estimator with application to dynamic
  analysis of integrated beta.
\newblock {\em Journal of Econometrics}, page 105810.

\bibitem[Pang et~al., 2014]{pang2014fastclime}
Pang, H., Liu, H., and Vanderbei, R.~J. (2014).
\newblock The fastclime package for linear programming and large-scale
  precision matrix estimation in r.
\newblock {\em Journal of Machine Learning Research}.

\bibitem[Park and Casella, 2008]{park2008bayesian}
Park, T. and Casella, G. (2008).
\newblock The bayesian lasso.
\newblock {\em Journal of the American Statistical Association},
  103(482):681--686.

\bibitem[Prest, 2018]{prest2018explanations}
Prest, B.~C. (2018).
\newblock Explanations for the 2014 oil price decline: Supply or demand?
\newblock {\em Energy Economics}, 74:63--75.

\bibitem[Ritter, 1991]{ritter1991long}
Ritter, J.~R. (1991).
\newblock The long-run performance of initial public offerings.
\newblock {\em The Journal of Finance}, 46(1):3--27.

\bibitem[Sharpe, 1964]{sharpe1964capital}
Sharpe, W.~F. (1964).
\newblock Capital asset prices: A theory of market equilibrium under conditions
  of risk.
\newblock {\em The Journal of Finance}, 19(3):425--442.

\bibitem[Shin and Kim, 2023]{shin2023robust}
Shin, M. and Kim, D. (2023).
\newblock Robust high-dimensional time-varying coefficient estimation.
\newblock {\em arXiv preprint arXiv:2302.13658}.

\bibitem[Tao et~al., 2013]{tao2013optimal}
Tao, M., Wang, Y., and Zhou, H.~H. (2013).
\newblock Optimal sparse volatility matrix estimation for high-dimensional
  {It\^o} processes with measurement errors.
\newblock {\em The Annals of Statistics}, 41(4):1816--1864.

\bibitem[Tibshirani, 1996]{tibshirani1996regression}
Tibshirani, R. (1996).
\newblock Regression shrinkage and selection via the lasso.
\newblock {\em Journal of the Royal Statistical Society: Series B
  (Methodological)}, 58(1):267--288.

\bibitem[Van~de Geer et~al., 2014]{van2014asymptotically}
Van~de Geer, S., B{\"u}hlmann, P., Ritov, Y., and Dezeure, R. (2014).
\newblock On asymptotically optimal confidence regions and tests for
  high-dimensional models.
\newblock {\em The Annals of Statistics}, 42(3):1166--1202.

\bibitem[Wang and Zou, 2010]{wang2010vast}
Wang, Y. and Zou, J. (2010).
\newblock Vast volatility matrix estimation for high-frequency financial data.
\newblock {\em The Annals of Statistics}, 38:943--978.

\bibitem[Wang et~al., 2014]{wang2014optimal}
Wang, Z., Liu, H., and Zhang, T. (2014).
\newblock Optimal computational and statistical rates of convergence for sparse
  nonconvex learning problems.
\newblock {\em Annals of Statistics}, 42(6):2164.

\bibitem[Y.~Chen and Zimmermann, 2022]{chen2021open}
Y.~Chen, A. and Zimmermann, T. (2022).
\newblock Open source cross-sectional asset pricing.
\newblock {\em Critical Finance Review}, 11(02):207--264.

\bibitem[Ye and Zhang, 2010]{ye2010rate}
Ye, F. and Zhang, C.-H. (2010).
\newblock {Rate minimaxity of the Lasso and Dantzig selector for the $\ell_q$
  loss in $\ell_r$ balls}.
\newblock {\em The Journal of Machine Learning Research}, 11:3519--3540.

\bibitem[Yuan and Lin, 2006]{yuan2006model}
Yuan, M. and Lin, Y. (2006).
\newblock Model selection and estimation in regression with grouped variables.
\newblock {\em Journal of the Royal Statistical Society: Series B (Statistical
  Methodology)}, 68(1):49--67.

\bibitem[Zhang and Zhang, 2014]{zhang2014confidence}
Zhang, C.-H. and Zhang, S.~S. (2014).
\newblock Confidence intervals for low dimensional parameters in high
  dimensional linear models.
\newblock {\em Journal of the Royal Statistical Society Series B: Statistical
  Methodology}, 76(1):217--242.

\bibitem[Zhang, 2011]{zhang2011estimating}
Zhang, L. (2011).
\newblock Estimating covariation: Epps effect, microstructure noise.
\newblock {\em Journal of Econometrics}, 160(1):33--47.

\bibitem[Zhang, 2006]{zhang2006information}
Zhang, X.~F. (2006).
\newblock Information uncertainty and stock returns.
\newblock {\em The Journal of Finance}, 61(1):105--137.

\bibitem[Zou, 2006]{zou2006adaptive}
Zou, H. (2006).
\newblock The adaptive lasso and its oracle properties.
\newblock {\em Journal of the American Statistical Association},
  101(476):1418--1429.

\bibitem[Zou and Hastie, 2005]{zou2005regularization}
Zou, H. and Hastie, T. (2005).
\newblock Regularization and variable selection via the elastic net.
\newblock {\em Journal of the Royal Statistical Society Series B: Statistical
  Methodology}, 67(2):301--320.

\bibitem[Zou and Zhang, 2009]{zou2009adaptive}
Zou, H. and Zhang, H.~H. (2009).
\newblock On the adaptive elastic-net with a diverging number of parameters.
\newblock {\em The Annals of Statistics}, 37(4):1733.

\end{thebibliography}
\end{spacing}

\newpage
\appendix

\counterwithin{figure}{section}
\counterwithin{table}{section}
\counterwithin{algorithm}{section}

\begin{center}
{\LARGE Supplement to ``High-Dimensional Time-Varying Coefficient Estimation in Diffusion Models''\par}
\end{center}
\vspace*{1\baselineskip}

\begin{spacing}{1.68}
\section{Implementation details of TED}\label{sec:implementation}
\begin{figure}[h!]
\centering
\includegraphics[width = 0.93\textwidth]{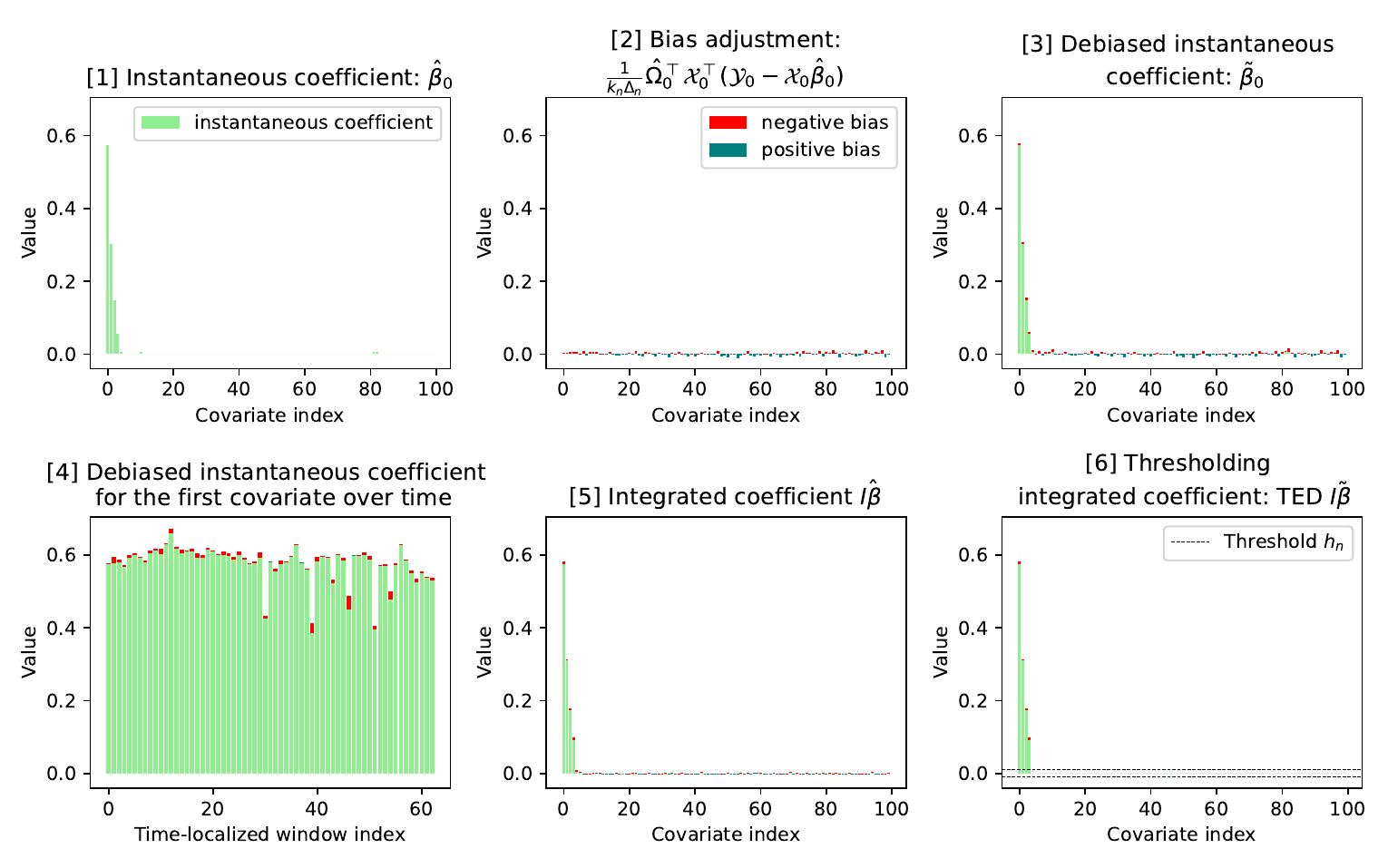}
\caption{Graphical illustration of the TED procedure.} \label{fig:graphical-example}
\end{figure}

In this section, we provide implementation details of the TED procedure, including the pseudocode in Algorithm \ref{TED-code}, graphical illustration in Figure \ref{fig:graphical-example}, and computational complexity analyses.
Specifically, Figure \ref{fig:graphical-example} illustrates the TED procedure with simulation data in Section \ref{SEC-4}.
The first panel displays the instantaneous coefficient $\hat{\bbeta}_0$ estimates obtained through the Dantzig selector.
We used the ``dantzig'' function in the fastclime package in R \citep{pang2014fastclime}.
The second panel shows the bias adjustment term, distinguishing between positive (blue) and negative (red) biases to debias the instantaneous coefficient.
For the CLIME estimator, we employed the ``fastclime'' function in the fastclime package.
The third panel presents the debiased instantaneous coefficient $\tilde{\bbeta}_0$ estimates, which are obtained by summing the estimates from the first two panels.
The fourth panel illustrates the iteratively calculated debiased instantaneous coefficient estimates over time.
The fifth panel shows the integrated coefficient estimates $\hat{I\beta}$, which is the integral of the debiased instantaneous coefficient estimates.
Finally, the last panel displays the TED estimates $\tilde{I\beta}$ after applying thresholding to the integrated coefficient estimates.

\begin{algorithm}[ht!]
\caption{TED pseudocode.}
\label{TED-code}
\begin{algorithmic}[1]

\Function{TED}{}

\State \textbf{input:} High frequency data $(\Delta^{n}_{i} Y, \Delta^{n}_{i} \bX)_{i=1,\ldots, n}$, tuning parameters $(k_n, \lambda_n, \tau_n, h_n)$, thresholding function $s(\cdot)$;

\State $N \leftarrow [1/(k_n \Delta_n)]$;

\For{$i=0, 1, \dots, N-1$}

\State $\mathcal{Y}_{i k_n} \leftarrow (\Delta_{i k_n + 1}^n Y, \Delta_{i k_n + 2}^n Y, \dots, \Delta_{(i+1) k_n}^n Y)^{\top}$;

\State $\mathcal{X}_{i k_n} \leftarrow (\Delta_{i k_n + 1}^n \bX, \Delta_{i k_n + 2}^n \bX, \dots, \Delta_{(i+1) k_n}^n \bX)^{\top}$;

\State $\hat{\bSigma}_{i k_n \Delta_n} \leftarrow \frac{1}{k_n \Delta_n} \mathcal{X}_{i k_n}^{\top} \mathcal{X}_{i k_n}$;

\State $\hat{\bSigma}_{XY, i k_n \Delta_n} \leftarrow \frac{1}{k_n \Delta_n} \mathcal{X}_{i k_n}^{\top} \mathcal{Y}_{i k_n}$;

\State $\hat{\bbeta}_{i k_n \Delta_n} \leftarrow \arg\min_{\bbeta} \|\bbeta\|_1 \quad \text{s.t.}\quad \left\|  \hat{\bSigma}_{i k_n \Delta_n} \bbeta - \hat{\bSigma}_{XY, i k_n \Delta_n}  \right\|_{\max}\leq \lambda_n$;

\State $\hat{\bOmega}_{i k_n \Delta_n} \leftarrow \arg\min_{\bOmega}\|\bOmega\|_1 \quad \text{s.t.}\quad \left\| \hat{\bSigma}_{i k_n \Delta_n}  \bOmega - \bI \right\|_{\max}\leq \tau_n$;

\State $\tilde{\bbeta}_{i k_n \Delta_n} \leftarrow \hat{\bbeta}_{i k_n \Delta_n} + \hat{\bOmega}_{i k_n \Delta_n}^{\top} \left(   \hat{\bSigma}_{XY, i k_n, \Delta_n} - \hat{\bSigma}_{i k_n \Delta_n} \hat{\bbeta}_{i k_n \Delta_n} \right) $;

\EndFor

\State $\hat{I\beta} \leftarrow \sum_{i=0}^{N-1}\tilde{\bbeta}_{i k_n \Delta_n} k_n \Delta_n$;

\For{$i=1,\ldots, p$}

\State $\tilde{I\beta}_i \leftarrow s(\hat{I\beta}_i) \mathbf{1}(|\hat{I\beta}_i| \geq h_n)$;

\EndFor

\State \textbf{Output:} $\tilde{I\beta} \leftarrow (\tilde{I\beta}_i)_{i=1,\dots,p}$;

\EndFunction

\end{algorithmic}
\end{algorithm}

When it comes to practical use, the time complexity of the TED procedure becomes an important consideration.
For the TED procedure, the most computationally intensive part is estimating the inverse covariance matrix using the CLIME estimator \citep{cai2011constrained}.
The CLIME estimator solves the following optimization problem:
\begin{equation*}
	\min \| \bOmega\|_1 \quad \text{s.t.} \quad  \|   \bSigma \bOmega - \bI  \|_{\max} \leq \tau_n, 
\end{equation*}
where $\bSigma = \frac{1}{k_n \Delta_n } \mathcal{X}_i  ^{\top}\mathcal{X} _i$.
The above optimization problem is equivalent to solving
\begin{eqnarray}\label{eq:clime-lp}
    \min \left( \bOmega_{\cdot i}^{+} +  \bOmega_{\cdot i}^{-} \right)^{\top} \1   \quad \text{s.t.} && \quad  \begin{pmatrix} \bSigma & - \bSigma \cr - \bSigma & \bSigma \end{pmatrix} \begin{pmatrix} \bOmega_{\cdot i}^{+}  \cr  \bOmega_{\cdot i}^{-} \end{pmatrix} \leq \begin{pmatrix} \tau_n \1 + e_i \cr \tau_n \1 - e_i  \end{pmatrix} \cr
    && \quad  \bOmega_{\cdot i}^{+} \geq 0 , \quad \text{and } \bOmega_{\cdot i}^{-} \geq 0 ,
\end{eqnarray}
for each $i=1, \ldots, p$ \citep{cai2011constrained}.
Using a primal-dual interior point method \citep{boyd2004convex}, each linear program can be solved in $O(p^3 \times \sqrt{p} \log {p})$ time.
Therefore, the total computational cost for estimating the inverse covariance matrix is $O(p^4  \sqrt{p} \log {p})$ time.
Dantzig selector \eqref{Dantzig} can be solved similarly using a primal-dual interior point method, but since there is only one optimization problem per time-localized window, the time complexity is $O(p^3  \sqrt{p} \log {p})$.
Since there are $O(n^{1/2})$ time-localized windows, the total time complexity for the TED procedure is $O(n^{1/2} p^4  \sqrt{p} \log {p})$.
Space complexity for the TED procedure is $O(p^2)$.
Therefore, the TED procedure can be implemented in polynomial time and space in terms of $p$ and $n$.

If $p$ is so large that the time complexity of TED is a critical issue in the application, we can use an alternative inverse covariance matrix estimator, such as Principal Orthogonal complEment Thresholding (POET) \citep{fan2013large}, or parallelize the TED procedure.
The POET procedure requires $O(p^2 r)$ time to compute the covariance and $O(p^3)$ time to compute the algebraic inverse, where $r$ is the number of common factors.
In this case, the total time complexity for the TED procedure becomes $O(n^{1/2} p^3  \sqrt{p} \log {p})$.

Alternatively, we can parallelize the TED procedure by computing each instantaneous coefficient independently.
Suppose $M$ processors are available.
Then, the total time complexity can be reduced to $O(n^{1/2}M^{-1}  p^4  \sqrt{p} \log {p} +  p^4  \sqrt{p} \log {p})$.
When the number of processors $M$ is large enough, specifically $M \gg n^{1/2}$, parallelizing the instantaneous coefficient estimations becomes inefficient.
In this case, it is preferable to parallelize the CLIME estimator, which is the most computationally intensive part of the instantaneous coefficient estimation.
Specifically, since each linear program \eqref{eq:clime-lp} can be solved independently, the CLIME estimator can be efficiently parallelized.
Then, each instantaneous coefficient estimator can be computed in $O( M^{-1} p^4 \sqrt{p} \log {p} +  p^3  \sqrt{p} \log {p})$ time.
In this case, the total time complexity becomes $O(n^{1/2} p^4 M^{-1} \sqrt{p} \log {p} + n^{1/2} p^3  \sqrt{p} \log {p})$.

\section{Conditions for extension of TED}\label{sec-extension}
Empirical data often exhibit specific features, such as microstructure noise or heavy-tailed distributions, that require robust estimation techniques.
While the TED procedure effectively handles high-dimensional and time-varying coefficients under regular conditions, practical applications might involve these additional complexities. %
If we can estimate instantaneous covariances with certain convergence rates, then we can directly utilize the TED framework by replacing the instantaneous covariance estimators $\frac{1}{k_n \Delta_n} \mathcal{X}_i^{\top} \mathcal{X}_i$ and $\frac{1}{k_n \Delta_n} \mathcal{X}_i^{\top} \mathcal{Y}_i$ with proper robust estimators $\hat{\bSigma}_{i k_n \Delta_n}$ and $\hat{\bSigma}_{XY, i k_n \Delta_n}$, respectively.
The instantaneous covariance estimators should satisfy the following error bounds with high probability:
\begin{eqnarray}\label{eq:two-errors}
    && \left\lVert \hat{\bSigma}_{i k_n \Delta_n} - \bSigma_{i k_n \Delta_n}  \right\rVert _{\max} \leq  C n^{-1/4} \sqrt{\log p }  \quad \text{and} \cr
    && \left\lVert  \hat{\bSigma}_{i k_n \Delta_n} \bbeta_{i k_n \Delta_n} - \hat{\bSigma}_{XY, i k_n \Delta_n}   \right\rVert _{\max} \leq  C s_p n^{-1/4} \sqrt{\log p} .
\end{eqnarray}
The first error is related to the estimation of the instantaneous covariance matrix $\bSigma_{i k_n \Delta_n}$.
The second error is associated with the smooth time-varying coefficient and the tail behavior of the residual process.
To robustly handle these empirical features, we can follow the ideas of \citet{ait2017using, barndorff2011multivariate, fan2018robust,kim2018adaptive, shin2023robust, zhang2011estimating}.
Under Assumption \ref{assumption1}, we can show that the instantaneous covariance estimators $\frac{1}{k_n \Delta_n} \mathcal{X}_i^{\top} \mathcal{X}_i$ and $\frac{1}{k_n \Delta_n} \mathcal{X}_i^{\top} \mathcal{Y}_i$ used in the TED procedure satisfy the error bounds \eqref{eq:two-errors}.
See \eqref{eq000} for details.

\section{Proofs} \label{SEC-proof}
\subsection{Proofs of Theorems \ref{Thm1} and \ref{Thm2}} 
Without loss of generality, it is enough to show the statements of Theorems \ref{Thm1} and \ref{Thm2} for fixed $i$. \\
   \textbf{Proof of Theorem \ref{Thm1}.} 
 We denote the true instantaneous coefficient  at time $i \Delta_n$ by $\bbeta_0$. 
 We have
 \begin{eqnarray*}
 	\Delta_{i+k} ^n Y  &=&   \int_{(i+k-1) \Delta_n } ^{(i+k)  \Delta_n} \bbeta_t ^{\top} d\bX_t + \int_{(i+k-1) \Delta_n }^{(i+k) \Delta_n}  d Z_t \cr
 		&=&      \bbeta_0 ^{\top} \Delta_{i+k} ^n  \bX + \Delta_{i+k} ^n  Z   +  \int_{(i+k-1) \Delta_n } ^{(i+k)  \Delta_n}   (\bbeta_t- \bbeta_0)^{\top} d\bX_t.
 \end{eqnarray*}
 Then, we have 
 \begin{equation}\label{eq:mathY-decomp}
  \mathcal{Y}_i =  \mathcal{X}_i \bbeta_0 + \mathcal{Z}_i + \mathcal{\tilde{X}}_i,
 \end{equation}
 where 
 \begin{eqnarray*}
   \mathcal{\tilde{X} }_i = 
\begin{pmatrix}
\int_{i \Delta_n } ^{ (i +1)  \Delta_n}   (\bbeta_t- \bbeta_0)  ^{\top} d\bX_t  \\ 
\int_{(i+1) \Delta_n } ^{(i+2)  \Delta_n}   (\bbeta_t- \bbeta_0)  ^{\top}  d\bX_t \\  
 \vdots \\ 
\int_{(i+k_n-1) \Delta_n } ^{(i+k_n)  \Delta_n}   (\bbeta_t- \bbeta_0) ^{\top}  d\bX_t   
\end{pmatrix} .
 \end{eqnarray*}
Since the instantaneous volatility and drift processes are bounded, $ \Delta_{i+k} ^n Z$ and $\Delta_{i+k} ^n X_j $ are sub-Gaussian. 
Then, similar to the proof of Theorem 1 \citep{kim2016sparse},  we can show, for some large $C$, 
  \begin{equation}\label{Thm1-eq1}
  P\( \max_{1 \leq j \leq p} \left|  \frac{1}{k_n \Delta_n}  \sum_{k=0}^ {k_n} \Delta_{i+k} ^n Z  \Delta_{i+k} ^n X_j \right| \geq C  \sqrt{   \log p / k_n}  \) \leq p ^{-1-a} ,
  \end{equation}
  where $X_j$ is the $j$th element of $\bX$.

  Consider $\frac{1}{k_n \Delta_n} \mathcal{X}_i ^{\top}  \mathcal{\tilde{X}}_i$. 
There exist standard Brownian motions, $W_{m t}^*$ and $B_{i t}^*$,  such that 
  \begin{equation*}
 d \beta_{mt} =  \mu_{\beta, mt} dt + \sqrt{ \Sigma_{\beta,mmt}} dW_{m t}^{*} \quad \text{and} \quad d X_{it} = \mu_{it} dt + \sqrt{ \Sigma_{iit}} dB_{i t} ^{*},
  \end{equation*}
 and let $\tilde{X}_{mt}= \int_{i \Delta_n } ^{t}   (\beta_{ms}- \beta_{m0} ) d X_{ms}$. 
 By It\^o's formula, we have
  \begin{eqnarray*}
    && \int_{(i+k-1) \Delta_n } ^{ (i+k)   \Delta_n}  d X_{jt}   \int_{(i+k-1) \Delta_n } ^{ (i+k)   \Delta_n}   (\beta_{mt}- \beta_{m0} ) d X_{mt} \cr
    &&=   \int_{(i+k-1) \Delta_n } ^{ (i+k)   \Delta_n}   (\beta_{mt}- \beta_{m0} )  \Sigma_{jm t}  d t \cr
    && \quad + \int_{(i+k-1) \Delta_n } ^{ (i+k)   \Delta_n} (\tilde{X}_{mt} -\tilde{X}_{m (i+k-1) \Delta_n})    d X_{jt} +    \int_{(i+k-1) \Delta_n } ^{ (i+k)   \Delta_n}   (X_{jt} - X_{j(i+k-1) \Delta_n  } )  d \tilde{X}_{mt} \cr 
    &&=   \int_{(i+k-1) \Delta_n } ^{ (i+k)   \Delta_n}   \int_{i \Delta_n} ^{t} \sqrt{ \Sigma_{\beta,mms}} dW_{m s}^{*}   \Sigma_{jm t}  d t   + \int_{(i+k-1) \Delta_n } ^{ (i+k)   \Delta_n}   \int_{i \Delta_n} ^{t} \mu_{\beta, m s} ds  \Sigma_{jm t} d t    \cr
   		&&\quad +  \int_{(i+k-1) \Delta_n } ^{ (i+k)   \Delta_n} (\tilde{X}_{mt} -\tilde{X}_{m (i+k-1) \Delta_n})   \sqrt{ \Sigma_{jjt}}  d B_{jt} ^{*}  +    \int_{(i+k-1) \Delta_n } ^{ (i+k)   \Delta_n} (\tilde{X}_{mt} -\tilde{X}_{m (i+k-1) \Delta_n}) \mu_{jt}   d t  \cr
   		&& \quad  + \int_{(i+k-1) \Delta_n } ^{ (i+k)   \Delta_n}   (X_{jt} - X_{j(i+k-1) \Delta_n  } )  \sqrt{\Sigma_{mmt}} (\beta_{mt}- \beta_{m0} )  d B_{mt} ^{*}  \cr
   		&& \quad +  \int_{(i+k-1) \Delta_n } ^{ (i+k)   \Delta_n}   (X_{jt} - X_{j(i+k-1) \Delta_n  } )    (\beta_{mt}- \beta_{m0} )  \mu_{mt} d t \cr
 	&&=  M_{1km} + D_{1 km}   +M_{2 km}  + D_{2 km}    +M_{3 km}  + D_{3 km}   \text{ a.s.}
  \end{eqnarray*}
   First, consider $D_{i km}$'s. 
By Assumption \ref{assumption1}(b)--(c), we have
  \begin{eqnarray*}
  \left | \frac{1}{k_n \Delta_n}  \sum_{m=1}^p  \sum_{k=1}^ {k_n}   D_{1 km}  \right |  \leq C s_p k_n \Delta_n   \text{ a.s.}
  \end{eqnarray*}
For $D_{2 km}$, by Assumption \ref{assumption1}(b)--(c), the process $\sum_{m=1}^p |\beta_{m t}- \beta_{m0} |$ has the sub-Gaussian tail and $\sum_{m=1}^p \sqrt{\Sigma_{\beta,mmt}} \leq C \sum_{m=1}^p |\Sigma_{\beta,mmt}|^{\delta/2} \leq C s_p$. 
Thus, we can show
  \begin{eqnarray} \label{eq001-thm1}
 &&P \(  \sup_{t \in [i \Delta_n,  \(i+k_n\) \Delta_n ] }  \sum_{m=1}^p |\beta_{m t}- \beta_{m0} |  \geq  C s_p \sqrt{ \Delta_n k_n \log p} \) \cr
 &&  \leq  C P \(  \sum_{m=1}^p |\beta_{m \(i+k_n\) \Delta_n }- \beta_{m0} |  \geq  C s_p \sqrt{ \Delta_n k_n \log p} \) \cr
 && \leq p^{-3-a-c/c_{1}}.   
  \end{eqnarray}
  Let 
\begin{equation*}
E= \left\{ \sup_{t \in [i \Delta_n,  \(i+k_n\) \Delta_n ] }  \sum_{m=1}^p |\beta_{m t}- \beta_{m0} |  \geq  C s_p \sqrt{ \Delta_n k_n \log p} \right\}.
\end{equation*}
  Then, we have
  \begin{eqnarray*}
  && P \( \sup_{t \in [ (i+k-1) \Delta_n,  (i+k) \Delta_n]} \sum_{m=1}^p |\tilde{X}_{mt} -\tilde{X}_{m (i+k-1) \Delta_n}|  \geq C  s_p \log p \sqrt{ \Delta_n ^2 k_n   } \)   \cr
  &&\leq C   P \(  \sum_{m=1}^p |\tilde{X}_{m  (i+k) \Delta_n} -\tilde{X}_{m (i+k-1) \Delta_n}|  \geq C  s_p \log p  \sqrt{ \Delta_n ^2 k_n  } \)  \cr
  && \leq C P \( \sum_{m=1}^p \left | \int_{(i+k-1) \Delta_n } ^{(i+k) \Delta_n}   (\beta_{mt}- \beta_{m0} ) d X_{mt}  \right |  \geq C  s_p \log p  \sqrt{ \Delta_n ^2 k_n  } ,E^c \) + p^{-3-a-c/c_{1}} \cr
  && \leq  Cp^{-3-a-c/c_{1}},
  \end{eqnarray*}
 which implies
   \begin{eqnarray}\label{eq002-thm1}
   P ( \sup_{1 \leq k \leq k_n}\sup_{t \in [ (i+k-1) \Delta_n,  (i+k) \Delta_n]} \sum_{m=1}^p |\tilde{X}_{mt} -\tilde{X}_{m (i+k-1) \Delta_n}|  \geq C  s_p \log p \sqrt{ \Delta_n ^2 k_n   } )   \leq Cp^{-3-a}.
  \end{eqnarray}
  Thus, we have, with probability at least $1- p^{-2-a}$, 
    \begin{equation*}
  \left | \frac{1}{k_n \Delta_n}  \sum_{m=1}^p  \sum_{k=1}^ {k_n}   D_{2  km}  \right |  \leq C s_p  \log p  \sqrt{ \Delta_n ^2 k_n  }.
  \end{equation*}
    Similarly, we can show, with probability at least $1- p^{-2-a}$, 
      \begin{equation*}
  \left | \frac{1}{k_n \Delta_n}  \sum_{m=1}^p  \sum_{k=1}^ {k_n}   D_{3  km}  \right |  \leq C s_p  \log p  \sqrt{ \Delta_n ^2 k_n  } .
  \end{equation*}

  Consider $M_{i km}$'s. 
  By Azuma-Hoeffding inequality, we have, with probability at least $1- p^{-2-a}$, 
      \begin{equation*}
  \left | \frac{1}{k_n \Delta_n}  \sum_{m=1}^p  \sum_{k=1}^ {k_n}  M_{1  km}  \right |  \leq C s_p    \sqrt{ \Delta_n  k_n  \log p }  .
  \end{equation*}
  For $M_{2km}$, let 
  \begin{equation*}
   \eta_{mt} = (\tilde{X}_{mt} -\tilde{X}_{m (i+k-1) \Delta_n})   \sqrt{ \Sigma_{jjt}} \quad \text{ and } \quad Q_t=\1_{\{\sum_{m=1}^p |\eta_{mt}| \leq  C  s_p \log p \sqrt{ \Delta_n ^2 k_n}\}}.
  \end{equation*}
  Then, by  Azuma-Hoeffding inequality, we have,  with probability at least $1-p^{-3-a}$,
  \begin{equation*}
  \left | \frac{1}{k_n \Delta_n}  \sum_{m=1}^p  \sum_{k=1}^ {k_n}   \int_{(i+k-1) \Delta_n } ^{ (i+k)   \Delta_n} \eta_{mt} Q_{t}  d B_{jt} ^{*}  \right |  \leq C s_p  \sqrt{\Delta_n}  ( \log p) ^{3/2}.    
  \end{equation*}
  Thus, by \eqref{eq002-thm1}, we have, with probability at least $1-p^{-2-a}$,  
      \begin{equation*}
  \left | \frac{1}{k_n \Delta_n}  \sum_{m=1}^p  \sum_{k=1}^ {k_n}   M_{2  km}  \right |  \leq C s_p  \sqrt{\Delta_n}  ( \log p) ^{3/2}.    
  \end{equation*}
  Similarly, we can show, with probability at least $1- p^{-2-a}$, 
  \begin{equation*}
  \left | \frac{1}{k_n \Delta_n}  \sum_{m=1}^p  \sum_{k=1}^ {k_n}   M_{3  km}  \right |  \leq C s_p  \sqrt{\Delta_n}  ( \log p) ^{3/2} .   
  \end{equation*}
  Therefore, we have, with probability at least $1- Cp^{-1-a}$, 
      \begin{eqnarray}\label{Thm1-eq2}
  	 && \max_{1 \leq j \leq p} \left | \frac{1}{k_n \Delta_n}  \sum_{k=1}^ {k_n} \Delta_{i+k} ^n X_j   \int_{(i+k-1) \Delta_n } ^{ (i+k)   \Delta_n}   (\bbeta_t- \bbeta_0)   ^{\top} d\bX_t   \right |  \cr
  	 &&\leq C    s_p    \sqrt{ \Delta_n  k_n  \log p }.   
  \end{eqnarray}

With probability greater than $1- p^{-1-a}$, we have
 \begin{eqnarray}\label{Thm1-eq3}
	\left \| \frac{d}{dt}   [\bX, \bX]_{i \Delta_n}  -   \frac{1}{k_n \Delta_n}   \mathcal{X} _i ^{\top} \mathcal{X}_i   \right  \| _{\max} &\leq&  \left \|  \bSigma_{i \Delta_n} -  \frac{1}{k_n \Delta_n} \int_{i \Delta_n}^{\(i+k_n\) \Delta_n}  \bSigma_s ds \right   \| _{\max}  \cr
	&& +  \left  \|  \frac{1}{k_n \Delta_n} \int_{i \Delta_n}^{\(i+k_n\) \Delta_n}  \bSigma_s ds -    \frac{1}{k_n \Delta_n}  \mathcal{X} _i ^{\top} \mathcal{X}_i \right  \| _{\max} \cr
&\leq& C \sqrt{ k_n \Delta_n} + C \sqrt{  \log p /k_n} ,
 \end{eqnarray}
  where $\frac{d}{dt}   [\bX, \bX] _t  = \bSigma_t$ and the last inequality can be showed similar to the proofs of Theorem 1 \citep{kim2016sparse}.
Thus, by \eqref{Thm1-eq1}, \eqref{Thm1-eq2}, and \eqref{Thm1-eq3}, we show the statement under the following statements:
\begin{eqnarray} \label{eq000}
	 && \left  \|  \frac{1}{k_n \Delta_n} \mathcal{X}_i ^{\top} \mathcal{Z}_i \right \|_{\max}  \leq C  \sqrt{   \log p / k_n}   , \cr
	 && \left  \| \frac{1}{k_n \Delta_n} \mathcal{X}_i ^{\top}  \mathcal{\tilde{X}}_i   \right \|_{\max} \leq C  s_p \sqrt{ k_n \Delta_n} \sqrt{\log p}, \cr
	 && \left \| \frac{d}{dt}   [\bX, \bX] _{i \Delta_n}  -   \frac{1}{k_n \Delta_n}  \mathcal{X} _i ^{\top} \mathcal{X}_i   \right  \| _{\max}  \leq C   ( \sqrt{ k_n \Delta_n} +   \sqrt{  \log p /k_n} ) .
\end{eqnarray}

 By \eqref{eq000}, we have, for some large $C_{\lambda,a}$, 
  \begin{eqnarray*}
  	 && \left \|  \frac{1}{k_n \Delta_n} \mathcal{X}_i ^{\top} \mathcal{X} _i  \bbeta _{0}   -   \frac{1}{k_n \Delta_n} \mathcal{X}_i ^{\top} \mathcal{Y}_i  \right  \|_{\max} \cr
  	  &&=   \left	 \|  \frac{1}{k_n \Delta_n}  \mathcal{X}_i ^{\top} \mathcal{X} _i  \bbeta _{0}   -  \frac{1}{k_n \Delta_n} \mathcal{X}_i ^{\top} \mathcal{X}_i  \bbeta_0  -  \frac{1}{k_n \Delta_n}\mathcal{X}_i ^{\top} \mathcal{Z}_i -  \frac{1}{k_n \Delta_n}\mathcal{X}_i ^{\top}  \mathcal{\tilde{X}}_i  \right  \|_{\max} \cr
  	 	&&\leq  \left  \|  \frac{1}{k_n \Delta_n} \mathcal{X}_i ^{\top} \mathcal{Z}_i \right \|_{\max} + \left  \| \frac{1}{k_n \Delta_n} \mathcal{X}_i ^{\top}  \mathcal{\tilde{X}}_i   \right \|_{\max} \cr
  	 	&&\leq \lambda_n.
  \end{eqnarray*}
 Thus, $\bbeta_0$ satisfies the constraint in \eqref{Dantzig}, which implies that
\begin{equation*}
	\| \hat{\bbeta}_{i \Delta_n} \|_1 \leq \| \bbeta_0\| _1.
\end{equation*}  
We have
\begin{eqnarray*}
	\|   \frac{1}{k_n \Delta_n}   \mathcal{X}_i ^{\top} \mathcal{X} _i  ( \hat{\bbeta}_{i \Delta_n} -\bbeta_0  )  \|_{\max}   &\leq&  \|  \frac{1}{k_n \Delta_n}   \mathcal{X}_i ^{\top} \mathcal{X} _i   \hat{\bbeta}_{i \Delta_n} -   \frac{1}{k_n \Delta_n}  \mathcal{X}_i ^{\top} \mathcal{Y}_i   \|_{\max}  \cr
	&& + \|  \frac{1}{k_n \Delta_n}   \mathcal{X}_i ^{\top} \mathcal{X} _i  \bbeta_0 -    \frac{1}{k_n \Delta_n} \mathcal{X}_i ^{\top} \mathcal{Y}_i   \|_{\max}   \cr
		&\leq& 2 \lambda_n  
\end{eqnarray*}
and
\begin{eqnarray*}
	\|  \bSigma _{i \Delta_n}  ( \hat{\bbeta}_{i \Delta_n} -\bbeta_0  )  \|_{\max}   &\leq&  \|   \frac{1}{k_n \Delta_n}    \mathcal{X}_i ^{\top} \mathcal{X} _i  ( \hat{\bbeta}_{i \Delta_n} -\bbeta_0  )  \|_{\max} \cr
	   && + \|  ( \bSigma_{i \Delta_n} -   \frac{1}{k_n \Delta_n}   \mathcal{X}_i ^{\top} \mathcal{X} _i  ) ( \hat{\bbeta}_{i \Delta_n} -\bbeta_0  )  \|_{\max} \cr
		&\leq& 2 \lambda_n + 2 \| \bbeta_0  \|_1 \| \bSigma_{i \Delta_n} -   \frac{1}{k_n \Delta_n}   \mathcal{X}_i ^{\top} \mathcal{X} _i   \| _{\max} \cr
			&\leq& 2 \lambda_n +C  s_p     ( \sqrt{ k_n \Delta_n } +   \sqrt{  \log p /k_n} )   \cr
			&\leq& C \lambda_n,
\end{eqnarray*}
where the third inequality is due to \eqref{eq000} and the sparsity condition \eqref{sparsity_beta}. 
Then, we have
\begin{eqnarray*}
	\| \hat{\bbeta}_{i \Delta_n} -\bbeta_0 \|_{\max}  & \leq&  \| \bSigma_{i \Delta_n} ^{-1} \|_1  	\|  \bSigma_{i \Delta_n} ( \hat{\bbeta}_{i \Delta_n} -\bbeta_0  )  \|_{\max}      \cr
 	&\leq& C \lambda_n,   
\end{eqnarray*}
where the last inequality is due to  Assumption \ref{assumption1}(b).

Now, we consider the $\ell_1$ norm error bound.
Let $a_n = 	\| \hat{\bbeta}_{i \Delta_n} -\bbeta_0 \|_{\max}$. 
Define 
\begin{eqnarray*}
	&& A= \hat{\bbeta}_{i \Delta_n} -\bbeta_0, \cr
	&& A_1= \( \hat{\beta}_{j i \Delta_n} 1 (  | \hat{\beta}_{j i \Delta_n} | \geq 2 a_n ) ; 1 \leq j \leq p \) ^{\top}   -\bbeta_0.
\end{eqnarray*}
Then, we have
\begin{equation*}
	\|\bbeta_0\|_1 - \| A_1 \|_1 + \| A- A_1\|_1 \leq \| A_1 +\bbeta_0\|_1 + \| A-A_1\|_1 = \|  \hat{\bbeta}_{i \Delta_n} \|_1 \leq \| \bbeta_0\|_1,
\end{equation*}
which implies
\begin{equation*}
	\| A\|_1 \leq \| A -A_1\|_1 + \|A_1\|_1  \leq 2 \|A_1\|_1.
\end{equation*}
Therefore, it is enough to investigate the convergence rate of  $\| A_1 \|_1$. 
We have
\begin{eqnarray} \label{eq-l1}
	\| A_1 \|_1 &=&  \sum_{j=1}^p  | \hat{\beta}_{j i \Delta_n} 1 (  | \hat{\beta}_{j i \Delta_n} | \geq 2 a_n )      -\beta_{j 0} |  \cr
	&\leq& \sum_{j=1}^p  | \hat{\beta}_{j i \Delta_n}   1 (  | \hat{\beta}_{j i \Delta_n} | \geq 2 a_n )    -\beta_{j 0}    1 (  | \beta_{j 0}  | \geq 2 a_n )  |    +  \sum_{j=1}^p    |\beta_{j 0}|  1 (  |\beta_{j 0}  | \leq 2 a_n )  \cr 
		&\leq& \sum_{j=1}^p  | \hat{\beta}_{j i \Delta_n}   1 (  | \hat{\beta}_{j i \Delta_n} | \geq 2 a_n )    -\beta_{j 0}    1 (  | \beta_{j 0}  | \geq 2 a_n )  |    +    C s_p a_n^{1-\delta} \cr 
		&\leq& a_n \sum_{j=1}^p    1 (  | \hat{\beta}_{j i \Delta_n} | \geq 2 a_n )    +  \sum_{j=1}^p | \beta_{j 0}| |     1 (  | \beta_{j 0}  | \geq 2 a_n ) -1 (  | \hat{\beta}_{j i \Delta_n} | \geq 2 a_n )   |    +    C s_p a_n^{1-\delta} \cr  
		&\leq&  a_n \sum_{j=1}^p   1 (  |\beta_{j 0}  | \geq   a_n )    +  \sum_{j=1}^p | \beta_{j 0}| |     1 (  | \beta_{j 0}  | \geq 2 a_n ) -1 (  | \hat{\beta}_{j i \Delta_n} | \geq 2 a_n )   |    +    C s_p a_n^{1-\delta} \cr 
		&\leq&     \sum_{j=1}^p | \beta_{j 0}| |     1 (  | \beta_{j 0}  | \geq 2 a_n ) -1 (  | \hat{\beta}_{j i \Delta_n} | \geq 2 a_n )   |    +    C s_p a_n^{1-\delta} \cr 
		&\leq&     \sum_{j=1}^p | \beta_{j 0}|    1 (   | | \beta_{j 0}  |- 2 a_n | \leq | | \beta_{j 0}  | -   | \hat{\beta}_{j i \Delta_n} ||  )    +    C s_p a_n^{1-\delta} \cr 
		&\leq&     \sum_{j=1}^p | \beta_{j 0}|     1 (   |  \beta_{j 0}  | \leq 3  a_n   )    +    C s_p a_n^{1-\delta} \cr  
		&\leq& C s_p a_n^{1-\delta}.  
\end{eqnarray}
\endpf

 \textbf{Proof of Theorem \ref{Thm2}.} 

We denote the inverse matrix of the true instantaneous volatility matrix at time $i \Delta_n$ by $\bOmega_0$. 
By \eqref{Thm1-eq3}, we have, with probability greater than $1-p^{-1-a}$,  
\begin{eqnarray*}
\left\|   \frac{1}{k_n \Delta_n }    \mathcal{X}_i  ^{\top}\mathcal{X} _i \bOmega_0 - \bI  \right\|_{\max} &=& \left\|   \(\frac{1}{k_n \Delta_n }    \mathcal{X}_i  ^{\top}\mathcal{X} _i  - \frac{d}{dt}   [\bX, \bX] _{i \Delta_n}  \) \bOmega_0  \right\|_{\max}  \cr
	&\leq& \left\| \bOmega_0 \right\| _1  \left\|   \(\frac{1}{k_n \Delta_n }    \mathcal{X}_i  ^{\top}\mathcal{X} _i  - \frac{d}{dt}   [\bX, \bX] _{i \Delta_n}  \)   \right\|_{\max} \cr
	&\leq& C   ( \sqrt{ k_n \Delta_n} +   \sqrt{  \log p /k_n} ).
\end{eqnarray*}
Thus, by the constraint in \eqref{CLIME}, we have
\begin{equation}\label{Thm2-eq1}
P \left(\left\| \hat{\bOmega}_{i \Delta_n} \right\| _1 \leq C\right) \geq 1-p^{-1-a}.
\end{equation}
Then, similar to the proofs of Theorem \ref{Thm1}, we can show that the statement holds.
\endpf

\subsection{Proof of Theorem \ref{Thm3}}
\textbf{Proof of Theorem \ref{Thm3}.} 
Consider \eqref{Thm3-result1-1} and \eqref{Thm3-result1-2}. 
Without loss of generality, it is enough to show \eqref{Thm3-result1-1} and \eqref{Thm3-result1-2} for fixed $i$.
We have
\begin{align}\label{eq:tbeta-bias}
\tilde{\bbeta}_{i \Delta_n} - \bbeta_{0,i \Delta_n} =&  \frac{1}{k_n \Delta_n}  \bOmega _{0,i \Delta_n}   \mathcal{X}_i ^{\top} \mathcal{Z}_i   + \frac{1}{k_n \Delta_n} \hat{\bOmega}_{i \Delta_n}^{\top}  \mathcal{X}_i ^{\top} \mathcal{\tilde{X}}_i  \notag \\
& - \( \frac{1}{k_n \Delta_n} \hat{\bOmega}_{i \Delta_n}^{\top}  \mathcal{X}_i ^{\top}   \mathcal{X}_i  -\bI  \)   (\hat{\bbeta} _{i \Delta_n} - \bbeta_{0,i \Delta_n}) + \frac{1}{k_n \Delta_n} ( \hat{ \bOmega}_{i \Delta_n}^{\top}  - \bOmega_{0,i \Delta_n}  )  \mathcal{X}_i ^{\top} \mathcal{Z}_i.
\end{align}
For $ \frac{1}{k_n \Delta_n} \hat{\bOmega}_{i \Delta_n}  \mathcal{X}_i ^{\top} \mathcal{\tilde{X}}_i $, by the proofs of \eqref{Thm1-eq2}, we have
\begin{eqnarray*}
  \frac{1}{k_n \Delta_n} \bOmega_{0,i \Delta_n}  \mathcal{X}_i ^{\top} \mathcal{\tilde{X}}_i  =  \frac{1}{k_n \Delta_n} \bOmega_{0,i \Delta_n}  \mathcal{A}_i   + R_{X,i} ,
\end{eqnarray*}
where $\|R_{X,i}\|_{\max} \leq C s_p  (\log p)^{3/2} \sqrt{\Delta_n} $ with probability greater than $1-Cp^{-1-a}$, and 
\begin{equation}\label{def-a}
 \mathcal{A}_i = \(  \sum_{m=1}^p    \int_{ i \Delta_n } ^{ \(i+k_n\) \Delta_n}   \int_{i \Delta_n} ^{t} \sqrt{\Sigma_{\beta,mms}} dW_{m s}^{*}    \Sigma_{jm t}  d t \)_{j=1, \ldots,p}.
\end{equation}
By Theorem \ref{Thm2} and \eqref{Thm1-eq2}, we  have, with probability greater than $1-Cp^{-1-a}$, 
\begin{eqnarray*}
 \left \|  \frac{1}{k_n \Delta_n}  \( \bOmega_{0,i \Delta_n}- \hat{\bOmega}_{i \Delta_n}^{\top}  \)  \mathcal{X}_i ^{\top} \mathcal{\tilde{X}}_i  \right \| _{\max} \leq C s_{\omega, p}  s_p  \(\sqrt{\log p  / n^{ 1/2} }  \) ^{2-q}.   
\end{eqnarray*}
Thus, we have, with probability greater than $1-Cp^{-1-a}$,
\begin{eqnarray}\label{Thm3-eq1}
 \frac{1}{k_n \Delta_n} \hat{\bOmega}_{i \Delta_n}^{\top}  \mathcal{X}_i ^{\top} \mathcal{\tilde{X}}_i=   \frac{1}{k_n \Delta_n} \bOmega_{0,i \Delta_n}  \mathcal{A}_i   + R_{X,i}^{'} ,
\end{eqnarray}
where  $\| R_{X,i}^{'} \|_{\max} \leq C \left \{s_{\omega, p}  s_p  \(  \sqrt{\log p  / n^{ 1/2} }  \) ^{2-q} + s_p  (\log p)^{3/2} \sqrt{\Delta_n}\right \}$. 

For  $\( \frac{1}{k_n \Delta_n} \hat{\bOmega}_{i \Delta_n}^{\top}  \mathcal{X}_i ^{\top}   \mathcal{X}_i  -\bI  \)   (\hat{\bbeta} _{i \Delta_n} - \bbeta_{0,i \Delta_n})$, we have, with probability greater than $1-p^{-1-a}$, 
\begin{eqnarray}\label{Thm3-eq2}
&&	\| \( \frac{1}{k_n \Delta_n} \hat{\bOmega}_{i \Delta_n}^{\top}  \mathcal{X}_i ^{\top}   \mathcal{X}_i  -\bI  \)   (\hat{\bbeta} _{i \Delta_n} - \bbeta_{0,i \Delta_n})  \|_{\max} \cr
&& \leq 	\|   \frac{1}{k_n \Delta_n} \hat{\bOmega}_{i \Delta_n}^{\top}  \mathcal{X}_i ^{\top}   \mathcal{X}_i  -\bI \| _{\max}     \| \hat{\bbeta} _{i \Delta_n} - \bbeta_{0,i \Delta_n}   \| _{1}  \cr
&& \leq  C \tau_n s_p  \lambda _n ^{ 1-\delta} \cr
&& \leq   C   s_p^{2-\delta}     (\log p )   ^{1-\delta/2}  \( \sqrt{k_n \Delta_n} + k_n ^{-1/2}   \)  ^{2-\delta},
\end{eqnarray}
where the second inequality is by  Theorem \ref{Thm1} and \eqref{CLIME}.
For the last term, we have
\begin{align}\label{Thm3-eq3}
	 \| \frac{1}{k_n \Delta_n} ( \hat{ \bOmega}_{i \Delta_n}^{\top}  - \bOmega_{0,i \Delta_n}  )  \mathcal{X}_i ^{\top} \mathcal{Z}_i  \|_{\max} &\leq  \| \hat{ \bOmega}_{i \Delta_n}^{\top}  - \bOmega_{0,i \Delta_n}  \|_{\infty}  	 \| \frac{1}{k_n \Delta_n}  \mathcal{X}_i ^{\top} \mathcal{Z}_i  \|_{\max}  \cr
	 	&\leq  C  s_{\omega, p} \tau_n ^{1-q} \sqrt{\log p/ k_n}    \cr
	&\leq C s_{\omega ,p}  ( \log p) ^{1-q/2} (\sqrt{ k_n \Delta_n} + k_n ^{-1/2}) ^{1-q} k_n ^{-1/2} ,
\end{align}
where the second inequality is due to Theorem \ref{Thm2} and \eqref{Thm1-eq1}, with probability greater than $1-Cp^{-1-a}$.
By \eqref{Thm3-eq1}, \eqref{Thm3-eq2}, and \eqref{Thm3-eq3}, we have, with probability greater than $1-p^{-a}$,
\begin{equation}\label{Thm3-eq4}
\tilde{\bbeta}_{i \Delta_n}  - \bbeta_{0,i \Delta_n}=  \frac{1}{k_n \Delta_n} \bOmega_{0,i \Delta_n}  \(   \mathcal{X}_i ^{\top}\mathcal{Z}_i  + \mathcal{A}_i \)+R_i ,
\end{equation}
where  
$$\| R_i \|_{\max}  \leq  C   \left\{   s_p ^{2-\delta}   ( \log p / n^{1/2}  )^{(2-\delta)/2}+  s_p s_{\omega,p}    ( \log p / n^{1/2}  )^{(2-q)/2}  + s_p  (\log p)^{3/2}  / n^{1/2} \right\}.$$

Consider \eqref{Thm3-result2}.
We have
\begin{eqnarray*}
   \hat{I\beta}  - I\beta_0  & =&    k_n \Delta_n    \sum_{i=0}^{[1/(k_n \Delta_n) ]-1} \( \tilde{\bbeta}_{i k_n \Delta_n}  - \bbeta_{0,i k_n \Delta_n} \)  \cr
   	&&    + \sum_{i=0}^{[1/(k_n \Delta_n) ]-1}    \int_{i k_n \Delta_n}  ^{(i+1)  k_n \Delta_n}   ( \bbeta_{0,i k_n \Delta_n} - \bbeta_{0,t} )  dt -\int_{[1/(k_n \Delta_n) ]k_n \Delta_n}^{1} \bbeta_{0,t}  dt.
\end{eqnarray*}
First, we consider the discretization error terms. 
Since the coefficient process has the sub-Gaussian tail, we can show, with probability greater than $1-p^{-1-a}$, 
\begin{eqnarray*}
 \left \|   \sum_{i=0}^{[1/(k_n \Delta_n) ]-1}    \int_{i k_n \Delta_n}  ^{(i+1)  k_n \Delta_n}   ( \bbeta_{0,i k_n \Delta_n} - \bbeta_{0,t} )  dt    \right \| _{\max}\leq C   \sqrt{  \log p / n}.
\end{eqnarray*}
Also, by Assumption \ref{assumption1}(b), we have
\begin{eqnarray*}
\left \| \int_{[1/(k_n \Delta_n) ]k_n \Delta_n}^{1} \bbeta_{0,t}  dt    \right \| _{\max} \leq Cn^{-1/2} \text{ a.s.}
\end{eqnarray*}
Consider $ \sum_{i=0}^{[1/(k_n \Delta_n) ]-1} \( \tilde{\bbeta}_{i k_n \Delta_n}  - \bbeta_{0,i k_n \Delta_n} \)$.
By \eqref{Thm3-result1-1}, we have
\begin{eqnarray*}
 \sum_{i=0}^{[1/(k_n \Delta_n) ]-1} \( \tilde{\bbeta}_{i k_n \Delta_n}  - \bbeta_{0,i k_n \Delta_n} \) = \sum_{i=0}^{[1/(k_n \Delta_n) ]-1} \frac{1}{k_n \Delta_n} \bOmega_{0,i k_n \Delta_n}  \(   \mathcal{X}_{i k_n } ^{\top}\mathcal{Z}_{i k_n }  + \mathcal{A}_{i k_n } \)+R_{i k_n },
\end{eqnarray*}
and, similar to the proofs of \eqref{Thm3-eq4}, we can show, with probability greater than $1-p^{-1-a}$, 
\begin{eqnarray*}
 && \left \| \sum_{i=0}^{[1/(k_n \Delta_n) ]-1}  R_{i k_n } \right \| _{\max}  \leq   \sum_{i=0}^{[1/(k_n \Delta_n) ]-1}   \left \|  R_{i k_n } \right \| _{\max}  \cr
 &&\leq    C  \frac{1}{ k_n \Delta_n}    \left \{   s_p ^{2-\delta}   ( \log p / n^{1/2}  )^{(2-\delta)/2}+  s_p s_{\omega,p}    ( \log p / n^{1/2}  )^{(2-q)/2}  + s_p  (\log p)^{3/2}  / n^{1/2} \right\}.
\end{eqnarray*}
Since the inverse matrix process $\bOmega_{0,i k_n \Delta_n} $ is bounded and $ \mathcal{X}$ and $\mathcal{Z}$ have sub-Gaussian tails, similar to the proofs of Theorem 1 \citep{kim2016sparse}, we can show, with probability greater than $1-p^{-1-a}$, 
\begin{eqnarray*}
 \left \| \sum_{i=0}^{[1/(k_n \Delta_n) ]-1}  \bOmega_{0,i k_n \Delta_n}     \mathcal{X}_{i k_n} ^{\top}\mathcal{Z}_{i k_n}   \right \| _{\max} \leq  C  \sqrt{ \log p / n} .
\end{eqnarray*}
Finally, we consider $  \bOmega_{0,i k_n \Delta_n} \mathcal{A}_{i k_n}  $ terms. 
Note that $\mathcal{A}_{i k_n}$'s are martingales with sub-Gaussian tails. Thus, by Azuma-Hoeffding inequality, we have, with probability greater than $1-p^{-1-a}$,
\begin{eqnarray*}
 \|  \sum_{i=0}^{[1/(k_n \Delta_n) ]-1}  \bOmega_{0,i k_n \Delta_n}   \mathcal{A}_{i k_n}   \| _{\max} \leq C  s_p  \sqrt{\log p} n^{-1/2}.
\end{eqnarray*}
Therefore, the statement \eqref{Thm3-result2} is shown.
\endpf

\subsection{Proof of Theorem \ref{Thm4}}

\textbf{Proof of Theorem \ref{Thm4}.}
By \eqref{Thm3-result2}, we can find $h_n$ such that, with   probability greater than $1-p^{-a}$,
\begin{equation}\label{Thm4-eq-1}
 \|  \hat{I\beta}  - I\beta_0 \| _{\max}  \leq h_n/2.
\end{equation}
We show that the statement holds under the event \eqref{Thm4-eq-1}. 
Similar to the proof of \eqref{eq-l1}, we can show
\begin{eqnarray*} 
	\|  \tilde{I\beta}  - I\beta_0 \|_1 \leq  C s_p h_n^{1-\delta}.  
\end{eqnarray*}
\endpf

\subsection{Proof of Theorem \ref{Thm5}}

\textbf{Proof of Theorem \ref{Thm5}.} 
Define
\begin{equation*}
 \mathcal{Y}_i^c  = 
\begin{pmatrix}
\Delta_{i+1} ^n Y^c \\ 
\Delta_{i+2} ^n Y^c \\  
 \vdots \\ 
\Delta_{i+k_n} ^n Y^c  
\end{pmatrix} \quad \text{and} \quad
 \mathcal{X}_i^c = 
\begin{pmatrix}
\Delta_{i+1} ^n \bX ^{c\top} \\ 
\Delta_{i+2} ^n \bX^{c\top} \\  
 \vdots \\ 
\Delta_{i+k_n} ^n \bX ^{c\top} 
\end{pmatrix}.
\end{equation*}
For some large constant $C>0$, let 
\begin{eqnarray*}
&& E_1=\left\{ \max_{i, j}  | \Delta_{i} ^n X_j^c |\leq   C \sqrt{ \log p} n^{-1/2} \right\} \cap \left\{\max_{i, j}  | \Delta_{i} ^n Y_j^c  |\leq   C s_p \sqrt{ \log p } n^{-1/2}\right\}, \cr
&& E_2=\left\{ \max_{i,j}\sum_{ k=1}^ {k_n} \1 _{\{ | \Delta_{i+k} ^n X_j | > v_{j,n}\} }  \leq C \log p \right\} \cap \left\{ \max_{i,j}  \int _{i \Delta_n} ^{\(i+k_n\) \Delta_n} d \Lambda_{j t}  \leq  C \log p \right\}.
\end{eqnarray*}
By the boundedness condition Assumption \ref{assumption1}(b) and sparsity condition \eqref{sparsity_beta}, we can show 
\begin{equation*}
P\( E_1 \) \geq 1- p^{-2-a}.
\end{equation*}
Under the event $E_1$, we have, for large $n$,
\begin{eqnarray*}
\max_{i,j}\sum_{ k=1}^ {k_n} \1 _{\{ | \Delta_{i+k} ^n X_j | > v_{j,n} \} } \leq \max_{i,j}  \int _{i \Delta_n} ^{\(i+k_n\) \Delta_n} d \Lambda_{j t},
\end{eqnarray*}
where $\int _{i \Delta_n} ^{\(i+k_n\) \Delta_n} d \Lambda_{j t}$ is a Poisson with the intensity $C k_n\Delta_n$  for some constant $C>0$, and 
\begin{equation*}
	P \( \max_{i,j}  \int _{i \Delta_n} ^{\(i+k_n\) \Delta_n} d \Lambda_{j t}  \geq  C \log p \) \leq  p^{-2-a}.
\end{equation*}
Thus, we have 
\begin{equation} \label{Jump-thm-eq1}
	P \( E_1 \cap E_2 \) \geq 1 - p^{-1-a}. 
\end{equation}
We have
\begin{eqnarray*}
&&\sum_{k=1}^{k_n}\Delta_{i+k} ^n X_j\, \1_{\{ | \Delta_{i+k} ^n X_j| \leq v_{j,n} \}} \Delta_{i+k} ^n X_l\, \1_{\{ | \Delta_{i+k} ^n X_l| \leq v_{l,n} \}} -\Delta_{i+k} ^n X_j^c  \Delta_{i+k} ^n X_l ^c  \cr
&&=\sum_{k=1}^{k_n}\Delta_{i+k} ^n X_j^c   \Delta_{i+k} ^n X_l^c  (\1_{\{ | \Delta_{i+k} ^n X_j| \leq v_{j,n} \}} \1_{\{ | \Delta_{i+k} ^n X_l| \leq v_{l,n} \}} -1)  \cr
&&\quad + \sum_{k=1}^{k_n} (\Delta_{i+k} ^n X_j^c\Delta_{i+k} ^n X_l^J  + \Delta_{i+k} ^n X_j^J\Delta_{i+k} ^n X_l^c + \Delta_{i+k} ^n X_j^J \Delta_{i+k} ^n X_l^J  )   \1_{\{ | \Delta_{i+k} ^n X_j| \leq v_{j,n} \}}    \1_{\{ | \Delta_{i+k} ^n X_l| \leq v_{l,n} \}} \cr
&&=  (I)_{ijl} + (II)_{ijl}.
\end{eqnarray*}
Under the event $E_1 \cap E_2 $, we have
\begin{eqnarray*}
\max_{i,j,l}\left | (I)_{ijl} \right | &\leq&  C \max_{i,j} |\Delta_{i} ^n X_j^c   |^2  \times \max_{i,j} \sum_{ k=1}^ {k_n} \1 _{\{ | \Delta_{i+k} ^n X_j | > v_{j,n}\} } \cr
&\leq&  C (\log p) ^2 /n 
\end{eqnarray*}
and
\begin{eqnarray*}
\max_{i,j,l}\left | (II)_{ijl} \right | &\leq&  C \sqrt{\log p} n^{-\rho} \max_{i,j} |\Delta_{i} ^n X_j^c   |  \times \max_{i,j}  \int _{i \Delta_n} ^{\(i+k_n\) \Delta_n} d \Lambda_{j t} \cr
&& +  C \(\sqrt{\log p} n^{-\rho} \)^2  \max_{i,j}  \int _{i \Delta_n} ^{\(i+k_n\) \Delta_n} d \Lambda_{j t}\cr
&\leq&  C (\log p )^2 n^{- 2\varrho}.
\end{eqnarray*}
Thus, by \eqref{Jump-thm-eq1}, we have, with probability greater than $1- p^{-1-a}$,
\begin{eqnarray}\label{Jump-thm-eq2}
 \max_{i} \frac{1}{k_n \Delta_n} \left \| \hat{ \mathcal{X}} _i ^{c \top} \hat{\mathcal{X}}_i^c   -  \mathcal{X} _i ^{c \top} \mathcal{X}_i^c   \right \| _{\max} &\leq& C (\log p )^2 n^{1/2-2\varrho} \cr
 &\leq& C \sqrt{\log p} n^{-1/4},
\end{eqnarray}
and similarly, we can show, with probability greater than $1- p^{-1-a}$,
\begin{eqnarray}\label{Jump-thm-eq3}
 \max_{i} \frac{1}{k_n \Delta_n} \left \|   \hat{\mathcal{X}}_i ^{c\top} \hat{\mathcal{Y}}_i^c  - \mathcal{X} _i ^{c\top}  \mathcal{Y} _i^c \right  \|_{\max} &\leq& C s_p (\log p )^2 n^{1/2-2\varrho} \cr
&\leq& C s_p \sqrt{\log p} n^{-1/4}. 
\end{eqnarray}
Consider $\left\| \sum_{i=0}^{[1/(k_n \Delta_n) ]-1} \hat{\bOmega}_{ik_n \Delta_n}^{\top} \left[  \hat{ \mathcal{X}} _{ik_n} ^{c \top} ( \hat{\mathcal{Y}}_{ik_n}^c  -  \hat{ \mathcal{X}} _{ik_n} ^{c} \hat{\bbeta}_{i k_n\Delta_n} )  - \mathcal{X} _{ik_n} ^{c \top} (\mathcal{Y} _{ik_n}^c - \mathcal{X}_{ik_n}^c  \hat{\bbeta}_{ik_n \Delta_n} ) \right] \right\|_{\infty}$.
By \eqref{Thm1-result1}, \eqref{Thm2-result1},  and \eqref{Thm2-eq1}, we have, with probability at least $1 - 2p^{-1-a}$,
\begin{eqnarray}\label{Jump-thm-eq4}
&& \left\| \sum_{i=0}^{[1/(k_n \Delta_n) ]-1} \hat{\bOmega}_{ik_n \Delta_n}^{\top} \left[  \hat{ \mathcal{X}} _{ik_n} ^{c \top} ( \hat{\mathcal{Y}}_{ik_n}^c  -  \hat{ \mathcal{X}} _{ik_n} ^{c} \hat{\bbeta}_{i k_n\Delta_n} )  - \mathcal{X} _{ik_n} ^{c \top} (\mathcal{Y} _{ik_n}^c - \mathcal{X}_{ik_n}^c  \hat{\bbeta}_{ik_n \Delta_n} ) \right] \right\|_{\infty} \cr
&& \leq \left\| \sum_{i=0}^{[1/(k_n \Delta_n) ]-1} \bOmega_{0,i k_n \Delta_n} \left[  \hat{ \mathcal{X}} _{ik_n} ^{c \top} ( \hat{\mathcal{Y}}_{ik_n}^c  -  \hat{ \mathcal{X}} _{ik_n} ^{c} \bbeta_{0,ik_n \Delta_n} )  - \mathcal{X} _{ik_n} ^{c \top} (\mathcal{Y} _{ik_n}^c - \mathcal{X}_{ik_n}^c  \bbeta_{0,ik_n \Delta_n} ) \right] \right\|_{\infty}  \cr
&& \quad +  \left\| \sum_{i=0}^{[1/(k_n \Delta_n) ]-1}  \left\| \hat{\bOmega}_{ik_n \Delta_n} -  \bOmega_{0,i k_n \Delta_n} \right\|_{1} \times   \left\|\hat{ \mathcal{X}} _{ik_n} ^{c \top} \hat{\mathcal{Y}}_{ik_n}^c  - \mathcal{X}_{ik_n} ^{c \top} \mathcal{Y}_{ik_n}^c    \right\|_{\max}  \right\|_{\infty}    \cr
&& \quad +  \left\| \sum_{i=0}^{[1/(k_n \Delta_n) ]-1}  \left\| \hat{\bOmega}_{ik_n \Delta_n} \right\|_{1} \times   \left\|\hat{ \mathcal{X}} _{ik_n} ^{c \top} \hat{ \mathcal{X}} _{ik_n} ^{c}  - \mathcal{X}_{ik_n} ^{c \top} \mathcal{X} _{ik_n} ^{c}  \right\|_{\max}  \times  \left\| \hat{\bbeta}_{ik_n \Delta_n} -  \bbeta_{0,ik_n \Delta_n}  \right\|_{1}  \right\|_{\infty}    \cr
&& \quad +  \left\| \sum_{i=0}^{[1/(k_n \Delta_n) ]-1}   \left\| \hat{\bOmega}_{ik_n \Delta_n} -  \bOmega_{0,i k_n \Delta_n} \right\|_{1} \times   \left\|\hat{ \mathcal{X}} _{ik_n} ^{c \top} \hat{ \mathcal{X}} _{ik_n} ^{c}  - \mathcal{X}_{ik_n} ^{c \top} \mathcal{X} _{ik_n} ^{c}  \right\|_{\max}  \times  \left\|  \bbeta_{0,ik_n \Delta_n}  \right\|_{1}  \right\|_{\infty}       \cr
&& \leq  \left\| \sum_{i=0}^{[1/(k_n \Delta_n) ]-1} \bOmega_{0,i k_n \Delta_n} \left[  \hat{ \mathcal{X}} _{ik_n} ^{c \top} ( \hat{\mathcal{Y}}_{ik_n}^c  -  \hat{ \mathcal{X}} _{ik_n} ^{c} \bbeta_{0,ik_n \Delta_n} )  - \mathcal{X} _{ik_n} ^{c \top} (\mathcal{Y} _{ik_n}^c - \mathcal{X}_{ik_n}^c  \bbeta_{0,ik_n \Delta_n} ) \right] \right\|_{\infty} \cr
&& \quad + C\left\{s_p ^{2-\delta}   ( \log p / n^{1/2}  )^{(2-\delta)/2} + s_p s_{\omega,p}    ( \log p / n^{1/2}  )^{(2-q)/2} \right\}. 
\end{eqnarray}
For the first term, let $\bu_t=\left(u_{1,t}, \ldots, u_{p,t} \right)^{\top}$  be any  $p$-dimensional process, which is independent of $\bLambda_t$ and satisfies $\left\| \bu_t \right\|_{1} \leq C$.
By the Jensen's inequality, we have
\begin{eqnarray*}
&& \E \Bigg[  \exp \( \bu_{i k_n \Delta_n}^{\top}  \int_{ik_n \Delta_n}^{\(i+1\)k_n \Delta_n}  d\bLambda_{t}  \) \Bigg | \FF_{i k_n \Delta_n} \Bigg] \cr
&&  \leq \sum_{j=1}^{p}  \dfrac{|u_{j,ik_n \Delta_n} |}{\left\| \bu_{ik_n \Delta_n} \right\|_{1}}  \E \Bigg[  \exp \(\left\| \bu_{ik_n \Delta_n} \right\|_{1} \int_{ik_n \Delta_n}^{\(i+1\)k_n \Delta_n} d \Lambda_{j t} \)     \Bigg | \FF_{i k_n \Delta_n} \Bigg] \cr
&& \leq \exp \(C n^{-1/2} \),
\end{eqnarray*}
where the last inequality is from the moment generating function of the Poisson distribution.
Then, by the Markov inequality, we have
\begin{eqnarray}\label{Jump-thm-eq5}
&& P \(\sum_{i=0}^{[1/(k_n \Delta_n) ]-1} \bu_{i k_n \Delta_n}^{\top} \int_{ik_n \Delta_n}^{\(i+1\)k_n \Delta_n}  d\bLambda_{t} \geq \(a+3\) \log p \) \cr
&& \leq p^{-3-a} \E \Bigg[ \prod_{i=0}^{[1/(k_n \Delta_n) ]-1}  \exp \( \bu_{i k_n \Delta_n}^{\top}  \int_{ik_n \Delta_n}^{\(i+1\)k_n \Delta_n}  d\bLambda_{t}  \) \Bigg] \cr
&& \leq Cp^{-3-a}.
\end{eqnarray}
Note that  $\bbeta_t/s_p$ and $\bOmega_t$ are bounded.
Hence, similar to the proofs of \eqref{Jump-thm-eq2}, we can show, with probability at least $1 - p^{-1-a}$,
\begin{equation}\label{Jump-thm-eq6}
\left\| \sum_{i=0}^{[1/(k_n \Delta_n) ]-1}  \bOmega_{0,i k_n \Delta_n} \left[  \hat{ \mathcal{X}} _{ik_n} ^{c \top} \hat{ \mathcal{X}} _{ik_n} ^{c}  - \mathcal{X} _{ik_n} ^{c \top} \mathcal{X}_{ik_n}^c  \right]  \bbeta_{0,ik_n \Delta_n} \right\|_{\infty} \leq Cs_p n^{-3/4}\sqrt{\log p},
\end{equation}
and similarly, we can show, with probability  at least $1- p^{-1-a}$,
\begin{equation}\label{Jump-thm-eq7}
 \left\| \sum_{i=0}^{[1/(k_n \Delta_n) ]-1} \bOmega_{0,i k_n \Delta_n} \left[  \hat{ \mathcal{X}} _{ik_n} ^{c \top}  \hat{\mathcal{Y}}_{ik_n}^c  - \mathcal{X} _{ik_n} ^{c \top}\mathcal{Y} _{ik_n}^c \right] \right\|_{\infty}  \leq Cs_p n^{-3/4}\sqrt{\log p}.
\end{equation}
From \eqref{Jump-thm-eq4}, \eqref{Jump-thm-eq6}, and \eqref{Jump-thm-eq7}, we have, with probability  at least $1- 4p^{-1-a}$,
\begin{eqnarray}\label{Jump-thm-eq8}
&& \left\| \sum_{i=0}^{[1/(k_n \Delta_n) ]-1} \hat{\bOmega}_{ik_n \Delta_n}^{\top} \left[  \hat{ \mathcal{X}} _{ik_n} ^{c \top} ( \hat{\mathcal{Y}}_{ik_n}^c  -  \hat{ \mathcal{X}} _{ik_n} ^{c} \hat{\bbeta}_{i k_n\Delta_n} )  - \mathcal{X} _{ik_n} ^{c \top} (\mathcal{Y} _{ik_n}^c - \mathcal{X}_{ik_n}^c  \hat{\bbeta}_{ik_n \Delta_n} ) \right] \right\|_{\infty} \cr
&& \leq C\left\{s_p ^{2-\delta}   ( \log p / n^{1/2}  )^{(2-\delta)/2} + s_p s_{\omega,p}    ( \log p / n^{1/2}  )^{(2-q)/2} \right\}.
\end{eqnarray}
By \eqref{Jump-thm-eq2}, \eqref{Jump-thm-eq3}, and \eqref{Jump-thm-eq8}, the effect of jumps is negligible. 
Thus, the statement can be shown by Theorem \ref{Thm4}. 
\endpf

\newpage

\section{A simulation setup}\label{sec:sim_setup}
We generated the data with frequency $1/n^{all}$ and considered the following time series regression jump diffusion model:
\begin{eqnarray*}
&& dY_t= \bbeta_{t} ^{\top}  d\bX_{t}^c +d Z_t +  J_t^y d \Lambda_t^y, \cr
&& d\bX_t=  d \bX_t ^c + d \bX_t^J,    \quad  d\bX_t^c= \bsigma_t d\bB_t, \quad d \bX_t^J =  \bJ_t  d \bLambda_t, \quad dZ_t= \nu_t d W_t,
\end{eqnarray*}
where $\bB_t$ and $W_t$ are $p$-dimensional and one-dimensional independent Brownian motions, respectively,
 $\bJ_t=\(J_{1t}, \ldots, J_{pt}\)^{\top}$ and $J_t^y$ are jump sizes,
  and $\bLambda_t=\(\Lambda_{1t}, \ldots, \Lambda_{pt} \)^{\top}$ and $\Lambda_t^y$ are the Poisson processes with the intensities $\(10, \ldots, 10\)^{\top}$ and $5$, respectively. 
The jump sizes $J_{it}$ and $J^y_{t}$ were independently generated from the Gaussian distribution with a mean of 0 and standard deviation of $0.05$.
We set the initial values $X_{i0}$ and $Y_0$ to 0, while $\nu_t$ follows the OU process
\begin{equation*}
	d \nu_t= 2\(0.12-\nu_t\)dt + 0.03 d \bW_t^{\nu},
\end{equation*}
where $\nu_0=0.15$ and $\bW_t^{\nu}$ is one-dimensional independent Brownian motion.
The instantaneous volatility process $\bsigma_t$ was taken to be a Cholesky decomposition  of $\bSigma_t=(\Sigma_{ijt})_{1 \leq i,j \leq p}$, where $\Sigma_{ijt}=\xi_t 0.7^{|i-j|}$ and $\xi_t$ satisfies
\begin{equation*}
	d \xi_t= 6\(0.3-\xi_t\)dt + 0.12 d \bW_t^{\xi},
\end{equation*}
where $\xi_0=0.5$ and $\bW_t^{\xi}$ is one-dimensional independent Brownian motion.
For the coefficient process $\bbeta_t$, we considered the time-varying coefficient and constant coefficient processes, where $[ s_p ]$ factors are only significant.
We first generated the time-varying coefficient process as follows:
\begin{equation*}
	d \bbeta_t= \bmu_{\beta, t} dt + \bnu_{\beta,t} d \bW_{t}^{\beta},
\end{equation*}
where $\bmu_{\beta, t}=\(\mu_{1,\beta,t}, \ldots, \mu_{p,\beta,t}\)^{\top}$,  $\bnu_{\beta,t}=\(\nu_{i,j,\beta,t}\)_{1 \leq i,j \leq p}$, and $\bW_{t}^{\beta}$ is a $p$-dimensional independent Brownian motion.
We set the process $\(\nu_{i,j,\beta,t}\)_{1 \leq i,j \leq [ s_p ]}$ as $\zeta_{t}\bI_{[ s_p ]}$,  where $\bI_{[ s_p ]}$ is the $[ s_p ]$-dimensional identity matrix and $\zeta_{t}$  was generated as follows:
\begin{equation*}
	d \zeta_{t}= 4\(0.5-\zeta_{t}\)dt + 0.2 d \bW_{t}^{\zeta},
\end{equation*}
where $\zeta_{0}=0.4$ and $\bW_{t}^{\zeta}$ is one-dimensional independent Brownian motion.
For  $i=1,\ldots, [ s_p ]$, we took the initial value $\beta_{i0}$ as $2^{3-i}$ and set $\mu_{i,\beta,t}= 0.05$ for $0 \leq t \leq 1$.
We set $\beta_{it}$, $i=[ s_p ]+1, \ldots, p$, as zero. 
In contrast, for the constant coefficient process,  we set $\beta_{it}=2^{3-i}$ for $i=1,\ldots, [ s_p ]$ and $0 \leq t \leq 1$, while the other $\beta_{it}$'s were set to 0. 
We chose $p=100$, $s_p=\log p$, $n^{all}=4000$, and we varied $n$ from $1000$ to $4000$.

\section{Visualization of TED procedure and simulated coefficient dynamics}\label{appendix:trajectory}
\begin{figure}[!ht]
\centering
\includegraphics[width = 1\textwidth]{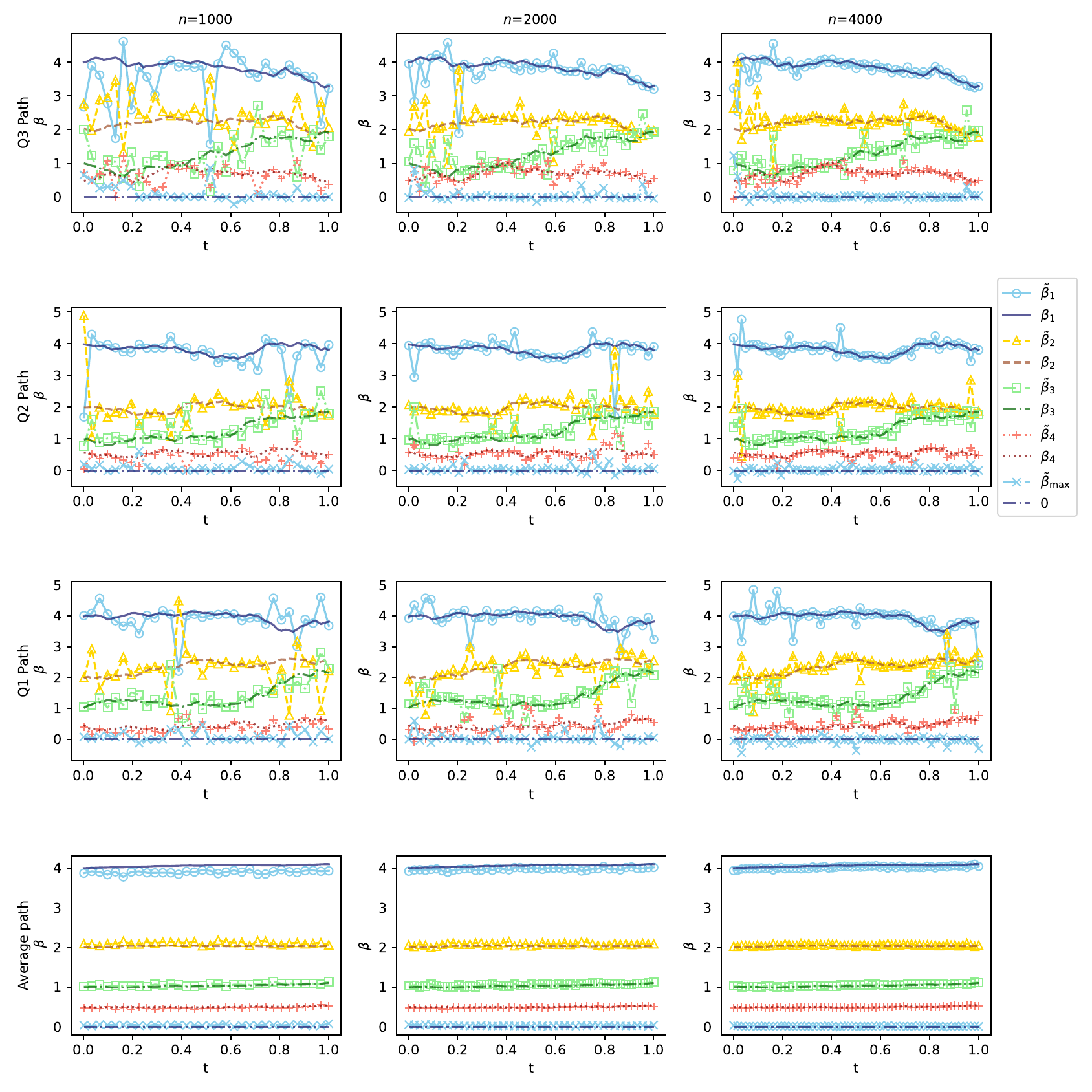}
\caption{Trajectory plots of the true time-varying coefficients and their corresponding TED estimates under different sample sizes $n=1000, 2000, 4000$.
Each row represents simulation paths selected based on the 75th percentile (Q3), median (Q2), and 25th percentile (Q1) of the $\ell_2$-norm errors at the largest sample size $n=4000$, along with the average path.
All nonzero coefficients and the largest estimated integrated coefficient among true zero coefficients before thresholding are presented.}
\label{fig:beta-path}
\end{figure}
\begin{figure}[!ht]
\centering
\includegraphics[width = 1\textwidth]{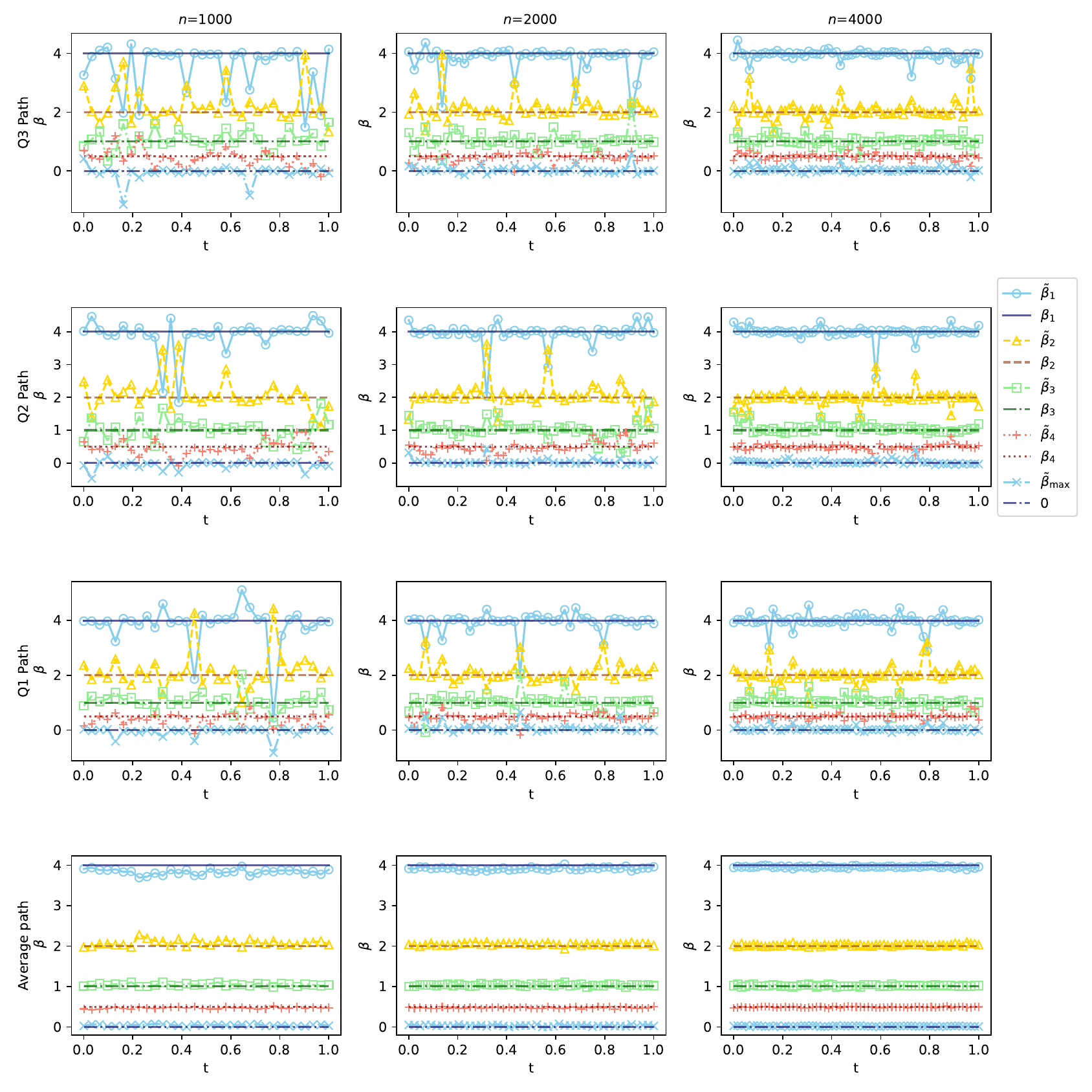}
\caption{Trajectory plots of the true constant coefficients and their corresponding TED estimates under different sample sizes $n=1000, 2000, 4000$.
Each row represents simulation paths selected based on the 75th percentile (Q3), median (Q2), and 25th percentile (Q1) of the $\ell_2$-norm errors at the largest sample size $n=4000$, along with the average path.
All nonzero coefficients and the largest estimated integrated coefficient among true zero coefficients before thresholding are presented.}
\label{fig:const-beta-path}
\end{figure}
To check whether TED accurately captures the time-varying behavior of the true coefficients, we present trajectory plots of the true coefficients and their debiased instantaneous estimates obtained from the TED procedure.
Figures \ref{fig:beta-path} and \ref{fig:const-beta-path} illustrate these coefficient dynamics for selected simulation paths under different sample sizes $n=1000, 2000, 4000$, corresponding to the time-varying coefficient and constant coefficient simulations, respectively.
We selected three representative simulation paths based on the 25th percentile (Q1), median (Q2), and 75th percentile (Q3) of the $\ell_2$-norm estimation errors of TED at the largest sample size $n=4000$, as well as the average path computed across all simulation paths.
We present the paths for all true nonzero coefficients and additionally include the path of the largest estimated integrated coefficient among the coefficients whose true values are zero, prior to thresholding.
From Figures \ref{fig:beta-path} and \ref{fig:const-beta-path}, we find that the estimated instantaneous coefficients capture the temporal dynamics more accurately as the sample size increases.

\section{Additional empirical analyses}\label{additional-empirical-analysis}
To investigate the impact of each tuning parameter on the performance of the TED estimator, we conducted sensitivity analyses by varying one parameter at a time while keeping the others fixed according to the scheme discussed in Section \ref{SEC-Tuning}.
\begin{figure}[!ht]
\centering
\includegraphics[width = 1\textwidth]{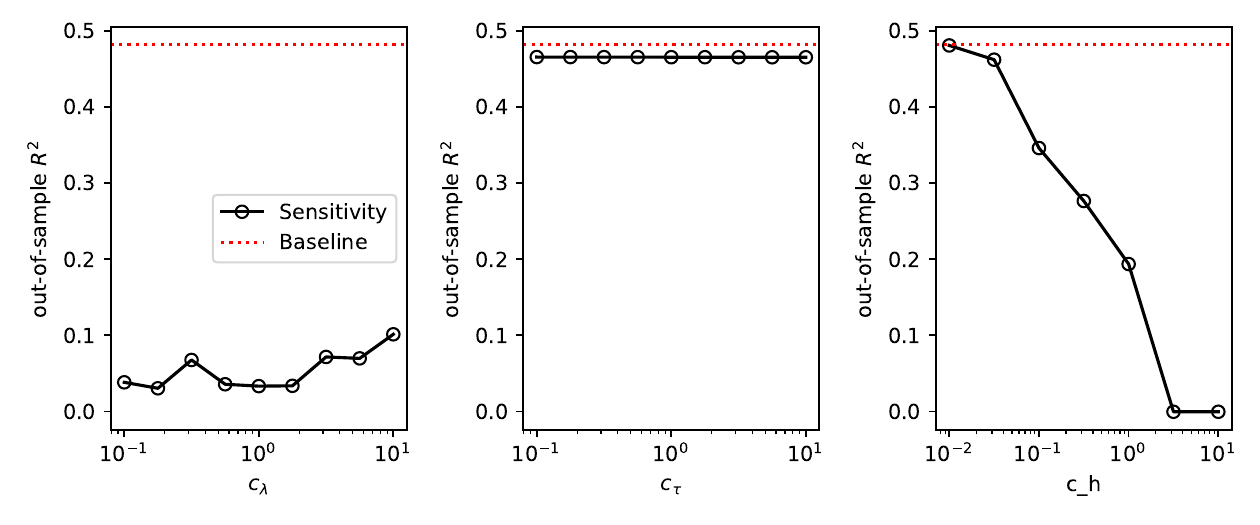}
\caption{Sensitivity analysis of the TED estimator with respect to tuning parameters $c_{\lambda}$, $c_{\tau}$, and $c_h$.
}
\label{fig:sensitivity}
\end{figure}
Figure \ref{fig:sensitivity} shows the sensitivity analysis results for the TED estimator with respect to the tuning parameters $c_{\lambda}$, $c_{\tau}$, and $c_h$.
The horizontal dashed line (baseline) indicates the result obtained by using the tuning parameter selection method described in Section \ref{SEC-Tuning}.
Each solid line corresponds to the results obtained by varying one tuning parameter while the others are chosen according to the baseline method.
We varied $c_{\lambda}$ and $c_{\tau}$ from 0.1 to 10, and $c_h$ from 0.01 to 10, using a logarithmically equidistant grid.
From Figure \ref{fig:sensitivity}, we find that selecting proper values for all three tuning parameters is crucial.
Specifically, using a single fixed value for $c_{\lambda}$ and $c_{\tau}$ results in worse performance.
This may be because of the time-varying sparsity of the coefficients and the inverse covariance matrix.
On the other hand, smaller values of $c_h$ tend to yield better performance.
This may be due to the presence of weak factors that have only minor impacts on individual assets.

Our theoretical results hold for any instantaneous coefficient estimators having the same convergence rate in \eqref{Thm1-result1} and the bias term in \eqref{Thm3-result1-1}.
As part of a comprehensive analysis, we replaced the Dantzig selector with alternative methods and compared their performances against our results.
\begin{table}[!ht]
\caption{The annual average out-of-sample $R^2$ of the TED framework using different penalized regression methods to estimate the instantaneous coefficients.}
\label{tab:penalty-outsample}
\centering
\resizebox{\textwidth}{!}{%
    \begin{tabular}{llrrrrrrrrrr}
    \toprule
    & & \multicolumn{10}{c}{Out-of-sample $R^2$} \\
    \cmidrule(lr){3-12}
    Period         & RV (\%) & TED   & ADAN  & LAS   & ALAS  & RID   & ARID  & ELA   & AELA  & SCA   & ASCA  \\
    \midrule
    whole          & 12.0    & 0.482 & 0.475 & 0.433 & 0.481 & 0.466 & 0.467 & 0.430 & 0.481 & 0.438 & 0.429 \\
    2013           & 7.9     & 0.613 & 0.602 & 0.539 & 0.613 & 0.596 & 0.591 & 0.546 & 0.613 & 0.523 & 0.525 \\
    2014           & 8.8     & 0.475 & 0.472 & 0.418 & 0.475 & 0.455 & 0.455 & 0.411 & 0.473 & 0.425 & 0.419 \\
    2015           & 10.2    & 0.505 & 0.496 & 0.473 & 0.504 & 0.485 & 0.488 & 0.475 & 0.504 & 0.475 & 0.460 \\
    2016           & 10.0    & 0.510 & 0.501 & 0.469 & 0.506 & 0.490 & 0.498 & 0.468 & 0.509 & 0.469 & 0.457 \\
    2017           & 5.5     & 0.398 & 0.391 & 0.353 & 0.399 & 0.396 & 0.383 & 0.350 & 0.399 & 0.354 & 0.355 \\
    2018           & 13.0    & 0.538 & 0.532 & 0.485 & 0.538 & 0.519 & 0.518 & 0.472 & 0.536 & 0.498 & 0.490 \\
    2019           & 9.1     & 0.461 & 0.454 & 0.407 & 0.460 & 0.446 & 0.444 & 0.398 & 0.462 & 0.420 & 0.412 \\
    2020           & 22.8    & 0.366 & 0.360 & 0.328 & 0.365 & 0.352 & 0.368 & 0.328 & 0.365 & 0.344 & 0.319 \\
    \bottomrule
    \end{tabular}
}
\end{table}
Table \ref{tab:penalty-outsample} shows the annual average out-of-sample $R^2$ of the TED framework using different penalized regression methods, such as LASSO (LAS), Ridge (RID), elastic net (ELA), and SCAD (SCA), as well as their adaptive versions including the adaptive versions of Dantzig (ADAN), LASSO (ALAS), Ridge (ARID), elastic net (AELA), and SCAD (ASCA), for estimating the instantaneous coefficients.
The adaptive methods impose different weights on the penalty terms for each coefficient.
Heavier penalties are assigned to smaller preliminary estimates, and lighter penalties to larger ones, where the preliminary estimates can be obtained from methods such as Ridge regression \citep{zou2006adaptive,zou2009adaptive}.
From Table \ref{tab:penalty-outsample}, we find that ALAS and AELA show comparable performance to the original TED, although TED achieves the best average performance.
This may be because those penalties address sparsity as well as the shrinkage bias.
Specifically, ALAS and AELA assign different penalty weights to each coefficient, inversely proportional to the magnitude of preliminary estimates.
This procedure effectively reduces bias for significant coefficients while shrinking insignificant coefficients more aggressively.

To check whether our results are sensitive to the inverse covariance estimation method, we replaced the CLIME estimator with an alternative method.
If covariates have a factor structure, we can estimate the inverse instantaneous volatility matrix using the POET method \citep{fan2013large}.
Specifically, we can apply the POET estimator to estimate the covariance matrix and then obtain the arithmetic inversion of this estimate.
We call this TED-POET.
This method is computationally more efficient than CLIME and achieves the $\ell_1$ convergence rate of $O_p(s'_{\omega,p} (\sqrt{\log p} n^{-1/4} + p^{-1/2})^{1-q})$, where $s'_{\omega,p} = \max_{i \leq p} \sum_{j\leq p} \left|\sigma_{u,ij}\right|^q$.
Under $n^{1/4} / \sqrt{p \log p}$ $\rightarrow 0$ and the same sparsity levels $s_{\omega,p}$ and $s'_{\omega,p}$, the convergence rate of the POET estimator is asymptotically the same as that of the CLIME estimator, $O_p (s_{\omega,p} (\sqrt{\log p} \times n^{-1/4})^{1-q})$.
\begin{table}[ht]
\centering
\caption{In-sample and out-of-sample $R^2$ for TED and TED-POET.}
\label{tab:poet}
\begin{tabular}{lrrrrr}
\toprule
& & \multicolumn{2}{c}{In-sample $R^2$} & \multicolumn{2}{c}{Out-of-sample $R^2$} \\
\cmidrule(lr){3-4} \cmidrule(lr){5-6}
Period    & RV (\%)       & TED   & TED-POET & TED   & TED-POET \\
\midrule  
whole     & 12.0    & 0.539 & 0.525    & 0.482 & 0.474    \\
2013      & 7.9           & 0.667 & 0.675    & 0.613 & 0.620    \\
2014      & 8.8           & 0.525 & 0.498    & 0.475 & 0.447    \\
2015      & 10.2          & 0.573 & 0.538    & 0.505 & 0.506    \\
2016      & 10.0          & 0.570 & 0.537    & 0.510 & 0.479    \\
2017      & 5.5           & 0.453 & 0.447    & 0.398 & 0.398    \\
2018      & 13.0          & 0.586 & 0.577    & 0.538 & 0.542    \\
2019      & 9.1           & 0.515 & 0.520    & 0.461 & 0.462    \\
2020      & 22.8          & 0.425 & 0.410    & 0.366 & 0.351    \\
\bottomrule
\end{tabular}
\end{table}
Table \ref{tab:poet} shows the in-sample and out-of-sample $R^2$ for TED and TED-POET.
From Table \ref{tab:poet}, we find that TED-POET shows comparable performance to TED, although TED performs better in a few subperiods.
This may be because the covariance matrix of the factor zoo does not always conform to a low-rank-plus-sparse representation across periods.

We also compared the performance of TED with the benchmarks using an alternative measure of predictive accuracy, the mean absolute error (MAE).
Specifically, we measured it as the average of the absolute differences between the observed high-frequency returns of the assets and the returns predicted by the estimated coefficient using the high-frequency factors for the next month.
\begin{table}[!ht]
\caption{The annual average out-of-sample MAE for the TED, AKX, AKX6, and various regression estimators across the five assets.}
\label{tab:mae-outsample}
\centering
\resizebox{\textwidth}{!}{%
    \begin{tabular}{lrrrrrrrrrrrrrrrr}
    \toprule
    & & \multicolumn{15}{c}{Out-of-sample MAE $\times 1000$} \\
    \cmidrule(lr){3-17}
    Period         & RV (\%) & TED    & AKX    & AKX6   & OLS6   & LAS    & ALAS   & RID    & ARID   & ELA    & AELA   & SCA    & ASCA   & BAY    & DAN    & ADAN   \\
    \midrule
    whole          & 12.0     & 0.598  & 0.740  & 0.878  & 0.799  & 0.627  & 0.608  & 0.613  & 0.609  & 0.611  & 0.616  & 0.616  & 0.749  & 0.670  & 0.623  & 0.641  \\
    2013           & 7.9      & 0.431  & 0.645  & 0.788  & 0.716  & 0.489  & 0.426  & 0.435  & 0.427  & 0.455  & 0.432  & 0.482  & 0.614  & 0.485  & 0.436  & 0.455  \\
    2014           & 8.8      & 0.535  & 0.668  & 0.769  & 0.707  & 0.563  & 0.545  & 0.541  & 0.541  & 0.546  & 0.553  & 0.554  & 0.672  & 0.595  & 0.555  & 0.571  \\
    2015           & 10.2     & 0.542  & 0.701  & 0.856  & 0.757  & 0.570  & 0.546  & 0.562  & 0.555  & 0.555  & 0.554  & 0.559  & 0.677  & 0.624  & 0.563  & 0.580  \\
    2016           & 10.0     & 0.502  & 0.668  & 0.814  & 0.732  & 0.536  & 0.515  & 0.524  & 0.518  & 0.520  & 0.522  & 0.527  & 0.626  & 0.589  & 0.526  & 0.547  \\
    2017           & 5.5      & 0.441  & 0.551  & 0.619  & 0.573  & 0.458  & 0.454  & 0.456  & 0.456  & 0.448  & 0.463  & 0.452  & 0.519  & 0.495  & 0.469  & 0.480  \\
    2018           & 13.0     & 0.616  & 0.776  & 0.968  & 0.854  & 0.649  & 0.624  & 0.632  & 0.625  & 0.632  & 0.629  & 0.632  & 0.804  & 0.673  & 0.643  & 0.657  \\
    2019           & 9.1      & 0.524  & 0.649  & 0.763  & 0.692  & 0.548  & 0.534  & 0.541  & 0.535  & 0.536  & 0.542  & 0.535  & 0.655  & 0.588  & 0.543  & 0.566  \\
    2020           & 22.8     & 1.181  & 1.257  & 1.438  & 1.350  & 1.192  & 1.205  & 1.197  & 1.201  & 1.180  & 1.216  & 1.177  & 1.411  & 1.294  & 1.232  & 1.254  \\
    \bottomrule
    \end{tabular}
    }
\end{table}
Table \ref{tab:mae-outsample} shows the annual average MAE for the TED, AKX, AKX6, and various regression estimators across the five assets.
We find that TED generally outperforms the other methods in terms of MAE.
This highlights the advantage of TED in accommodating the time-varying nature of coefficients for more accurate predictions.

We analyzed how TED performance changes with different sampling frequencies---10-min and 30-min log price data---compared to our baseline frequency (5-min).
We note that sampling frequencies higher than 5-min are inappropriate due to microstructure noise \citep{ait2019hausman}.
\begin{table}[!ht]
\caption{The annual average in-sample and out-of-sample $R^2$ for the TED, AKX, AKX6, and various regression estimators across the five assets using 10-min sampling frequency data.}
\label{tab:loocv-outsample-10}
\centering
\resizebox{\textwidth}{!}{%
\begin{tabular}{l*{16}{r}}
\toprule
 & & \multicolumn{15}{c}{In-sample $R^2$} \\
\cmidrule(lr){3-17}
Period & RV (\%) & TED & AKX & AKX6 & OLS6 & LAS & ALAS & RID & ARID & ELA & AELA & SCA & ASCA & BAY & DAN & ADAN \\
\midrule
whole & 12.0 & 0.490 & 0.324 & 0.086 & 0.216 & 0.474 & 0.555 & 0.570 & 0.572 & 0.515 & 0.557 & 0.478 & 0.273 & 0.496 & 0.553 & 0.547 \\
2013  & 7.9  & 0.590 & 0.333 & 0.050 & 0.194 & 0.550 & 0.696 & 0.716 & 0.718 & 0.621 & 0.703 & 0.557 & 0.270 & 0.654 & 0.695 & 0.696 \\
2014  & 8.8  & 0.475 & 0.291 & 0.062 & 0.193 & 0.461 & 0.552 & 0.565 & 0.569 & 0.505 & 0.554 & 0.465 & 0.248 & 0.491 & 0.549 & 0.544 \\
2015  & 10.2 & 0.532 & 0.361 & 0.094 & 0.247 & 0.516 & 0.593 & 0.603 & 0.609 & 0.548 & 0.596 & 0.525 & 0.348 & 0.529 & 0.588 & 0.585 \\
2016  & 10.0 & 0.529 & 0.337 & 0.087 & 0.226 & 0.499 & 0.584 & 0.598 & 0.600 & 0.540 & 0.585 & 0.508 & 0.332 & 0.510 & 0.582 & 0.578 \\
2017  & 5.5  & 0.402 & 0.224 & 0.026 & 0.149 & 0.395 & 0.472 & 0.490 & 0.492 & 0.434 & 0.475 & 0.396 & 0.229 & 0.431 & 0.470 & 0.467 \\
2018  & 13.0 & 0.545 & 0.406 & 0.142 & 0.278 & 0.526 & 0.590 & 0.609 & 0.608 & 0.560 & 0.593 & 0.528 & 0.313 & 0.534 & 0.591 & 0.584 \\
2019  & 9.1  & 0.468 & 0.304 & 0.069 & 0.205 & 0.454 & 0.524 & 0.541 & 0.542 & 0.493 & 0.525 & 0.457 & 0.249 & 0.470 & 0.523 & 0.514 \\
2020  & 22.8 & 0.377 & 0.334 & 0.155 & 0.236 & 0.390 & 0.426 & 0.441 & 0.440 & 0.413 & 0.424 & 0.385 & 0.193 & 0.348 & 0.426 & 0.408 \\
\midrule
 & & \multicolumn{15}{c}{Out-of-sample $R^2$} \\
\cmidrule(lr){3-17}
Period & RV (\%) & TED & AKX & AKX6 & OLS6 & LAS & ALAS & RID & ARID & ELA & AELA & SCA & ASCA & BAY & DAN & ADAN \\
\midrule
whole & 12.0 & 0.449 & 0.301 & 0.079 & 0.209 & 0.442 & 0.439 & 0.463 & 0.449 & 0.465 & 0.422 & 0.442 & 0.260 & 0.343 & 0.438 & 0.384 \\
2013  & 7.9  & 0.546 & 0.316 & 0.049 & 0.193 & 0.514 & 0.597 & 0.614 & 0.610 & 0.570 & 0.589 & 0.515 & 0.261 & 0.511 & 0.596 & 0.571 \\
2014  & 8.8  & 0.437 & 0.275 & 0.057 & 0.192 & 0.434 & 0.434 & 0.455 & 0.448 & 0.458 & 0.422 & 0.437 & 0.244 & 0.357 & 0.430 & 0.391 \\
2015  & 10.2 & 0.499 & 0.349 & 0.091 & 0.252 & 0.492 & 0.492 & 0.500 & 0.487 & 0.511 & 0.471 & 0.497 & 0.333 & 0.361 & 0.480 & 0.413 \\
2016  & 10.0 & 0.478 & 0.311 & 0.081 & 0.212 & 0.456 & 0.472 & 0.483 & 0.472 & 0.482 & 0.455 & 0.460 & 0.312 & 0.355 & 0.460 & 0.425 \\
2017  & 5.5  & 0.367 & 0.202 & 0.024 & 0.144 & 0.361 & 0.341 & 0.379 & 0.362 & 0.383 & 0.323 & 0.359 & 0.220 & 0.253 & 0.341 & 0.283 \\
2018  & 13.0 & 0.513 & 0.379 & 0.125 & 0.269 & 0.503 & 0.509 & 0.520 & 0.517 & 0.525 & 0.495 & 0.506 & 0.306 & 0.433 & 0.509 & 0.465 \\
2019  & 9.1  & 0.421 & 0.282 & 0.070 & 0.195 & 0.407 & 0.388 & 0.430 & 0.407 & 0.428 & 0.372 & 0.403 & 0.223 & 0.301 & 0.400 & 0.338 \\
2020  & 22.8 & 0.344 & 0.310 & 0.141 & 0.229 & 0.369 & 0.307 & 0.351 & 0.320 & 0.377 & 0.283 & 0.354 & 0.191 & 0.218 & 0.315 & 0.230 \\
\bottomrule
\end{tabular}%
}
\end{table}
\begin{table}[!ht]
\caption{The annual average in-sample and out-of-sample $R^2$ for the TED, AKX, AKX6, and various regression estimators across the five assets using 30-min sampling frequency data.}
\label{tab:loocv-outsample-30}
\centering
\resizebox{\textwidth}{!}{%
\begin{tabular}{l*{16}{r}}
\toprule
 & & \multicolumn{15}{c}{In-sample $R^2$} \\
\cmidrule(lr){3-17}
Period & RV (\%) & TED & AKX & AKX6 & OLS6 & LAS & ALAS & RID & ARID & ELA & AELA & SCA & ASCA & BAY & DAN & ADAN \\
\midrule
whole & 12.0 & 0.438 & 0.321 & 0.115 & 0.218 & 0.474 & 0.519 & 0.535 & 0.536 & 0.502 & 0.518 & 0.474 & 0.233 & 0.470 & 0.511 & 0.507 \\
2013  & 7.9  & 0.505 & 0.331 & 0.076 & 0.194 & 0.549 & 0.656 & 0.679 & 0.685 & 0.603 & 0.663 & 0.550 & 0.241 & 0.599 & 0.647 & 0.648 \\
2014  & 8.8  & 0.419 & 0.294 & 0.088 & 0.195 & 0.462 & 0.510 & 0.530 & 0.529 & 0.495 & 0.510 & 0.466 & 0.210 & 0.460 & 0.503 & 0.499 \\
2015  & 10.2 & 0.486 & 0.357 & 0.131 & 0.250 & 0.514 & 0.556 & 0.564 & 0.573 & 0.536 & 0.557 & 0.522 & 0.284 & 0.509 & 0.551 & 0.541 \\
2016  & 10.0 & 0.459 & 0.330 & 0.115 & 0.226 & 0.498 & 0.549 & 0.562 & 0.568 & 0.527 & 0.550 & 0.501 & 0.282 & 0.497 & 0.538 & 0.535 \\
2017  & 5.5  & 0.362 & 0.228 & 0.042 & 0.150 & 0.395 & 0.437 & 0.457 & 0.456 & 0.423 & 0.434 & 0.393 & 0.214 & 0.392 & 0.431 & 0.425 \\
2018  & 13.0 & 0.496 & 0.402 & 0.183 & 0.280 & 0.529 & 0.566 & 0.578 & 0.581 & 0.554 & 0.566 & 0.529 & 0.238 & 0.504 & 0.562 & 0.556 \\
2019  & 9.1  & 0.425 & 0.306 & 0.100 & 0.210 & 0.455 & 0.494 & 0.506 & 0.509 & 0.482 & 0.491 & 0.456 & 0.195 & 0.442 & 0.482 & 0.482 \\
2020  & 22.8 & 0.355 & 0.321 & 0.187 & 0.234 & 0.388 & 0.382 & 0.399 & 0.391 & 0.398 & 0.374 & 0.373 & 0.205 & 0.354 & 0.377 & 0.369 \\
\midrule
 & & \multicolumn{15}{c}{Out-of-sample $R^2$} \\
\cmidrule(lr){3-17}
Period & RV (\%) & TED & AKX & AKX6 & OLS6 & LAS & ALAS & RID & ARID & ELA & AELA & SCA & ASCA & BAY & DAN & ADAN \\
\midrule
whole & 12.0 & 0.404 & 0.301 & 0.107 & 0.211 & 0.440 & 0.416 & 0.448 & 0.430 & 0.454 & 0.401 & 0.437 & 0.226 & 0.380 & 0.419 & 0.393 \\
2013  & 7.9  & 0.467 & 0.315 & 0.074 & 0.192 & 0.508 & 0.556 & 0.590 & 0.584 & 0.546 & 0.553 & 0.500 & 0.224 & 0.521 & 0.545 & 0.537 \\
2014  & 8.8  & 0.391 & 0.278 & 0.083 & 0.194 & 0.429 & 0.405 & 0.443 & 0.424 & 0.440 & 0.392 & 0.430 & 0.208 & 0.370 & 0.404 & 0.392 \\
2015  & 10.2 & 0.461 & 0.348 & 0.130 & 0.257 & 0.487 & 0.451 & 0.474 & 0.453 & 0.497 & 0.439 & 0.490 & 0.284 & 0.412 & 0.458 & 0.420 \\
2016  & 10.0 & 0.418 & 0.308 & 0.109 & 0.213 & 0.459 & 0.435 & 0.466 & 0.445 & 0.472 & 0.424 & 0.462 & 0.275 & 0.385 & 0.429 & 0.404 \\
2017  & 5.5  & 0.332 & 0.205 & 0.038 & 0.144 & 0.363 & 0.341 & 0.379 & 0.360 & 0.376 & 0.322 & 0.357 & 0.208 & 0.288 & 0.344 & 0.317 \\
2018  & 13.0 & 0.467 & 0.378 & 0.164 & 0.271 & 0.506 & 0.491 & 0.507 & 0.502 & 0.519 & 0.483 & 0.508 & 0.233 & 0.443 & 0.503 & 0.465 \\
2019  & 9.1  & 0.378 & 0.282 & 0.098 & 0.200 & 0.410 & 0.397 & 0.420 & 0.407 & 0.425 & 0.384 & 0.398 & 0.179 & 0.359 & 0.397 & 0.379 \\
2020  & 22.8 & 0.332 & 0.304 & 0.173 & 0.231 & 0.368 & 0.284 & 0.331 & 0.297 & 0.369 & 0.248 & 0.344 & 0.205 & 0.289 & 0.294 & 0.267 \\
\bottomrule
\end{tabular}%
}
\end{table}
Tables \ref{tab:loocv-outsample-10} and \ref{tab:loocv-outsample-30} report the annual average out-of-sample $R^2$ results for the 10- and 30-min sampling frequencies, respectively.
From Tables \ref{Table1}, \ref{tab:loocv-outsample-10}, and \ref{tab:loocv-outsample-30}, we find that the TED estimator consistently shows improved performance as the sampling frequency becomes higher.
In contrast, the performance gains from using higher-frequency data are relatively smaller for the benchmarks.
This may be because the TED estimator captures the time-varying structure more effectively as the sampling interval becomes shorter, while the benchmarks fail to adequately account for this time-varying behavior.

\begin{landscape}
\begin{figure}
    \centering
    \includegraphics[width=0.82\linewidth]{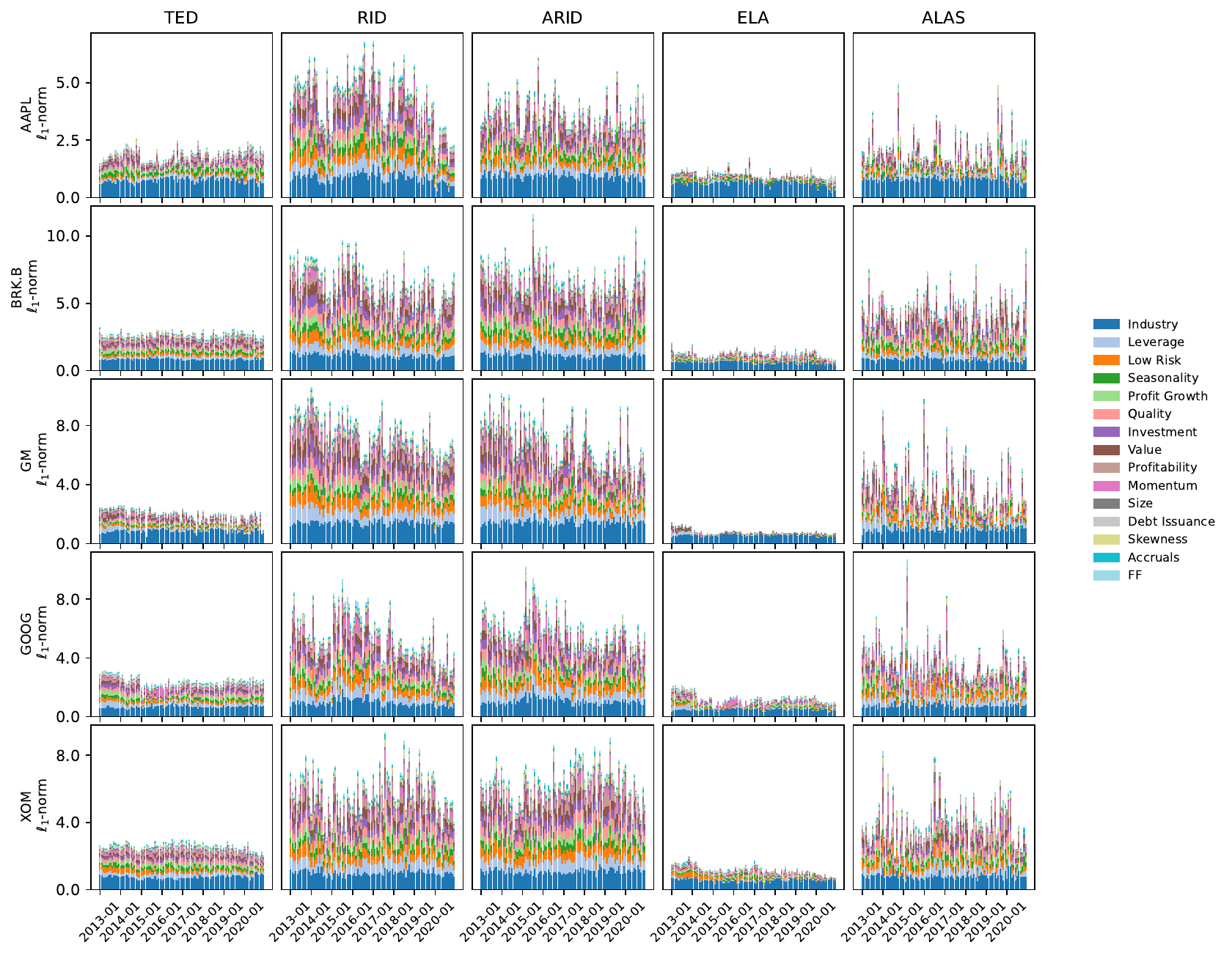}
    \caption{Stacked bar charts of the absolute sums of the monthly integrated coefficients from TED, RID, ARID, ELA, and ALAS within each of the 15 factor clusters for the five assets.
    The five estimators are ordered by their average out-of-sample $R^2$ in descending order.
    }
    \label{fig:top5stack}
\end{figure}
\end{landscape}

\begin{landscape}
\begin{figure}
\centering
\includegraphics[width=\linewidth]{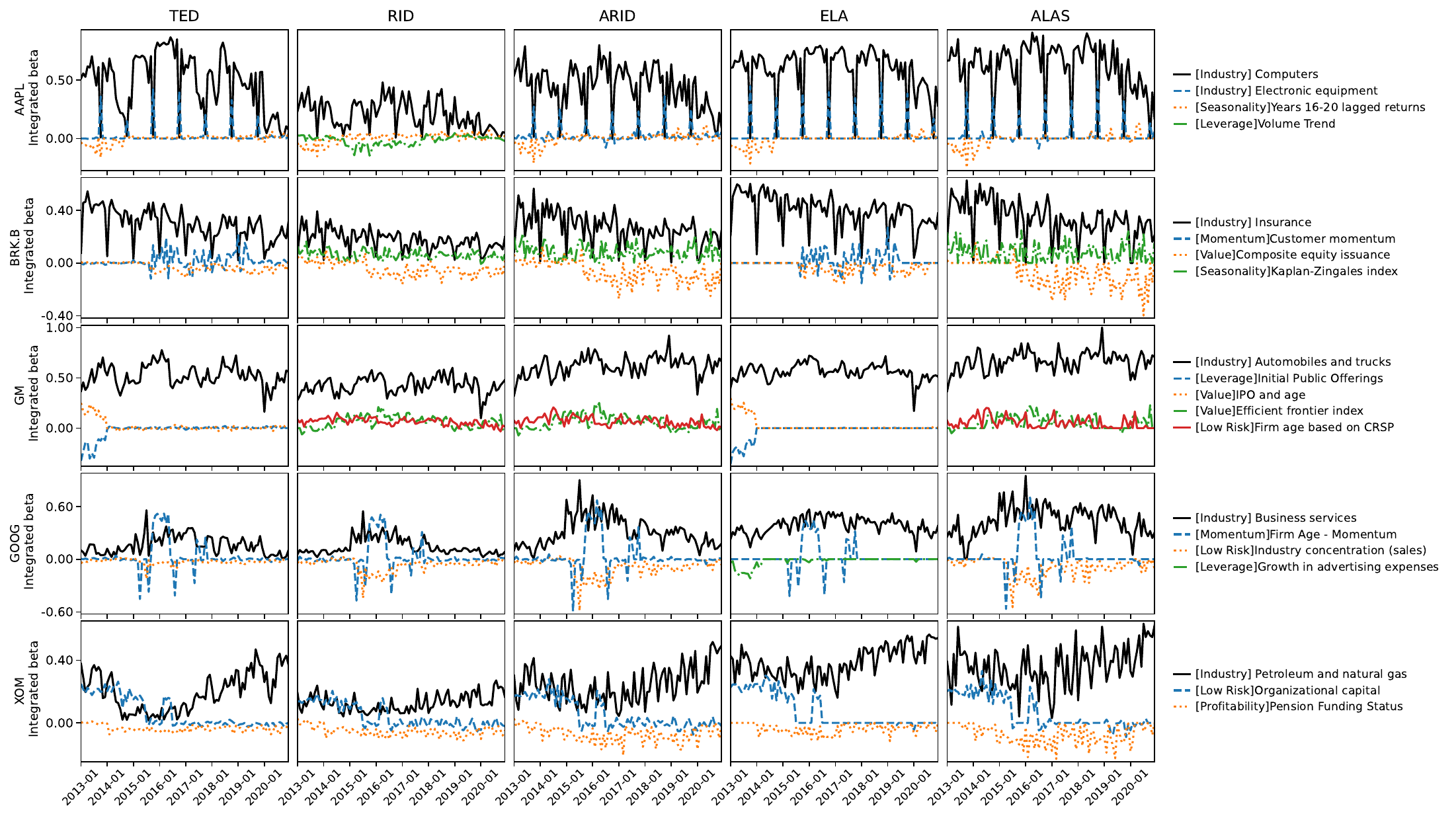}
\caption{Integrated coefficients from TED, RID, ARID, ELA, and ALAS for the three factors with the largest absolute sums of integrated coefficients over time for each of the five assets.}
\label{fig:top5ib}
\end{figure}
\end{landscape}

We compare the time-varying dynamics of the integrated coefficients from TED with those from benchmark estimators.
Specifically, we select the five estimators---TED, RID, ARID, ELA, and ALAS---that exhibit the highest average out-of-sample $R^2$ values in Table \ref{Table1}.
Figure \ref{fig:top5stack} displays the absolute sums of the monthly integrated coefficients from these estimators within each of the 15 factor clusters (see Section \ref{SEC-5}) for the five assets across time.
Figure \ref{fig:top5ib} plots the integrated coefficients of the three factors with the largest absolute sums of integrated coefficients over time for each of the five assets from these estimators.
From Figure \ref{fig:top5stack}, we observe that within each factor cluster, the absolute sums of the monthly integrated coefficients obtained from RID, ARID, and ALAS are substantially larger than those from TED and exhibit more pronounced variation over time.
For ELA, the absolute sum of the monthly integrated coefficients exhibits less variation over time compared to RID, ARID, and ALAS, and its magnitude is considerably smaller than that of TED.
From Figure \ref{fig:top5ib}, we observe that for factors with large integrated coefficients, RID, ARID, and ALAS produce results similar to TED; however, for factors with smaller integrated coefficients, they select different factors from those selected by TED.
ELA consistently selects nearly the same factors as TED---those with the largest integrated coefficients---for all assets except for GOOG, although the magnitudes of these coefficients differ.
From these results, we may conclude that, among the estimators with high out-of-sample $R^2$, RID, ARID, and ALAS tend to produce denser coefficients than TED, which leads to differences in the selection of prominent coefficients.
In contrast, ELA is close to TED in terms of selecting prominent coefficients but may not accurately estimate their magnitudes, which may lead to its lower out-of-sample $R^2$.

\begin{figure}[h]
\centering
\includegraphics[width = 0.9\textwidth]{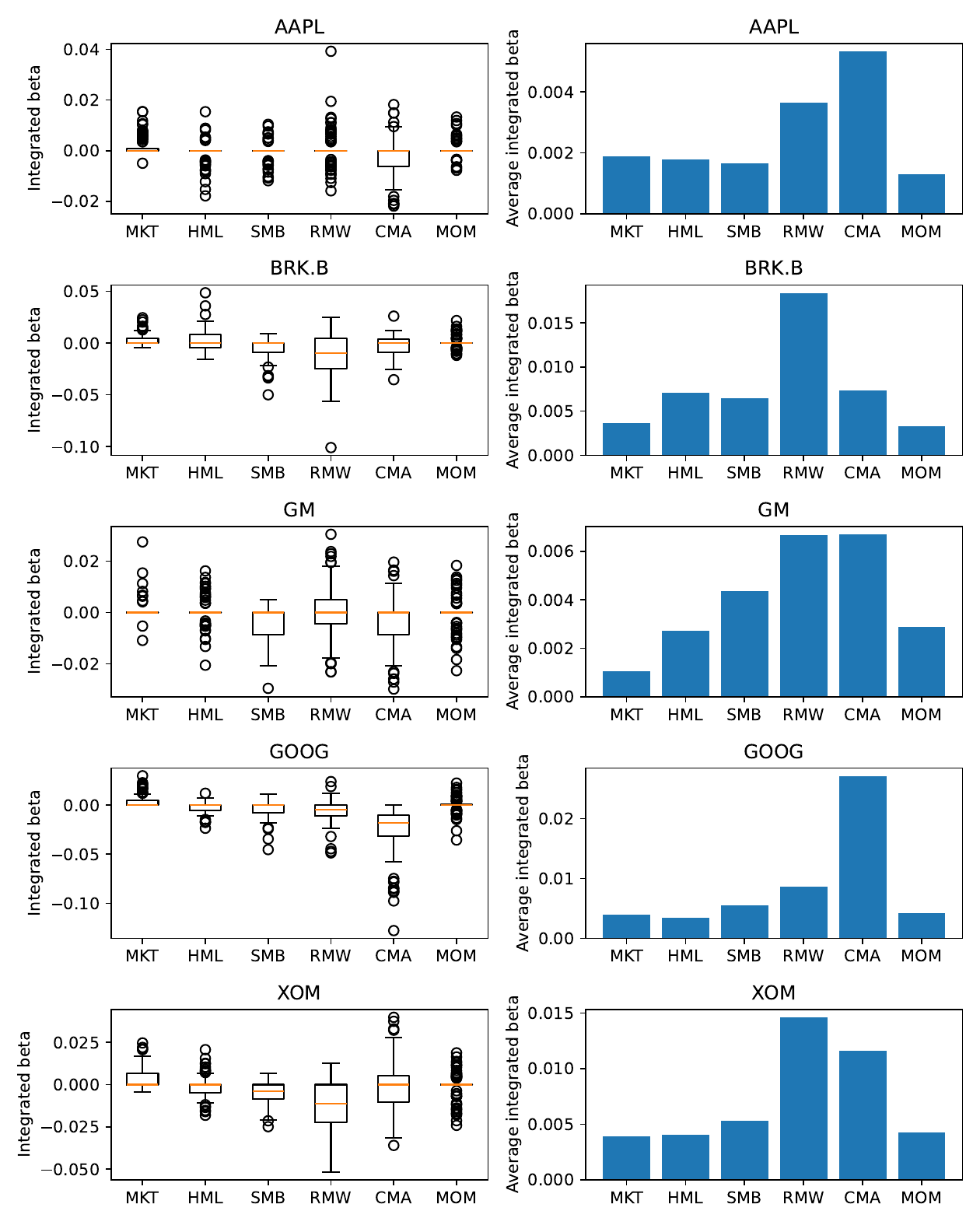}
\caption{Boxplots of the integrated coefficient estimates from the TED estimation procedure (left) and the average $\ell_1$-norm (right) for the six factors, MKT, HML, SMB, RMW, CMA, and MOM. 
}
\label{fig:emp_6factors}
\end{figure}

In financial practice, the six factors (the Fama-French five factors and the momentum factor) are among the most frequently used canonical asset pricing factors.
Thus, we investigate their integrated coefficient behaviors.
Figure \ref{fig:emp_6factors} depicts the estimates of the monthly integrated coefficients for the widely used MKT, HML, SMB, RMW, CMA, and MOM factors.
We find that these factors are generally not significant for most periods and assets.
This may be because the industry factors dominate the explanatory power of the six factors. %

\clearpage
\section{Notation table}

\begin{table}[!htp]
\caption{Definitions of notations.}
\label{tab:TED_notation}
\centering
\begin{tabular}{ll} \hline
Notation & Meaning \\ \hline
$p$ & dimension of covariate process \\
$n$ & number of high-frequency observations \\
$Y_t$ & dependent process \\
$\bX_t$ & $p$-dimensional covariate process \\
$\bbeta_t$ & time-varying coefficient process \\
$Z_t$ & residual process \\
$\bmu_t$ & drift process of $\bX_t$ \\
$\bsigma_t$ & volatility matrix process of $\bX_t$ \\
$\bSigma_t$ & instantaneous covariance, $\bsigma_t\bsigma_t^\top$ \\
$\bSigma_{XY,t}$ & instantaneous cross-covariance, $\tfrac{d}{dt}[\bX,Y]_t$ \\
$\nu_t$ & volatility process of $Z_t$ \\
$\bB_t$, $W_t$ & standard Brownian motions of $\bX_t$ and $Z_t$\\
$\bmu_{\beta,t}$, $\bnu_{\beta,t}$ & drift and diffusion processes of $\bbeta_t$ \\
$s_p$ & sparsity level (number of nonzero coefficients) \\
$\Delta_i^n A$ & increment $A_{i\Delta_n}-A_{(i-1)\Delta_n}$ \\
$\Delta_n$ & sampling interval  \\
$k_n$ & window size (observations per time-localized window) \\
$\lambda_n$ & tuning parameter for time-localized Dantzig selector \\
$\tau_n$ & tuning parameter for time-localized CLIME estimator \\
$h_n$ & thresholding level \\
$\hat{\bbeta}_{i\Delta_n}$ & time-localized Dantzig selector \\
$\tilde{\bbeta}_{i\Delta_n}$ & debiased instantaneous coefficient estimator \\
$\hat{\bOmega}_{i\Delta_n}$ & time-localized CLIME estimator \\
$I\beta$ & integrated coefficient \\
$\hat{I\beta}$ & estimator of integrated coefficient before thresholding \\
$\tilde{I\beta}$ & TED estimator \\
$s(\cdot)$ & thresholding function \\
$\delta$, $q$ & exponent in sparsity condition for $\bbeta_t$ and $\bOmega_t$\\ \hline
\end{tabular}
\end{table}

\end{spacing}
\end{document}